\pdfoutput=1

\documentclass [10pt,twocolumn]{IEEEtran}
\usepackage{algorithm,algorithmic,amsbsy,amsmath,amssymb,epsfig,bbm,mathrsfs,multirow,amsthm}
\usepackage[T1]{fontenc}
\usepackage[latin9]{inputenc}
\usepackage{amsmath}
\usepackage{fixltx2e}
\usepackage{algorithm}
\usepackage{array,multirow,graphicx}

\newtheorem{theorem}{Theorem}

\newtheorem{definition}{Definition}

\begin{document}

\title{Applications of Repeated Games in Wireless Networks: A Survey}
\author{\IEEEauthorblockN{Dinh Thai Hoang\IEEEauthorrefmark{1}, Xiao Lu\IEEEauthorrefmark{1}, Dusit Niyato\IEEEauthorrefmark{1}, Ping Wang\IEEEauthorrefmark{1}, and Zhu Han\IEEEauthorrefmark{2}}

\IEEEauthorblockA{\IEEEauthorrefmark{1}School of Computer Engineering, Nanyang Technological University, Singapore}	\\
\IEEEauthorblockA{\IEEEauthorrefmark{2}Electrical and Computer Engineering, University of Houston, Texas, USA}
}

\maketitle
\begin{abstract}
A repeated game is an effective tool to model interactions and conflicts for players aiming to achieve their objectives in a long-term basis. Contrary to static noncooperative games that model an interaction among players in only one period, in repeated games, interactions of players repeat for multiple periods; and thus the players become aware of other players' past behaviors and their future benefits, and will adapt their behavior accordingly. In wireless networks, conflicts among wireless nodes can lead to selfish behaviors, resulting in poor network performances and detrimental individual payoffs. In this paper, we survey the applications of repeated games in different wireless networks. The main goal is to demonstrate the use of repeated games to encourage wireless nodes to cooperate, thereby improving network performances and avoiding network disruption due to selfish behaviors. Furthermore, various problems in wireless networks and variations of repeated game models together with the corresponding solutions are discussed in this survey. Finally, we outline some open issues and future research directions.
\end{abstract}


\section{Introduction}
\label{sec:Intro}

Game theory is a branch of applied mathematics, which is used to study interactions among ``intelligent'' and ``rational'' participants (i.e., players) in a multi-agent decision making process (i.e., a game). In a game, the players want and are able to choose optimal strategies to maximize their benefits or payoffs. Game theory is widely used in many areas such as economics, computer science, military, evolution biology, and so on. In general, game theory can be divided into two main types, i.e., static and dynamic games. Static games model an interaction among players when they take actions only once in a single period. By contrast, dynamic games are applied when players take actions over multiple periods. In particular, the dynamic game is played repeatedly. Therefore, in the dynamic games, the players can observe the behaviors of the other players in the past and are able to adjust their strategies to achieve their goal. In this paper, we focus on one of the most important types of dynamic games, namely repeated games and their applications in wireless networks. 

Although there are some existing materials and resources presenting the applications of game theory in wireless networks such as~\cite{Han2012BookGame, Charilas2010Asurvey}, there exists no survey specifically for the repeated game models developed for wireless networks. This motivates us to deliver the survey with the objective to provide the necessary and fundamental information about repeated game models in wireless networks. Hence, through this article, the readers will understand how repeated games can be used to address different issues in wireless networks. Moreover, the rapid development of wireless networks brings many benefits for human beings; however, it also brings many challenges for researchers. The future wireless networks often have unique characteristics that differ from conventional wireless networks. For example, in mobile cloud computing~\cite{Hoang2013ASurvey}, we need to take not only the impact of wireless transmissions, but also cloud computing into account. As a result, in order to apply repeated games for problems in the next generation wireless networks, we have to consider the special characteristics of each network and then find models and appropriate solutions for using repeated games. This paper will be the first step to encourage researchers to explore applications of repeated games for future wireless networks.

\begin{figure*}[tbh]
\centering
\includegraphics[scale=0.7]{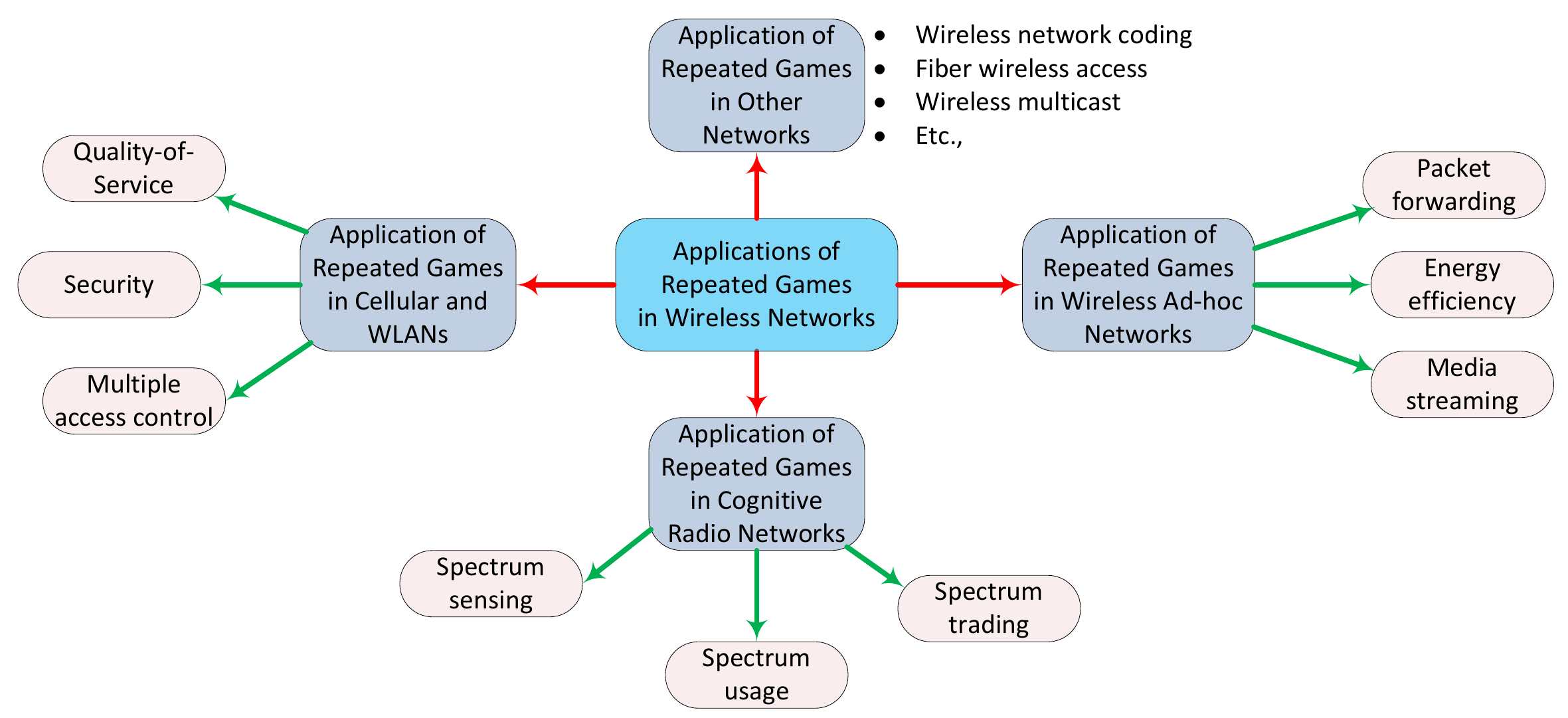}
\caption{Taxonomy of applications of repeated games in wireless networks.}
\label{fig:chap2_Taxonomy}
\end{figure*}

In addition, there are many advantages of using repeated games in wireless networks which will create favorable conditions for researches in this area. 
\begin{enumerate} 
\item 	Interactions among nodes and users in wireless networks often happen repeatedly over multiple time periods. Thus, the nodes are able to observe actions of their opponents in the past. Repeated games allow players to adjust their actions and adopt a certain strategy in response to other players' behaviors to optimize their long-term benefits which cannot be done through using a static game. 
\item 	In wireless networks in particular, the selfishness nature of players who aim to achieve their own objectives is common. Consequently, they will act responsively only to their interest without concerning about social welfare or network-wide performance. This can cause deleterious effects to all players involved. To encourage cooperation, we can impose rules and mechanisms (e.g., punishment) to self-enforce the players. Such rules and mechanisms can be modeled in repeated games in which the players are aware of potential benefits from cooperation through long-time interactions.
\item 	In wireless networks, communication is not perfect due to channel variations such as noise, fading, and signal attenuation. Additionally, network nodes are limited by hardware and energy supply. Repeated games offer a complete framework for handling noisy, incomplete and imperfect information about the network. Additionally, repeated games can support distributed decision making by using local information, thereby avoiding communication overhead and minimizing energy consumption.
\item 	Repeated games support diverse equilibrium solution concepts which are suitable for different requirements of wireless networks. System designers have choices to implement rules and mechanisms to achieve a desirable outcome of the games developed to resolve problems in existing and emerging wireless networks. In addition, any feasible solution can be theoretically maintained by the repeated game, which is proved by Folk's theorem. 
\item 	With the vigorous development of mathematical tools for solving complex problems of repeated games, we can apply many variations of repeated game models to address specific and characteristics issues of wireless communication environment. For example, to deal with problems with noise or imperfect information which are common in wireless environment, we can apply the model of repeated game under the noise or imperfect information as shown in the next sections. 
\end{enumerate}

In~Fig.~\ref{fig:chap2_Taxonomy}, we present the taxonomy of repeated game applications for wireless networks which is organized based on network models. Basically, we consider three major wireless networks, i.e., structured networks (i.e., cellular and wireless local area networks), unstructured networks (i.e., wireless ad hoc networks) and cognitive radio networks. Additionally, there are other wireless networks with unique characteristics which will be also introduced in this paper. The use of this taxonomy stems from the fact that different networks have distinct properties that lead to different problems and issues to be solved. 

The rest of this paper is organized as follows. Section~\ref{sec:AnOverview} introduces the background and basic of repeated games. In Sections~\ref{sec:CWLANs}-\ref{sec:CRNs}, the reviews of different existing work are given. We highlight trends and summarize current research in Section~\ref{sec:FutureResearch}. Moreover, we outline some open issues and present some research directions in Section~\ref{sec:FutureResearch}. Finally, we summarize and conclude the paper in Section~\ref{sec:Conclusion}.

\section{Fundamental of Repeated game}
 \label{sec:AnOverview}
The repeated game theory provides a formal framework to model a multi-player sequential decision making process and to explore the potential cooperation in the long run. In this section, we provide some preliminary of a repeated game and its variants. We then provide some comparison with other optimization and game theoretic approaches.

\subsection{Definitions and Fundamental Concepts}
The basic component of a repeated game is called the {\em stage game} $G$, which is a finite $N$-player simultaneous-move game in a strategic form, also called a normal form. The stage game can be represented by $<\mathcal{N}, (A_{i}), (u_{i}) >$ with a finite action space $A_{i}$, and payoff function $u_{i}$ for player $i \in \mathcal{N}=\{1,2, \ldots ,N\}$, where $N$ is the total number of players. In a repeated game, denoted by $G^{T}$, the players play the same stage game for $T$ rounds or periods (possibly $T \to \infty$). At the end of each round, the same structure repeats itself. In other words, the game in the current stage does not change the structure of the next stage. Usually, repeated games assume the availability of a common entity that produces a publicly observed signal, uniformly distributed and independent across periods\footnote{In this paper, we use round, period and game stage interchangeably.}, so that the perfect public information is available at all players. 

Repeated games can be broadly classified into two types, depending on whether the time horizon, i.e., the number of periods, is finite or infinite. The infinite time horizon game is suitable for the situations where players always presume the game to be played one more round with a high probability (i.e., without a known termination point). The finite time horizon game describes the situation where the information about termination of playing game is commonly known. 
 
Let $a^{t} \equiv \{a^{t}_{1},a^{t}_{2}, \ldots, a^{t}_{n}\}$ denote the actions used by all players in round $t$. Suppose that the game begins in round $t=1$, for $t \geq 2$, let $h^{t}=\{a^{1},a^{2}, \ldots ,a^{t-1}\}$ denote the actions used at all rounds before $t$, and let $H^{t}=(A)^{t}$ be the space of all possible round-$t$ histories. In each period, the players can  observe all history in each previous period. Let $\mathbb{A}_{i}$ be the space of probability distribution over $A_{i}$. The infinitely repeated game is formally defined as following.

\begin{definition} \label{infinite_def}
Let $A^{\infty}$ represent the set of infinite sequences of action profiles. An infinitely repeated game of a stage game $G$ is an extensive game with simultaneous form moves based on perfect information  $<N,H,P,(\succeq^{\star}_{i})>$, where $H=\{ \varnothing\}  \cup \big ( \bigcup^{\infty}_{t=1} A^{t} \big) \cup A^{\infty}$, $P$ is a profile that maps every non-terminal history $h \in H$ to each player,  $\succeq^{\star}_{i}$ is a preference relation on $A^{\infty}$ that satisfies the following notion of week separability: if $(a^{t}) \in A^{\infty}$, $a\in A$, $a^{\prime} \in A$, and $u_{i}(a) > u_{i}(a^{\prime})$, then for all $t$, we have $(a^{1},\ldots,a^{t-1},a,a^{t+1},\ldots) \succeq^{\star}_{i}$ $(a^{1},\ldots,a^{t-1},a^{\prime},a^{t+1},\ldots)$. 
\end{definition}
Note that in a repeated game, a strategy for player $i$ assigns an action $A_{i}$ for every finite sequence of outcomes in $G$.

Similarly, finitely repeated games with a fixed known time horizon $T$ is formally defined as following.

\begin{definition}
A T-round finitely repeated game of $G$ is an extensive form game of perfect information satisfying all the conditions of Definition \ref{infinite_def}, with $\infty$ replaced by $T$. The preferences can be represented by the mean payoff expressed as follows: $ \sum^{T}_{t=1}u_{i}(a^{t})/T$.

\end{definition}

The overall payoff of a player $i \in \mathcal{N}$ is a weighted average of the payoffs in each period, which is given by
\begin{equation}
u_{i}=(1-\delta_{i})\sum^{T}_{t=1}\delta^{t}_{i} u_{i}(a^{t}).
\end{equation}
The factor $(1-\delta_{i})$ normalizes an overall payoff so that the payoffs in the repeated game are on the same scale as in the stage game. The discount factor $\delta_i$ can be interpreted in two ways. Firstly, it denotes how much a future payoff is valued at the current period. Secondly, a player is patient about the future, but the game ends at any round with probability 1-$\delta_{i}$. 
 
\begin{definition}
In a repeated game, since all players observe $h^{t}$, a {\bf pure strategy} $s_{i}$ for player $i$ is a sequence of maps $s^{t}_{i}$. This maps possible round-$t$ history $h^{t} \in H^{t}$ to actions $a_{i} \in A_{i}$. A {\bf mixed strategy} $\sigma_{i}$ is a sequence of maps $\sigma^{t}_{i}$ from $H^{t}$ to mixed actions $\alpha_{i} \in \mathbb{A}_{i}$. 
\end{definition}
 
A repeated game is suitable for modeling long-term interactions and relationship among multiple players, and it can achieve a good equilibrium outcome. Unlike a single stage game, repeated interaction among players is induced in a repeated game which may lead to the cooperation instead of defection. This is due to the threat of potential punishment by other players in future rounds of the game. In particular, in the design of a repeated game, a strategy can be chosen to compensate the player for his instantaneous payoff loss due to cooperation, by rewarding the player later using an additional payoff that will be more than the current loss. Thus, punishment rules are required to avoid the player deviating from a socially optimum point.

\subsection{Equilibrium Concepts and Strategies in Repeated Games}

\subsubsection{Equilibrium Concepts}
In a normal-form stage game, each player is concerned only about its own payoff, and will choose the strategy accordingly. The most commonly used solution concept of the stage game is the Nash equilibrium.

\begin{definition}
A strategy profile $\sigma$ is a {\bf Nash equilibrium} if for every player $i$ and every strategy $\sigma^{\prime}_{i}$,
\begin{equation}
 u_{i}(a(\boldsymbol \sigma)) \geq u_{i}(a(\sigma_{-i},\sigma^{\prime}_{i}))	.
\end{equation} 
\end{definition}

At the Nash equilibrium, none of the players can improve its payoff by a unilateral deviation. The Nash equilibrium strategy is the best for a player if the others also use their Nash equilibrium strategies. However, the Nash equilibrium has some shortcomings. Firstly, there may be no Nash equilibrium in a game. Secondly, there may exist multiple Nash equilibria in a game. Therefore, some equilibrium may be better than other equilibria and the players have to choose one of them. Third, compared with the centralized optimization, the Nash equilibrium might have poor performance, which is called the Price of Anarchy. 

For an extensive-form repeated game, strategies are played repeatedly to guarantee the Nash equilibrium at any subgame of the initial state game. The outcome is the subgame perfect equilibrium.  
\begin{definition}
A strategy profile $\sigma$ is a {\bf subgame perfect equilibrium} if it is a Nash equilibrium, and for every history $h(t)$, every player $i$, and every alternative strategy $\sigma^{\prime}_{i}$, 
\begin{equation}
 u_{i}(a(\boldsymbol \sigma,h(t))) \geq u_{i}(a(\sigma_{-i},\sigma^{\prime}_{i},h(t)))  .
\end{equation} 
\end{definition}
For a repeated game with a finite number of stages, backward induction is commonly used to obtain the subgame perfect equilibrium. 

Typically, there are more than one equilibrium in the game. Therefore, it is important to define and evaluate a preference among equilibrium outcomes. A common approach is to measure an optimality, which is called Pareto optimality. The Pareto optimality is the payoff profile that no strategy can make at least one player better off without making any other player worse.

\begin{definition}
Let $\Sigma \subseteq \mathbb{R}^{N}$ be a set of possible payoffs. For $\boldsymbol \sigma$, $\boldsymbol \sigma^{\prime} \in \Sigma$, if $\exists$ $\sigma^{\prime}_{i} > \sigma_{i} $ and $\nexists$ $\sigma^{\prime}_{i} < \sigma_{i} $, then  $\boldsymbol \sigma^{\prime} $ Pareto dominates  $\boldsymbol \sigma$. Then $ \boldsymbol \sigma  \in \Sigma$ is {\bf Pareto optimal} if there exists no $\boldsymbol \sigma^{\prime} \in \Sigma$ for which $\sigma_{i}^{\prime}> \sigma_{i}$ for all $i \in \mathcal{N}$. And $\boldsymbol \sigma \in \Sigma$ is {\bf strongly Pareto optimal} if there exists no $\boldsymbol \sigma^{\prime} \in \Sigma$ for which $\sigma_{i}^{\prime} \geq \sigma_{i}$ for all $i \in \mathcal{N}$ and $\sigma_{i}^{\prime}>\sigma_{i}$ for some $i \in \mathcal{N}$.

\end{definition}  
Pareto optimality serves as the approach to restrict a large set of repeated game equilibria when the players are patient. The {\em Pareto frontier} is referred to as the set of all $ \boldsymbol{\sigma} \in \Sigma$ that are {\em Pareto optimal}. For the case with multiple equilibrium outcomes, the ones on the Pareto frontier are superior to others.

\begin{table*}
\footnotesize
\centering
\caption{\footnotesize  Common Strategies in Repeated Games.} \label{tab:strategy}
\begin{tabular}{|l|p{15cm}|} 
\hline
\footnotesize  Strategy & Description \\
\hline
Always cooperate & Cooperates on every move \\
\hline
Always defect  & Defects on every move \\
\hline
Random play & Random (defect/cooperate) on every move \\
\hline
Grim & Cooperate until one of others defects then defect forever  \\
\hline
Tit-for-tat	(TFT) & Start by cooperating and repeatedly select the last strategy played by the opponent \\
\hline
Generous TFT & Same as TFT, except that it cooperates with a probability $q$ when the opponent defects.  \\
\hline
Tit for two tats & Cooperates on the first move and defects only when the opponent defects two times \\
\hline
Two tits for tat	 & Same as TFT except that it defects twice when the opponent defects \\
\hline
Suspicious TFT & Same as TFT, except that it defects on the first move \\
\hline
Contrite TFT & Same as TFT when no noise. However, in a noisy environment, once it receives a wrong signal because of error, it will choose cooperation twice in order to recover mutual cooperation. \\
\hline
Adaptive strategy	 & An adaption rate r is used to compute a continuous variable ``world'' according to the history moves of the opponent.  \\
\hline
Cartel maintenance & In this strategy, players first compute a cooperation point that has better payoff than Nash equilibrium points. If any player deviates from the cooperation while others still playing the cooperative strategy, all other players will defect in next game stages.  \\
\hline
Forgiving strategy	 & A player starts with cooperation and he plays cooperation (C) as long as everybody played C in this past. If anybody plays defection (D) at any period, then the player plays D for the next k periods. After k periods of punishment he returns to playing C until someone deviates again.  \\
\hline
Cheat-proof &  In game theory, an asymmetric game where players have private information is said to be cheat-proof (of truthful mechanism) if there is no incentive for any of the players to lie about or hide their private information.  \\
\hline
\end{tabular}
\end{table*}

\subsubsection{Strategies}
The above equilibrium concepts describe the properties of an equilibrium outcome, but they do not address the problem how to reach the equilibrium outcome of a game. Indeed, as the repeated games are the extensive form of the stage games, the strategies in the repeated games are highly dependent on the best actions in the stage games. We therefore introduce some principles in the design of strategies for repeated games.  

A simple approach is to check whether a given strategy profile of a repeated game constitutes a subgame perfect equilibrium. This is {\em one-shot deviation principle}. 

\begin{theorem}
{\bf One-Shot Deviation Principle}: 
A strategy profile for a finitely repeated game is a subgame perfect equilibrium if and only if there is no history such that the player can increase its payoff in the subgame following that history by choosing a different action at the beginning of the subgame while leaving the remainder of its strategy unchanged.
\end{theorem} 
By applying the one-shot deviation principle, a remarkable reduction in complexity of finding equilibrium strategies can be achieved.

For finitely repeated game, the following theorem holds.  
\begin{theorem}
If the stage game $G$ has a unique Nash equilibrium $a^{*}$. Then, for any $T$, the unique subgame perfect equilibrium in the T-horizon repeated game $G^{T}$ is given by the strategy profile in which every player after every non-terminal history chooses $a^{\star}_{i}$. If the state game $G$ has multiple Nash equilibria, then any outcome $h^{T} = (a_{1},\ldots,a_{T})$, in which for every t ($1 \leq  t \leq T$) the action profile $a^{t}$ is one of the Nash equilibria of $G$, can be obtained as the outcome of the subgame perfect equilibrium of $G^{T}$.
\end{theorem}

For infinitely repeated game, the following two theorems hold. 
\begin{theorem} 

{\bf Nash Folk Theorem}: For every feasible payoff vector $v$ with $v_{i} > v_{-i}$ for all players $i$, there exists a discount factor $\delta < 1$ such that for all $\delta \in (\delta,1)$ there exists a Nash equilibrium of $G(\delta)$ with payoffs $v$.
\end{theorem}
This theorem indicates that when the players are patient enough, any one-stage gain with a finite value is outweighed by even a small loss in payoff in every future round.

\begin{theorem}

{\bf Subgame Perfect Folk Theorem}:
Let $a^{\star}$ be a static Nash equilibrium of the stage game with
payoffs $v^{\star}$. For any feasible payoff $u$ with $u_{i} > v_{i}^{\star}$, for all $i \in \mathcal{N}$, there exists some $\underline{\delta} < 1$ such that for all $\delta > \underline{\delta}$, there exists the subgame perfect equilibrium of $G^{\infty}(\delta)$ with payoffs $u$.
\end{theorem}

The subgame perfect Folk theorem shows that any payoff above the static Nash equilibrium payoffs can be the payoff of the subgame perfect equilibrium of the repeated game.

Typically, repeated games employ trigger strategies to incentivize cooperation and punish any defecting player if a certain level of defection (i.e., the trigger) is observed. Let $a^{D}_{i}$ denote the defection payoff of player $i$, given that other players play the equilibrium strategy, $a^{\star}_{-i}$. Let $a^{P}_{i}$ represent the punishment payoff of player $i$. Generally, we have $a^{D}_{i}>a^{\star}_{i}>a^{P}_{i}$. To design the trigger strategy for a repeated game, the following condition should hold, i.e.,
\begin{equation}
\frac{a^{\star}_{i}}{1-\delta} \geq a^{D}_{i} + \frac{\delta a^{P}_{i}}{1-\delta},  
\end{equation}
which gives, 
\begin{equation}
\delta \geq \frac{a^{D}_{i}-a^{\star}_{i}}{a^{D}_{i}-a^{P}_{i}}	.
\end{equation}
Note that the level of punishment and the sensitivity of the trigger vary with different trigger strategies. In Table~\ref{tab:strategy}, we have summarized some of the well-known strategies in repeated games applied to wireless networks.

\subsection{Variations and Extensions}

Next, we briefly discuss some variations and extensions of a repeated game that are often used in wireless networks.

\subsubsection{Repeated Game under Noise}

The above introduced repeated game can be regarded as the game played in a noise-free environment. Namely, all direct or indirect observations in the repeated games are assumed to be correct. However, in reality, there exist errors in observing other's behavior by the players. A repeated game under noise is an alternative tool that can model the situations when the players have wrong knowledge about their opponents. In particular, two types of error could occur during the iterations of play~\cite{A.1995Nowak}. The first type of error, namely a perception error, accounts for the wrong observation of other players. The second type of error explains the wrongly taken action instead of the intended one due to the interference in the environment, which is called an implementation error. For example, a relay node in a multihop network may misunderstand the behavior of other nodes in the same network, i.e., due to the perception error, if the relay node makes the decision to take a cooperation action (i.e., to relay packets of other nodes). However, due to temporary poor channel condition, the packets may not be successfully forwarded. Inadvertently, this is possibly observed as a selfish defection instead of a cooperative relay action. This is the implementation error.

Usually, noise appears in the form of random errors, which consequently makes the possible outcomes in each stage unpredictable. Thus, the players can only make responses based on expected stage payoffs. The existence of noise in an environment considerably increases the complexity of the repeated games. Moreover, cooperation becomes much more difficult to maintain~\cite{Bereby-Meyer2006}. Typically, there exist three approaches to address the noise issues in a repeated game~\cite{Wu1995How}. The first approach is called ``generosity'' which lets players tolerate some noncooperative behavior and with some certain degree the players do not punish a deviating player. The second approach is called ``contrition''. This approach allows contrition in a reciprocating strategy to avoid responding to the other player's defection aroused by its own unintended action. The third approach, namely ``win-stay, lose-shift'', adopts the strategy that repeats the same action if the latest payoff is high, but changes the action otherwise.

\subsubsection{Repeated Game with Imperfect Public Monitoring} 

In repeated games with imperfect public monitoring, players cannot directly observe the other players' strategies, but can observe imperfect and public signals about them. The players' information is a stochastic public signal, the distribution of which is dependent on the strategy profiles chosen by the players. Similar to perfect public monitoring, there is also a recursive structure in the case with imperfect public monitoring. However, there is often no proper subgame, and thus a subgame perfect equilibrium may not be effective. To address this problem, the concept of {\em perfect public equilibrium}~\cite{Abreu1990Toward} has been introduced. In this concept, the players can make {\em public strategies} based on {\em public history}, which is a sequence of realization of the public signal. The Folk theorem for imperfect public information has also been proposed in \cite{Fudenberg1994thefolk}. 

\subsubsection{Repeated Game with Imperfect Private Monitoring} 

Repeated games with imperfect private monitoring deal with the situations that the public information of players are not openly available, and each player can only obtain imperfect private information about the other players through its own direct or indirect monitoring. ,The difficulties associated with private monitoring lie in two aspects~\cite{M.2002Kandori}. Firstly, the games lack recursive structure which yields the equilibria that do not possess a simple characterization. Secondly, at each round, players must conduct statistical inference on what others are about to do. There are mainly two model settings that bypass the above two difficulties. One is called {\em No Discounting Model}, or {\em $\delta$-Rationality}, in which there is no discounting loss, i.e., $\delta$=1, or the discounting loss in the average discounted payoff is tolerated. Another one is {\em Communication Model}~\cite{O.1998Compte}, which introduces communication in the repeated games. At each stage, the players are asked to reveal their private signals, but they can tell a lie if that is beneficial. By constructing equilibria where a player's report is used to force other players, truth-telling can be used among players and equilibrium strategies can be devised based on the publicly observable history of communication.

\subsubsection{Repeated Game with Incomplete Information} 

In a repeated game, when some players lack information of the others, they are said to be with {\em incomplete information}, and their games are accordingly called repeated games with incomplete information. This type of game is developed to capture widely existent situations in which a variety of features of the environment may not be commonly known by all the involved players. Unlike a repeated game with imperfect (public or private) monitoring, the player with incomplete information might not have common knowledge of the followings: i) payoffs of himself and other players; ii) who/what types the other players are; iii) what actions are possible for himself and other players; iv) how the action affects the outcome; v) what are the preferences of the other players; vi) what the other players know about what he knows. 

A variant for Folk theorem has also been introduced in \cite{Fudenberg1986Thefolk} for a repeated game with incomplete information. It is proven that any payoffs that Pareto-optimality dominates the Nash equilibrium can be sustained at an equilibrium of a finitely repeated game with incomplete information. With incomplete information, the {\em Bayesian Nash equilibrium} is a general solution concept~\cite{Francoise_Forges}. In a Bayesian repeated game, the players make the best response action based on the beliefs about the other players' strategies. During the repetition of the game, the players can iteratively update their beliefs through learning~\cite{H.1997Nachbar}.

\subsection{Comparisons with other tools}

We provide some brief comparison among a repeated game and other commonly used optimization or game approaches.

\subsubsection{Markov decision process (MDP)}

An MDP~\cite{puterman94} is an approach to make optimal sequential decisions under uncertainty. It is designed for a player making a decision based on a state and interacting with a system or environment. An MDP model is composed of a state, action, and reward. The policy is a mapping of a state to an action. When an action is taken at any state, a player receives a reward. An optimal policy is to maximize long-term reward, which could be discounted or average reward~\cite{J.1993White}. However, the MDP is for decision making of a single player. An interaction among multiple players over multiple time periods cannot be modeled using an MDP.

\subsubsection{Stochastic game}

Stochastic game~\cite{A.2003Neyman} is a generalization of an MDP with some applications in wireless networks. Strategies in stochastic games are made based on the history statistic of the interactions. Specifically, in each round, a player makes decisions based on a competitive policy. After the action is taken, the current state transits to the next state. By contrast, in a repeated game, there is no need for a state transition. Additionally, a player's strategy is not necessarily dependent on the past values of its opponent's randomizing probabilities. Instead, it can depend only on the past values of its opponent's payoff. This major difference differentiates the applications of repeated games and stochastic games.

\subsubsection{Coalition formation game}

A coalition formation game \cite{W.2009Saad} is a type of a cooperative game that models an interaction of players by forming coalitions to improve their individual payoffs. The strategies for forming a stable coalitional structure among the players can be broadly classified into two types: myopic and far-sighted. The former allows the players to adapt their strategies given the current state of the coalition, while the latter lets the players make their strategies by learning, and predicting future strategies of the other players. A repeated game is similar to the far-sighted coalition formation game in that both games capture long-term payoffs. The main difference is that a repeated game can model different strategies (e.g., punishment), not necessary cooperation. 

\subsubsection{Differential game}

A differential game~\cite{basar1999} is an extension of an optimal control framework which aims to find an optimal dynamic control strategy for the system with multiple agents and single agent, respectively. In differential games, the strategy is a continuous function of time. The solution of differential games can be the open-loop or the close-loop Nash equilibrium. Again, unlike repeated games, the differential game lacks a trigger and punishment cannot be implemented. Furthermore, differential games need an ordinary differential equation and the utility function is typical linear quadratic, which limits the scope of applications a lot.

\section{Applications of Repeated Games in Cellular and Wireless Local Area Networks}	
\label{sec:CWLANs}
In this section, we review repeated game models developed to solve problems in cellular and wireless local area networks (WLANs). The repeated games have been used to address the following issues.
\begin{itemize}
\item 	\textbf{Multiple Access Control}: In cellular and WLANs, wireless nodes need to connect to common access points, and this leads to a contention in accessing a common radio resource. Repeated games are used to control the access and also enhance network performance through encouraging the nodes to cooperate. 
\item 	\textbf{Security}: Instead of competing with each other which can cause the damage to all nodes, wireless nodes can choose a cooperation solution to improve their benefits. However, this cooperation is based on mutual trust, and thus if a player is selfish or malicious, the partnership may be severed. Repeated games are used to detect deviations and enforce players into the cooperation. 
\item 	\textbf{Quality-of-service (QoS) Management}: Service providers need to manage the relation with other partners, e.g., customers or other providers to optimize their profits while guaranteeing QoS for the customers. Repeated games are used by the service providers to manage the interaction with other partners to minimize the cost and maximize the revenue.
\end{itemize}

In the following, we provide the reviews for related work based on aforementioned issues.


\subsection{Multiple Access Control}
\label{subsection:MAC}
\begin{figure}[htb]
\centering
\includegraphics[scale=0.7]{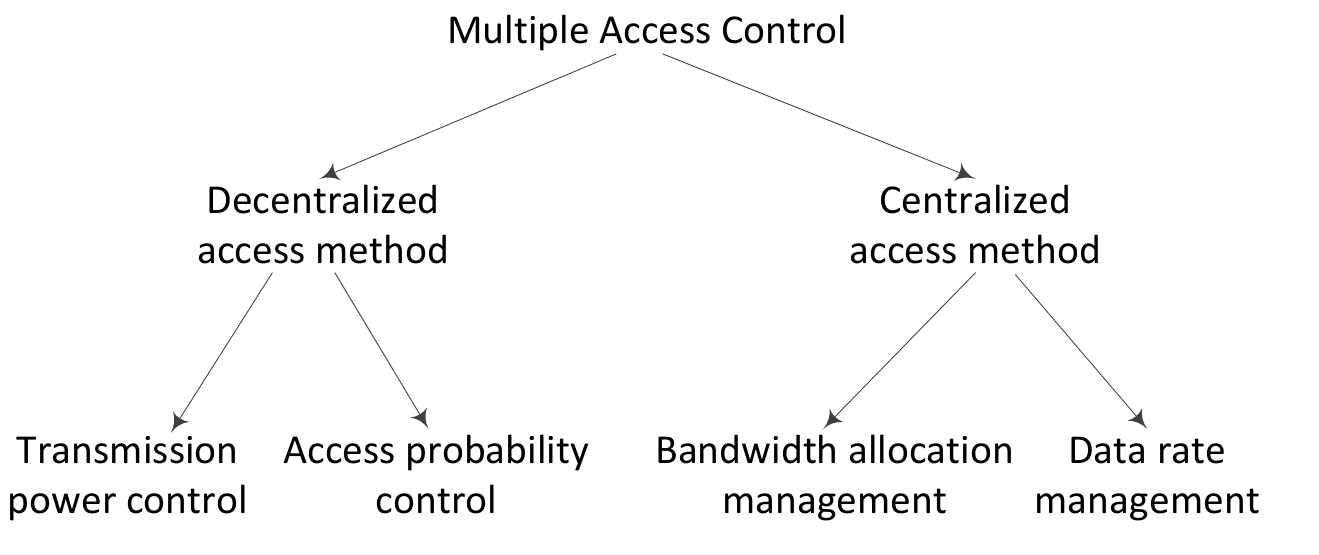}
\caption{Solutions for the multiple access control problems in cellular and WLANs.}
\label{fig:chap3_MAC}
\end{figure}

In cellular and wireless local area networks (WLANs), wireless users communicate with each other or with the Internet through access points and base stations. Therefore, one of the challenges is the access control for multiple users. This section will review the repeated game models for multiple access control problems under two access methods: decentralized and centralized access. Fig.~\ref{fig:chap3_MAC} outlines the corresponding approaches.

\subsubsection{Decentralized access method} 

\paragraph{Transmission power control}

MacKenzie~\cite{MacKenzie2004Gametheory} is one of the first pioneers in applying repeated games to power control problems in wireless networks. There are multiple users sharing the same spectrum allocated by a base station. In the repeated game, the players are the users. Each player chooses a non-negative power level to transmit signals to the base station. The payoff of each player is the number of bits successfully transmitted per time slot. The authors introduced the trigger strategy in which at the beginning of the game the players cooperate by transmitting signals at the desirable received power. If anyone uses more energy, which causes interference to the other users, in the next time slot all other users will increase their power levels to the noncooperative Nash equilibrium to punish the deviating player. Such a punishment lasts for one shot-game after the deviating user is detected. After that, the users cooperate again by using mild power levels. 

Similarly, Han~\textit{et al.}~\cite{Han2004Dynamicdistributed} also considered the power control. The players are the nodes, but the actions are transmission rates instead of power levels as in~\cite{MacKenzie2004Gametheory}. The payoff for each player is a function of profits obtained for each successful transmission minus the cost for link usage. To encourage players to cooperate in the repeated game, the authors used the Cartel maintenance strategy~\cite{Zhong2003Sprite}. Specifically, at an initial time, the players will transmit data at a cooperative transmission rate predeterminedly agreed. The players then compute the successful transmission probability and compare it with a threshold. If the computed successful transmission probability is lower than the threshold, they choose the noncooperative strategy in the next few game stages. Afterward, all the players will re-cooperate. To determine the optimal parameters, the authors used the policy gradient method as presented in~\cite{Zhong2003Sprite}. Under the proposed strategy, it is proven in~\cite{Han2008Acartel} that the solution of this game is a perfect public equilibrium~\cite{Fudenberg1991GameThery}.

Different from~\cite{MacKenzie2004Gametheory} and~\cite{Han2004Dynamicdistributed} where wireless users can transmit data to the base station simultaneously, Auletta~\textit{et al.}~\cite{Auletta2008Interferencegames} assumed that there is at most one successful transmission at each game stage. The player successfully transmits data if and only if its signal strength is higher than the total signal from other players and the noise. Since there is at most one player successfully transmitting data, the authors applied the alternating transmission strategy. At each game stage, only one player is allowed to transmit. If the player transmits data, but the transmission fails, i.e., someone deviates from the cooperation, the punishment will be implemented in the next few stages. In the punishment phase, all the players will transmit data with the highest power level. 

In the above work~\cite{MacKenzie2004Gametheory},~\cite{Han2004Dynamicdistributed}, and~\cite{Auletta2008Interferencegames}, the channel between nodes and the base station is assumed to be constant over time. However, in some cases, this assumption is inapplicable. Therefore, the authors in~\cite{Treust2010Implicitcooperation} investigated two cases with channel variations, namely, fast power control (FPC) and slow power control (SPC). In FPC, the channel gain is constant over the game stage. In SPC, the channel gain can be varied over the game stages. The payoff is a ratio of the transmission rate multiplied with the successful transmission rate and the transmission power level. It is shown that the static one-shot game has a non-saturated Nash equilibrium. The authors also proposed a trigger strategy for the nodes and it is proven that the repeated game has a subgame perfect equilibrium. The proof for the existence of the equilibrium can be found in~\cite{Treust2010Arepeated}. To detect any deviating player, the authors proposed the following detecting mechanism based on the public signal. If players cooperate, the public signal will be constant. However, if someone deviates, the public signal will not be constant, and thus players will change to noncooperative strategy. 

For the channels with SPC, since the channel gain can vary, the payoff of each player depends not only on the joint actions, but also on the channel gains in each game stage. The repeated game therefore becomes a stochastic repeated game (SRG) that is modeled through the irreducible state transition probability. In SRG, the Folk theorem is no longer useful since the stage game changes over time, and thus the authors applied the extended theory of the Folk theorem which is proven for the stochastic games with public information~\cite{Horner2010Recursive}. With the extended Folk theorem, it is demonstrated that there exists a perfect public equilibrium strategy of the stochastic game. 

\paragraph{Access probability control}

Cagalj~\textit{et al.}~\cite{Cagalj2005Onselfish} studied the multiple access problem in CSMA/CA networks. The CSMA/CA networks operate based on the assumption that the users are honest, and they follow the protocol predefined in a standard (e.g., IEEE 802.11 DCF~\cite{LANMANstandards1999}). However, selfish users can modify the rules defined by the protocol to receive more benefits. Consequently, the honest users will experience poor performance and unfairness in using a common radio resource. In the repeated game, the players are wireless users, their actions are selections of channel access probabilities, and the payoff function is achievable throughput. The authors developed the distributed learning algorithm for the users to adjust their behaviors such that their strategies converge to the Pareto optimal point. However, with the proposed algorithm, the player may not follow the designed rules and deviate from the cooperation. Thus, the authors proposed the detection and penalizing mechanism. To detect deviating users, the authors assumed that the users are able to measure the throughput of all other users in the network. Thus, if any user has its throughput that is different from other users, it will be treated as a deviating player. Then, the penalizing mechanism is executed that the packet transmitted by the deviating user will be jammed by the other users in the cooperation.

While the solution proposed in~\cite{Cagalj2005Onselfish} requires the modification in the MAC layer of wireless nodes, the proposed approach in~\cite{Konorski2006Agame} can protect wireless users from selfish users without any change to the MAC protocol. With the same players, their actions are the selections of backoff configurations. The authors defined three configurations for the nodes, i.e., to choose a standard backoff configuration predefined by IEEE 802.11, a greedy configuration which will access the channel once a collision happens, and a selfish configuration that will access channel after a collision happens in two time slots. The payoff is the throughput. The payoff at every game stage is equally weighted (i.e., there is no discounting for future payoffs). Thus, the payoff can be quantified by the liminf-type asymptotic~\cite{Knoblauch1994Computable}. The authors then proposed a cooperation strategy via randomized inclination to selfish/greedy play (CRISP) to detect and punish selfish nodes. By using CRISP, all the strategies of the nodes are the subgame perfect equilibrium. However, the condition for the existence of the subgame perfect equilibrium is that the payoff function of players must be liminf-type asymptotic~\cite{Knoblauch1994Computable}. 

\subsubsection{Access management based on a central node} 
In the following, we will review two schemes for network access through using a central node. In the first scheme, the central node will decide the amount of bandwidth that each node uses. In the second scheme, the central node will indicate the transmission rate for each node. 

\paragraph{Bandwidth allocation management}

The authors in~\cite{Riedel2006Achieving} examined a non-monetary mechanism introduced in~\cite{Liao2003Wireless} for the bandwidth allocation problem with the aim to achieve the Pareto optimal equilibrium. The authors adopted the concept of service purchasing power which is interpreted as a ``constant budget''. The purchasing power is determined based on the closed contrast between the customer and the service provider. Based on the users' service purchasing power, the base station will calculate the allocated bandwidth for each customer given a predefined rule. In the repeated game, the players are the nodes and they can select two actions, namely, cooperate or defect. The player defects if it uses the maximal allocated bandwidth, while the player cooperates if it utilizes the allocated bandwidth only if its marginal utility exceeds marginal social costs. To enforce the players to cooperate, the tit-for-tat (TFT) strategy is applied. To detect and punish the defecting user, the base station will monitor the users' behaviors and compute their weighted ratios. If the player is cooperative, the base station will improve the purchasing power of this player. Consequently, this player may receive more bandwidth in the next game stage. By contrast, if a player defects, the base station will reduce the purchasing power, and thus this player will have less allocated bandwidth. 

Unlike in~\cite{Riedel2006Achieving}, the authors in~\cite{Tran2010Onselfish} considered the bandwidth allocation based on the traffic demand of users. The mobile users first declare their traffic demands to the base station, and then the base station will decide the amount of bandwidth allocated to the users. However, the users may report fake demand, and thus repeated games are used to prevent selfish users and enforce them into cooperation. The authors applied a cost function in the payoff of the players. Specifically, when the users submit traffic demand, the base station will charge a fee depending on the demand. The payoff is the profit obtained from transmitting packets successfully minus the cost. With the proposed truthful mechanism, it is shown that the selfish users cannot gain higher payoff than that of honest users, and hence the selfish users have no incentive to deviate from truthful declaration.

While the bandwidth allocation problem is solved based on the contract as in~\cite{Riedel2006Achieving} and the traffic demand as in~\cite{Tran2010Onselfish}, in~\cite{Hajj2011SIRA} the decision relies on the channel condition reported from users. After the users send channel state information to the base station, the base station decides the leading player on each channel based on the signal-to-noise ratio (SNR) of the users. Apparently, weak users with a poor channel condition may have no opportunity to transmit/receive data. Therefore, repeated games is used to allow the user to cooperate and bring opportunities for weak users. In this game, users who achieve high profit but do not cooperate will be punished by adding cost in the next game stage. Thus, selfish users will have no incentive to deviate. 

\paragraph{Data transmission rate management}

The authors in~\cite{Kong2010Efficient} studied the packet scheduling problem in a wireless mesh network in which Mesh clients communicate with a mesh router. The mesh clients report their channel conditions to the mesh router. Based on this information, the mesh router selects and allows one mesh client to transmit in the current time slot. However, mesh clients could be selfish and report bogus channel information to the mesh router. Consequently, the selfish client can gain benefits while the overall network performance will be adversely affected. To address this problem, the authors proposed a trigger strategy, namely ``Striker,'' to detect and punish selfish users when they report fake information. In this strategy, if the player is detected to defect, all other players will report the highest channel condition to the mesh router in the next game stage. This punishment lasts for a certain number of periods. To detect the defector, each mesh client will measure the average data rate of other mesh clients and compare with a predefined threshold. 

In~\cite{Shen2014Universalnon}, the authors studied the allocation that allows multiple users to transmit data to the access point simultaneously. The authors built the framework for multiple accesses by proposing two new entities, namely, the \textit{regulator} and the \textit{system optimizer}. The users send their service requirements to the \textit{regulator} and report channel state information to the \textit{system optimizer}. Based on the received information, the \textit{regulator} will compute the pricing parameter, and the \textit{system optimizer} will determine and send the optimal power allocation to the users. The selfish user can gain benefits by transmitting misinformation about its channel state to the \textit{system optimizer}. Therefore, a trigger strategy is applied at the \textit{system optimizer} to detect and punish the selfish user. For detection, after observing the actual rates of the users and comparing them with the reported channel, the \textit{system optimizer} can identify the cheater that will be removed from the cooperation. With this strategy, the honest users always gain benefits, while the selfish user will be punished and receive low overall payoff. However, the framework cannot apply when there are multiple cheaters. 

In the same context, the authors in~\cite{Lai2008Thewater} also studied the data transmission rate control problem. However, the base station is also treated as a player in the game. Therefore, the players are wireless nodes and the base station. The actions of the wireless nodes are the transmission power levels and the action of the base station is the decoding order. The players make their decisions simultaneously at each game stage. The payoff of the wireless nodes is the achievable rate, while that of the base station is the revenue that the wireless nodes pay per unit rate. The authors then designed the strategy to enforce nodes to cooperate. At the beginning, the base station announces its rate reward vector to the players, and then based on the rate reward each player determines the optimal rate control that maximizes sum rate achievable by using the optimal centralized control policy as presented in~\cite{Tse1998Multiaccess}. If the player is detected to be a deviator, the base station will decode for this player first for some periods and other players will do water-filling algorithm during these periods. It is proven that the achievable transmission rate of the deviating node will be reduced, while that of others will be maximized. Based on the Fudenberg and Maskin theorem~\cite{Fudenberg1986Thefolk}, it is shown that under the proposed strategies, when the repeated game is infinitely played, all the boundary points of the capacity region are achievable and the obtained equilibrium of the repeated game is subgame perfect equilibrium. 

\subsection{Security} 
\begin{figure*}[h]
\begin{center}
$\begin{array}{ccc} 
\epsfxsize=2.1 in \epsffile{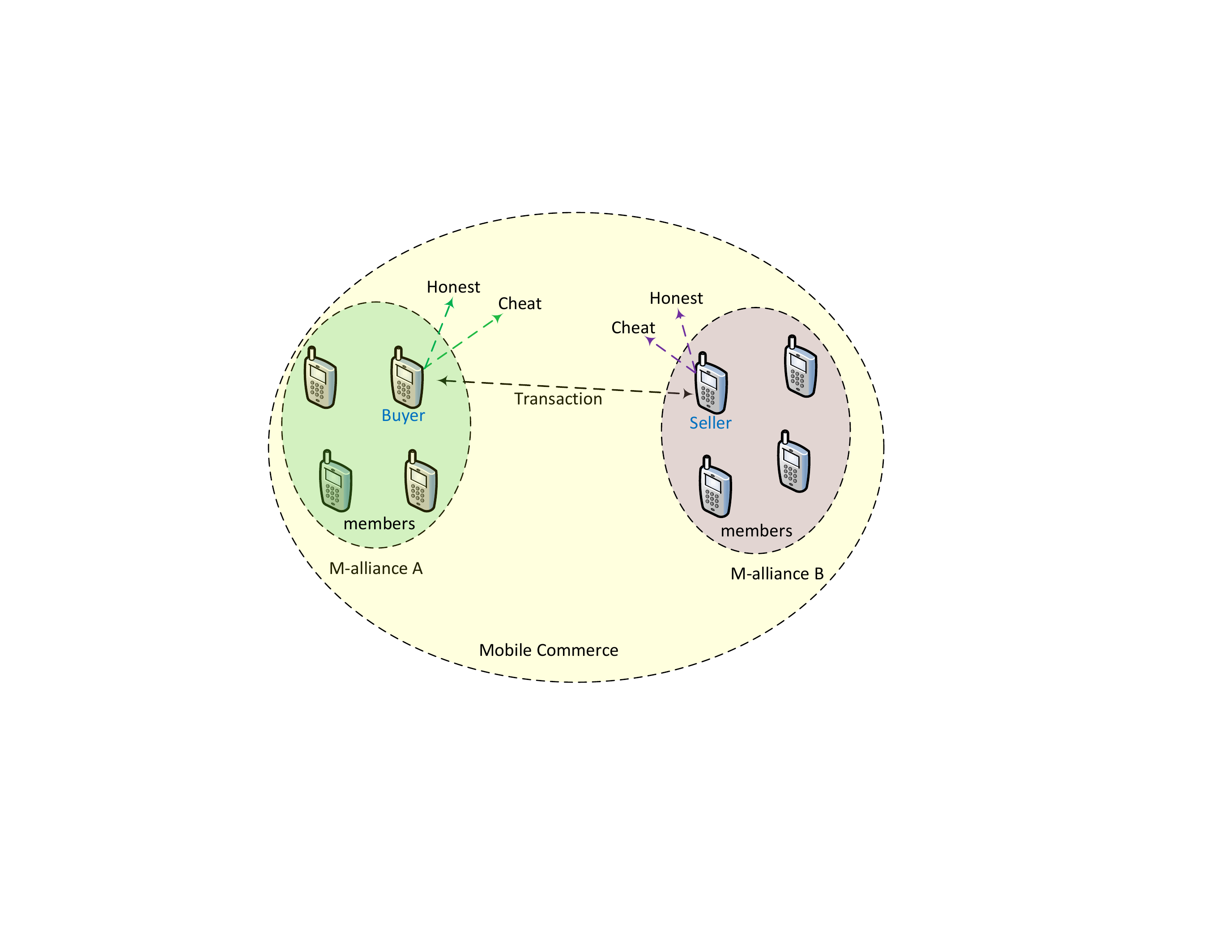}	&
\epsfxsize=2.1 in \epsffile{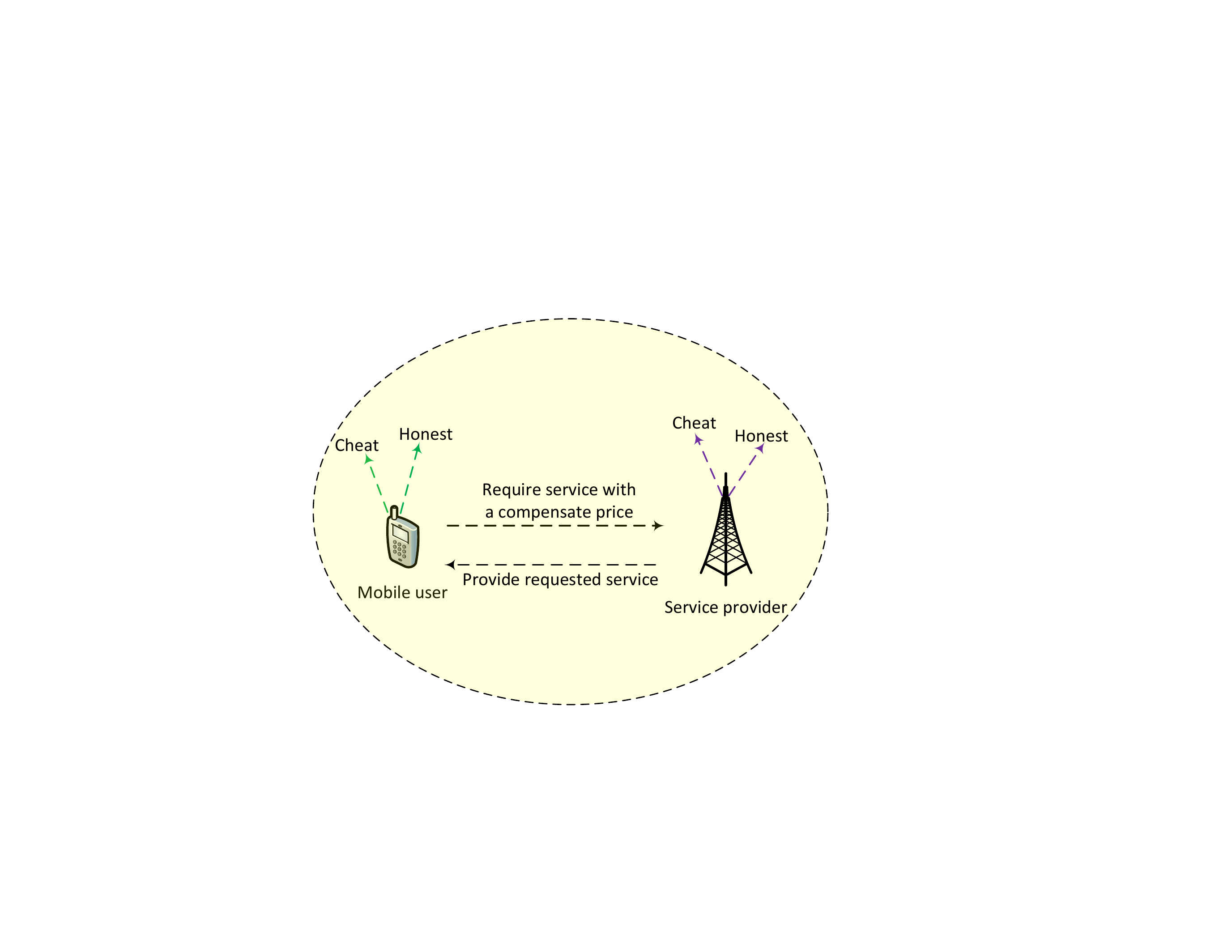}	&
\epsfxsize=2.6 in \epsffile{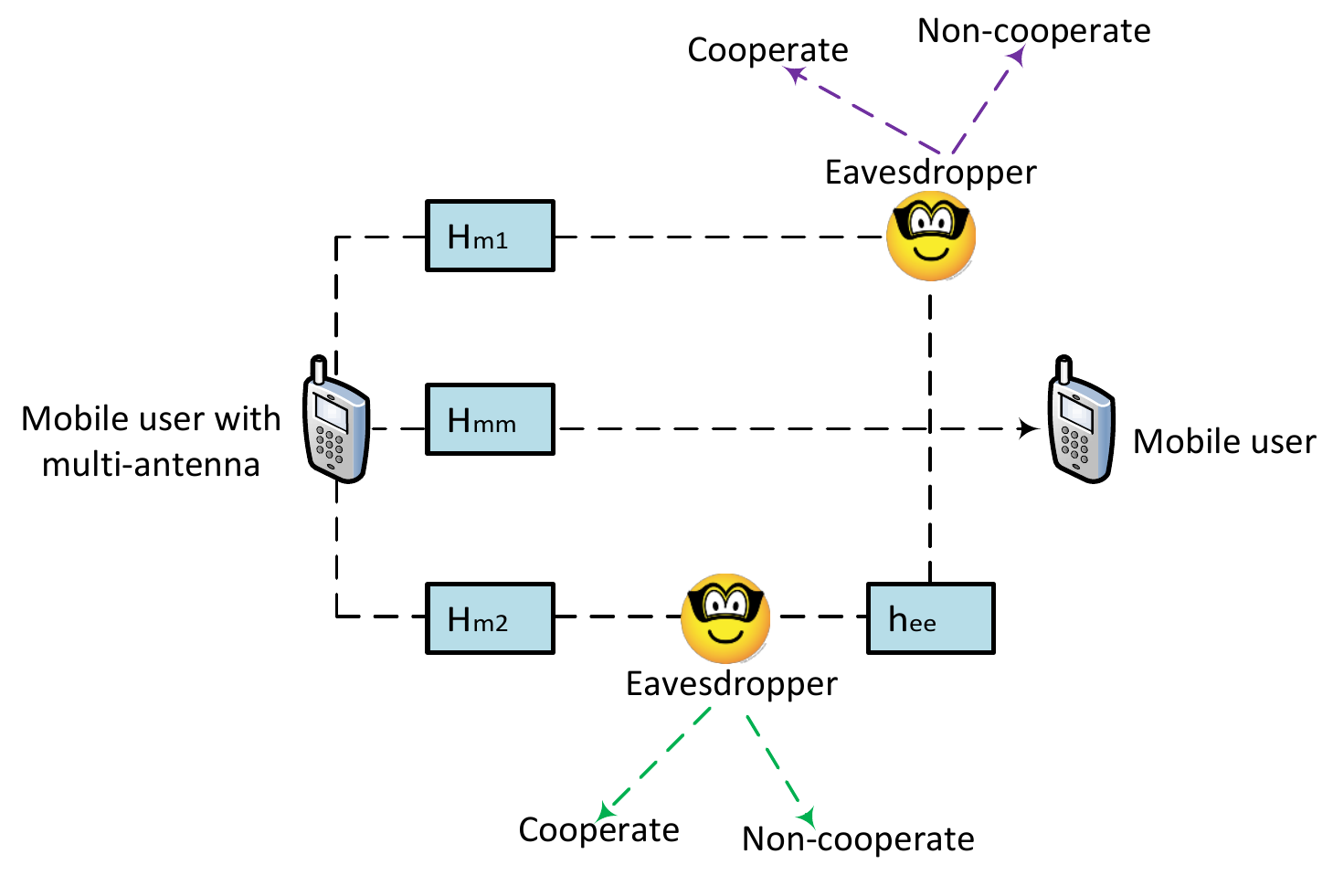}		\\ [-0.2cm]
(a)	& (b) & (c) 
\end{array}$
\caption{Interaction between (a) user and user, (b) user and service provider, and (c) user and eavesdropper.}
\label{fig:chap3_security}
\end{center}
\end{figure*}

In this section, we review the repeated game models to address some security issues. We classify them into three types based on the interaction of users with their partners as shown in Fig.~\ref{fig:chap3_security}. 

\subsubsection{Interaction between user and user} 

In mobile commerce (M-commerce) markets, transactions are performed online and free-control by authorities and thus buyers can cheat sellers by agreeing buying/using goods/services offered by sellers, but do not pay fees as agreed. Therefore, a repeated game model was introduced in~\cite{Hu2009Atransaction} to address this problem. The authors first defined two terms, i.e., member and M-alliance. The member is a trader (seller or buyer) who takes part in one M-alliance to perform transactions, and each member belongs to only one M-alliance. Each transaction is considered as a stage game, and if the game is played only once, the seller and buyer will always choose a cheating strategy that leads both of them to gaining nothing. However, if the game is played repeatedly, there is an incentive for the players to cooperate by choosing an honest strategy. In the repeated game, the players are the seller and buyer, and they can choose one of two actions, namely, honest or cheat for each transaction. The payoff is the profit that the players gain after each transaction. The authors then proposed a trigger strategy to enforce players to cooperate. If any cheater is detected (player A), the cheated player (player B) will notify to its M-alliance (MB). The MB then checks whether the complaint is true or not. If the complaint is true, the MB will ask the MA (M-alliance of player A) for compensation. After verifying the information, the MA will ask player A to compensate for player B. If player A does pay the compensation, the game is still in cooperation. However, if player A does not compensate, the MA will revoke its membership and send a warning message to other M-alliances. Now, if the MA compensates for player B on behalf of player A, then the MA still keeps its reputation. However, if MA does not compensate, its reputation will be reduced that will impact to transactions of all members in its alliance. The authors then designed the conditions for the punishment such that the selfish players do not have any incentive to deviate from cooperation. 
\subsubsection{Interaction between user and provider} 

While in~\cite{Hu2009Atransaction} the repeated game is used to deal with the cheating problem between a user and another user, in~\cite{Antoniou2011Networkselection}, Antoniou~\textit{et al.} developed the repeated game model to handle a cheating problem between a mobile user and a network provider. In particular, the mobile user sends a request for QoS support with a compensate price to the provider. After receiving the request, the provider will deliver QoS to the user. However, the user can pay less than the price that it offers. Similarly, the provider can be honest by proving QoS as promised or cheat by providing lower QoS. The players make decisions simultaneously, and they know the actions of each other only after the game ends. Thus, if the game is played only once, the players will choose to cheat. However, if the interaction between them is repeated over multiple periods, the cooperation by making honest actions is the best strategy. Additionally, to punish the deviating player, the authors consider many punishment strategies, e.g., grim, TFT, and leave-and-return. Based on the analysis, it is found that the best strategy for the mobile user is leave-and-return (i.e., cooperate as long as the provider cooperates and defect for one period if the provider defects) and the best strategy for the provider is TFT. The paper can be extended by considering the optimal pricing and resource allocation to support QoS.

\subsubsection{Interaction between user and eavesdropper} 

In~\cite{Cho2011Agame}, the authors studied the security problem in multiple-input single-output (MISO) wireless networks. The main goal is to prove that the eavesdroppers' noncooperation assumption, which is often used in the literature, is not always true. The authors considered the scenario in which there are two eavesdroppers that want to know the information about the transmission between an authentic transmitter Alice and an authentic receiver Bob. The transmitter Alice is assumed to be equipped with MISO technology. The authors indicated that the eavesdroppers can cooperate if the interaction between them is repeated. The players of the game are the eavesdroppers and their actions are to choose to cooperate or not cooperate. If they are noncooperative, they will choose zero power for relay signal. Alternatively, if they cooperate, the signal power will be greater than zero. The payoff of the eavesdropper is the mutual information between the transmitter Alice and itself. If the game is a one-shot game, it is shown that optimal strategies for both eavesdroppers are noncooperative. However, if the game is played repeatedly for a sufficiently many periods and under a set of appropriate conditions for channels and payoff functions, it is proven that there is an equilibrium point in which the eavesdroppers will cooperate in the repeated game.

\subsection{Quality-of-Service Management} 

\begin{figure*}[htb]
\begin{center}
$\begin{array}{ccc} 
\epsfxsize=2.1 in \epsffile{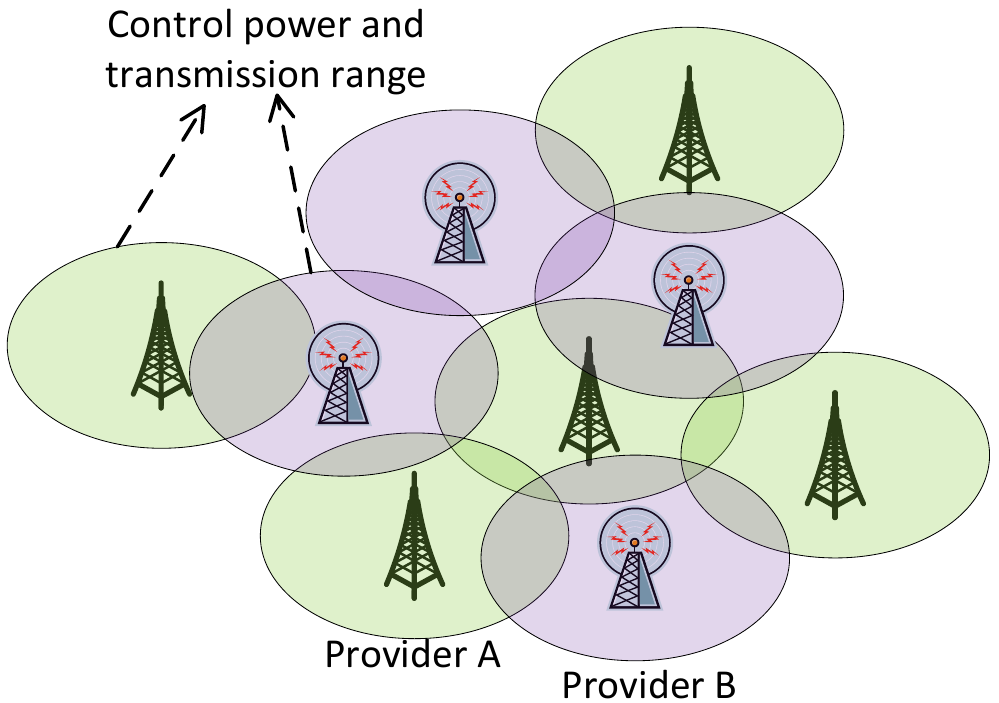}	&
\epsfxsize=2.1 in \epsffile{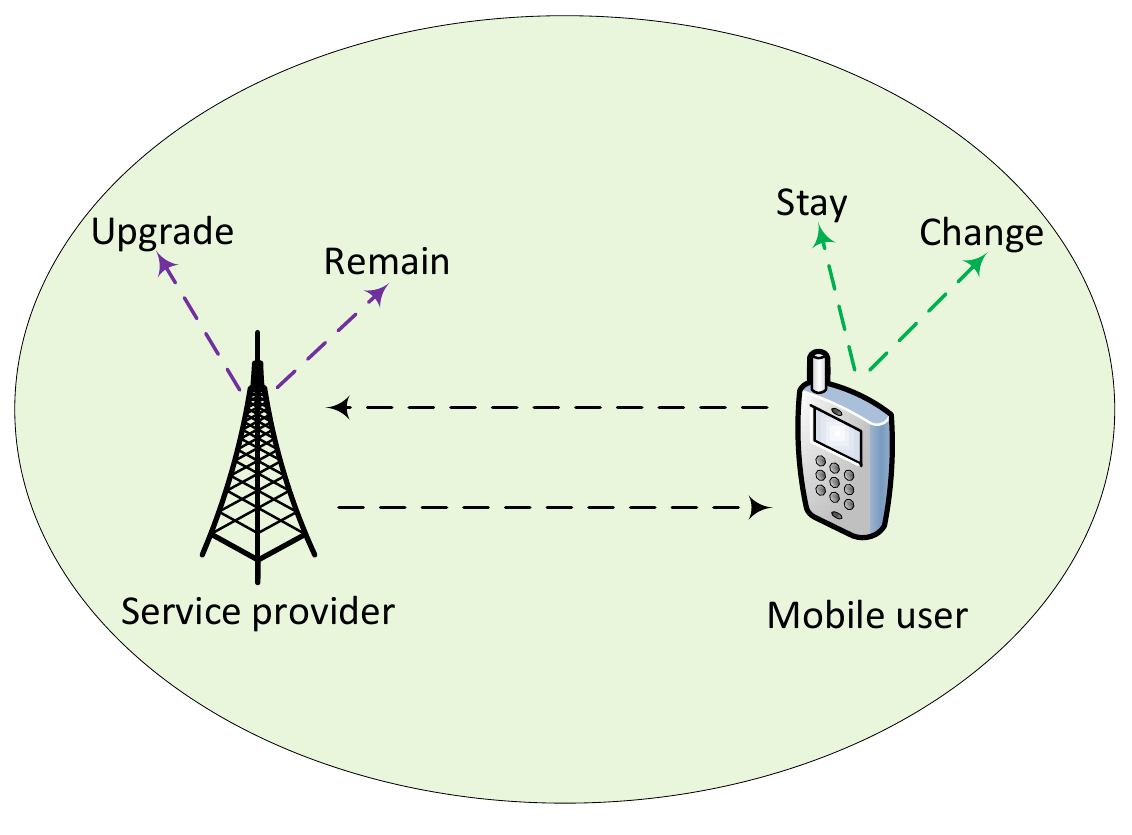}	&
\epsfxsize=2.5 in \epsffile{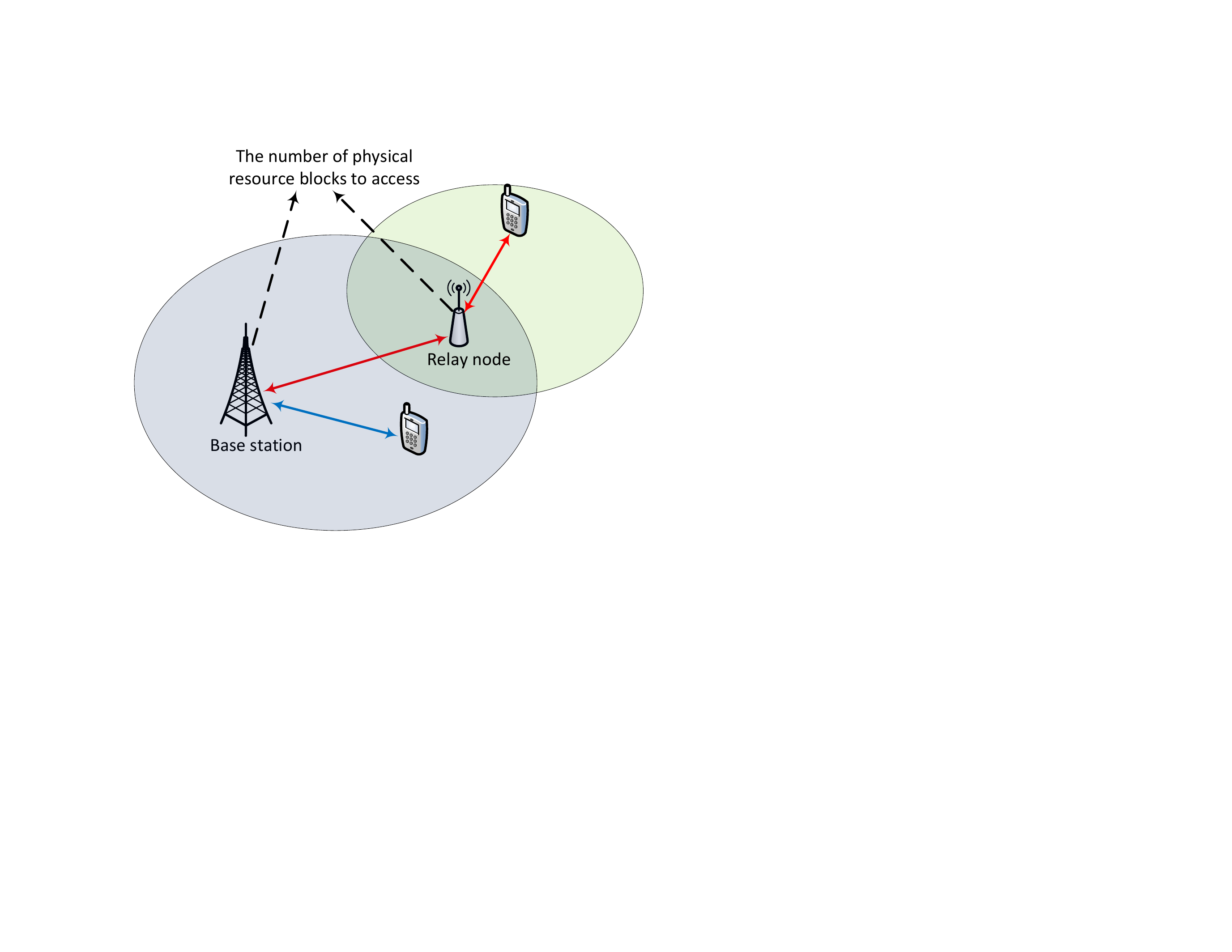}		\\ [-0.2cm]
(a)	& (b) & (c) 
\end{array}$
\caption{Interaction between (a) provider and provider, (b) provider and user, and (c) provider and relay node.}
\label{fig:chap3_qos}
\end{center}
\end{figure*}

Next, we review the work related to the Quality-of-Service (QoS) management from the perspective of service providers. Specifically, the service providers aim to achieve maximum revenues while the QoS for their customers is still guaranteed. There are three cases that are summarized in Fig.~\ref{fig:chap3_qos}. We consider three scenarios corresponding to the interactions between a provider and another provider, a mobile user, and a relay node.

\subsubsection{Operation management of base stations} 

In~\cite{Felegyhazi2006Wireless}, the authors studied the scenario with two operators as the players and each operator has a set of base stations. The operators want to maximize the number of users attached to their base stations by increasing the transmission range of their base stations. However, when the transmission range is increased, the interference among them is also rising. Therefore, the actions for the players are to choose the transmission range for their base stations such that the coverage area is maximized while the interference is minimized. For a one-stage game, the players will choose the maximum transmission range for their base stations, causing severe interference. However, if the game is repeated, there is an incentive for them to cooperate by reducing the transmission range to an optimal value (i.e., a cooperation point). The authors then applied a trigger strategy for the players. At the beginning, the players cooperate by setting the transmission range at the cooperation point. Then, if any user deviates, the players will set the maximum transmission range for their base stations in the next few periods until the deviating players cooperate again. Under the proposed strategy, it is proven that the operators can achieve the Pareto optimal equilibrium. However, the authors did not show how to detect the deviating players. 

While in~\cite{Felegyhazi2006Wireless} the authors considered the transmission range control problem of base stations, in~\cite{Treust2010Converage}, the authors studied the transmission power control problem. Similar to~\cite{Felegyhazi2006Wireless}, a repeated game is also used to model the interaction among base stations. However, in~\cite{Treust2010Converage}, the actions of the base stations are to select the transmit power levels, and the payoff is impacted by a density function of mobile nodes. The authors first used the Kalai-Smorodinsky bargaining~\cite{Kalai1975Other} to find the cooperative point of the game. Then, a trigger strategy is used to enforce the nodes into cooperation. To detect deviating nodes, the suspect detection procedure~\cite{Kalai1985Solutions} is developed and and it is shown that the players' strategies converge to an optimal point.

\subsubsection{Infrastructure upgrading} 
The authors in~\cite{Kamhoua2012Gametheoretic} used game theory to analyze the interaction between mobile users and service providers. In each period (e.g., a monthly billing cycle), the provider needs to make a decision to invest money to improve or to keep its infrastructure, while the users will make a decision to stay with the same provider or change to another provider. In the one-shot game, both of them will choose the noncooperative strategy, i.e., the provider will not invest and the user will change the provider due to higher cost and better performance, respectively. However, if the interaction between them lasts for multiple periods, there will be an incentive for them to cooperate. By applying the grim trigger strategy, it is shown that there exists a subgame perfect equilibrium for the users and provider in which the provider will invest money to improve the infrastructure and the users will not change the provider. Additionally, the authors considered the case with imperfect monitoring. Specifically, the authors assumed that the users' actions are public and monitored by the provider. However, the provider's action is private, and thus the users do not know the actions of the provider. Thus, this can reduce the profit of the provider because in some cases though the provider improves the network, mobile users still receive the low quality of service due to the other factors (e.g., shadowing), and thus they will change the provider. The authors then established the condition for the provider to continue to cooperate under the patience and the monitoring accuracy of the users. However, the practicability has to be further evaluated since it may be intractable to measure the patience as well as monitoring accuracy of users.


\subsubsection{Bandwidth sharing with relay nodes} 

In~\cite{Chaabane2012Anew}, the authors considered the interaction between a base station (eNB) and a relay node (RN) in a downlink dual-hop LTE network. In the repeated game, the players are the base station and RN. Their actions are to choose the number of physical resource blocks (PRBs) to access. It is clear that assigning more PRBs to the node can improve throughput. However, since PRBs are limited, the interference between the nodes sharing the same PRB can increase. Therefore, the node can choose one of two strategies, either to cooperate by choosing PRBs with high channel gains and letting its opponent use these PRBs, or to not cooperate by utilizing all PRBs assigned. To make player cooperate in the repeated game, a trigger strategy is used that will punish the deviating node for $T$ periods if this node does not cooperate. Furthermore, to reduce the number of punishment periods, the authors proposed using a penalty factor in the payoff function of the players. This factor reduces the additional achieved throughput if the player is punished. Simulation results then show that the proposed solution can improve the total achievable throughput.

\begin{table*}
\caption{Applications of repeated games in cellular and WLAN networks (SPE = subgame perfect equilibrium, BS = base station, TFT = tit-for-tat, CRISP  = cooperation strategy via randomized inclination to selfish/greedy play)} 
\label{table_sec3_sum}

\begin{centering}
\begin{tabular}{|>{\centering\arraybackslash}m{1cm}|>{\centering\arraybackslash}m{1cm}|>{\centering\arraybackslash}m{2cm}|>{\centering\arraybackslash}m{3cm}|>{\centering\arraybackslash}m{3.3cm}|>{\centering\arraybackslash}m{2.6cm}|>{\centering\arraybackslash}m{1.7cm}|}
\hline 
\textbf{Problem} & \textbf{Article} & \textbf{Players} & \textbf{Actions} & \textbf{Payoff} & \textbf{Strategy} & \textbf{Solution}\tabularnewline
\hline 
\hline 
\parbox[t]{2mm}{\multirow{20}{*}{\rotatebox[origin=c]{90}{Multiple Access Control}}} 
& \cite{MacKenzie2004Gametheory} & wireless nodes & transmission power levels & successful transmission data & Cartel maintenance & SPE 		\tabularnewline 	\cline{2-7} 
& \cite{Han2004Dynamicdistributed} & wireless nodes & transmission rates & profits minus cost & Cartel maintenance & SPE	\tabularnewline	\cline{2-7} 
& \cite{Han2008Acartel} & wireless nodes & transmission rates & profits minus cost & Cartel maintenance & perfect public 	\tabularnewline	\cline{2-7} 
& \cite{Auletta2008Interferencegames} & wireless nodes & transmission power levels & profits minus cost & forgiving & SPE	\tabularnewline 	\cline{2-7} 
& \cite{Treust2010Implicitcooperation}~\cite{Treust2010Arepeated} & wireless nodes & transmission power levels & a function of transmission rate and power level & Cartel maintenance & perfect public		\tabularnewline	\cline{2-7} 
& \cite{Cagalj2005Onselfish} & wireless nodes & channel access probability & throughput & adaptive & Pareto optimal		\tabularnewline	\cline{2-7} 
& \cite{Konorski2006Agame} & wireless nodes & backoff configuration & throughput & CRISP & Pareto optimal	 \tabularnewline	\cline{2-7} 
& \cite{Riedel2006Achieving} & mobile users & the use of bandwidth & throughput & TFT & Pareto optimal 		 \tabularnewline	\cline{2-7} 
& \cite{Tran2010Onselfish} & mobile users & report traffic demand & successful data transmission & cheat-proof & Pareto optimal 	\tabularnewline	\cline{2-7} 
& \cite{Hajj2011SIRA} & wireless nodes & report channel condition & a function of data transmission rate & punishment trigger & Pareto optimal 	\tabularnewline	\cline{2-7} 
& \cite{Kong2010Efficient} & mesh clients & report maximal data rate & achievable data rate & forgiving & SPE	\tabularnewline	\cline{2-7} 
& \cite{Shen2014Universalnon} & wireless users & channel information and service requirement & profits minus cost & cheat-proof & Pareto optimal 	\tabularnewline	\cline{2-7} 
& \cite{Lai2008Thewater} & wireless nodes and BS & nodes: power levels, BS: decoding order & achievable rate for wireless nodes and revenue for BS & forgiving & SPE		\tabularnewline	\cline{2-7} 
\hline 
\parbox[t]{2mm}{\multirow{6}{*}{\rotatebox[origin=c]{90}{Security}}} 
& \cite{Hu2009Atransaction} & mobile seller and buyer & honest or cheat & profits & cheat-proof & SPE	\tabularnewline \cline{2-7} 
& \cite{Antoniou2011Networkselection} & mobile users and network provider & honest or cheat & profits & users: leave-and-return provider:  TFT & SPE 	\tabularnewline  \cline{2-7} 
&\cite{Cho2011Agame} & eavesdroppers & cooperate or noncooperate & mutual information & punishment trigger & SPE 	\tabularnewline \cline{2-7} 
\hline 
\parbox[t]{2mm}{\multirow{8}{*}{\rotatebox[origin=c]{90}{QoS}}} 
& \cite{Felegyhazi2006Wireless} & BSs & transmission range of BSs & cover area & Cartel maintenance & Pareto optimal	\tabularnewline 	\cline{2-7} 
& \cite{Treust2010Converage} & BSs & transmission power level & a function of achievable rate and density function & Cartel maintenance & Pareto optimal	\tabularnewline 	\cline{2-7} 
& \cite{Kamhoua2012Gametheoretic} & mobile users and service provider & users: stay/leave, provider: improve/remain & QoS for users and profits for provider & Grim & SPE	\tabularnewline	\cline{2-7} 
& \cite{Chaabane2012Anew} & BS and a relay node & the number of physical resource blocks & throughput & forgiving & Pareto optimal 	\tabularnewline
\hline 
\end{tabular}
\par\end{centering}
\end{table*}

\begin{figure*}[htb]
\centering
\includegraphics[scale=0.65]{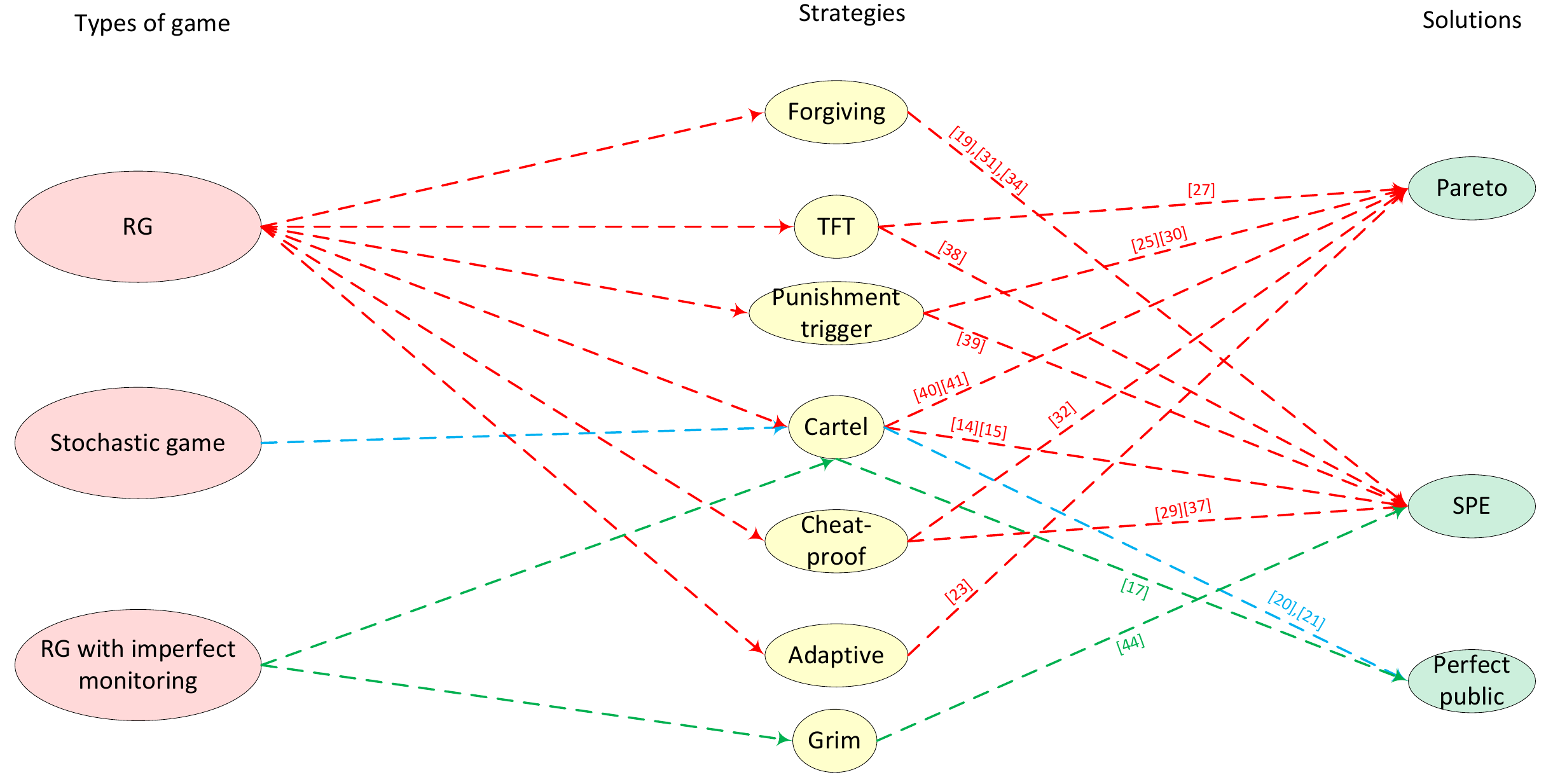}
\caption{Relation among types of repeated games, strategies, and solutions (RG = repeated game, SPE = subgame perfect equilibrium, TFT = Tit-for-Tat) in cellular and WLAN networks.}
\label{fig:chap5_Celluarl_WLAN_sum2}
\end{figure*}

\textbf{Summary:} In this section, we have identified three main issues in cellular and WLAN networks (CWLANs) and reviewed applications of repeated games for these networks. We summarize the issues along with references in Table~\ref{table_sec3_sum}. From the table, we observe that many papers investigate the multiple access control problem, while security and quality-of-service (QoS) problems are less studied. Additionally, Fig.~\ref{fig:chap5_Celluarl_WLAN_sum2} shows the relation among different types of repeated game models, their strategies and solutions. From Fig.~\ref{fig:chap5_Celluarl_WLAN_sum2}, it is found that while conventional repeated games (i.e., with perfect information and monitoring) are mostly used, their variations (e.g., a repeated game with imperfect monitoring) are not much adopted.

\section{Applications of Repeated Games in Wireless Ad Hoc Networks}
\label{sec:WAHN}

\begin{figure}[h]
\centering
\includegraphics[scale=0.35]{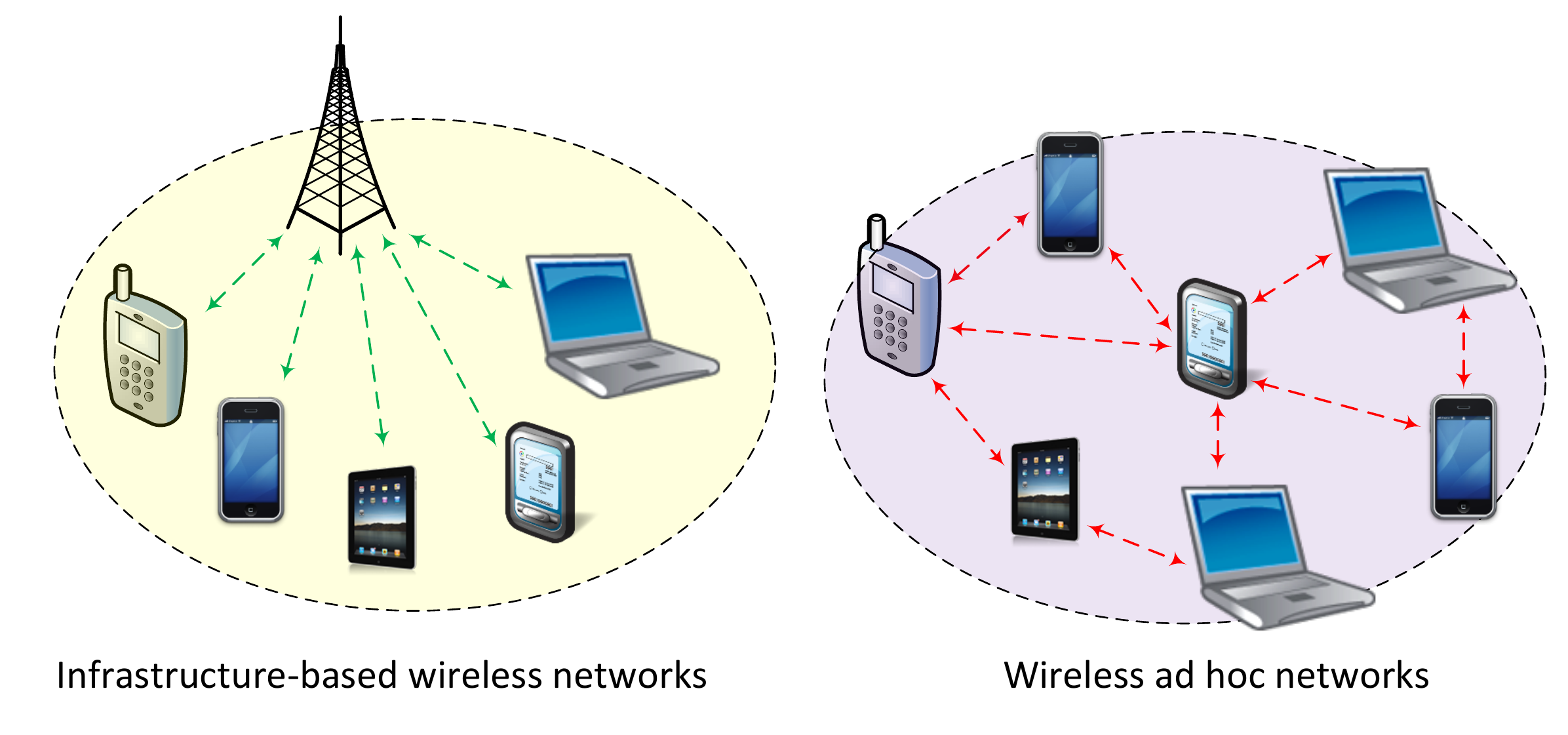}
\caption{Comparison between infrastructure networks (cellular network and WLAN) and wireless ad hoc networks}
\label{fig:chap5_CompareNetworks}
\end{figure}

In this section, we review the repeated game models developed for wireless ad hoc networks, which are known as infrastructureless networks. Such networks are multihop networks, sensor networks, cooperative transmission networks, and peer-to-peer networks. In wireless ad hoc networks, wireless nodes can communicate directly with each other without relying on any infrastructure or pre-configuration as illustrated in Fig.~\ref{fig:chap5_CompareNetworks}. Thus, there are many advantages and applications of wireless ad hoc networks in practice as discussed in~\cite{Rubinstein2006Asurvey, Sesay2004Asurvey}. However, lack of infrastructure poses many issues such as interference, decentralization, limited-range, and insecurity~\cite{Rubinstein2006Asurvey, Sesay2004Asurvey}. In this section, we review the applications of repeated games in wireless ad hoc networks in the following aspects. 
\begin{itemize}
\item 	\textbf{Packet forwarding}: In multihop networks, wireless nodes transmit packets to a distant destination through the support from other intermediate nodes. However, the nodes may belong to different authorities and forwarding process consumes a certain amount of resource of forwarders or relays. Therefore, repeated games are used to motivate the nodes for the cooperation. 
\item 	\textbf{Cooperative transmission}: Cooperative transmission or simply called relay networks can help improving performance in terms of speed and reliability through amplify-and-forward and decode-and-forward techniques. However, energy is a crucial resource and has to be efficiently utilized. Repeated games are also used to encourage the cooperation for cooperative transmission networks. 
\item 	\textbf{Resource sharing in peer-to-peer (P2P) networks}: In a P2P network, users help each other to acquire the desired services, for example, content distribution. Repeated games are used in this network to avoid free-riding behaviors of users and enforce them into cooperation. 
\item 	\textbf{Miscellaneous issues}: Besides aforementioned issues, repeated games are also used to address other issues in wireless ad hoc networks such as clustering, spectrum accessing, and security. 
\end{itemize}

\subsection{Packet Forwarding} 

\begin{figure*}[htb]
\centering
\includegraphics[scale=0.65]{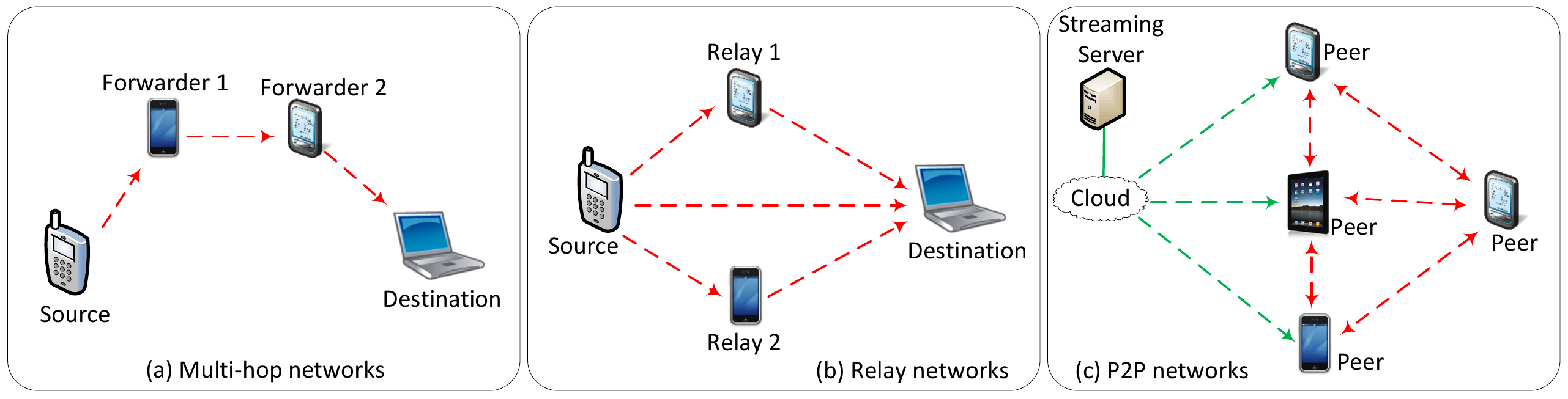}
\caption{Illustration of (a) Multihop networks (b) Relay networks, and (c) Peer-to-peer networks}
\label{fig:chap5_CompareNetworks2}
\end{figure*}

Due to the cost of forwarding packets over multihop networks (Fig.~\ref{fig:chap5_CompareNetworks2}(a)), we need to design mechanisms to encourage nodes to cooperate and enhance network performance. In general, there are three available mechanisms used to incentivize nodes for cooperation, i.e., credit-based, reputation-based, and game-based methods. For credit-based methods, e.g., Nuglet~\cite{Butt2003Stimulating} and Sprite~\cite{Zhong2003Sprite}, nodes receive the payment if they accept to forward packets for others. However, this method needs a centralized accounting server to manage payments. Consequently, it may not be suitable for decentralized networks such as ad hoc networks. Therefore, in the following, we will discuss two approaches, i.e., reputation-based and game theoretic. 

\subsubsection{Reputation-based Schemes} 

Different from the credit-based approach which needs a central authority for payment management, the nodes in reputation-based multihop networks are able to monitor and observe their neighbors' behaviors autonomously and independently. In the reputation-based networks, the node updates reputation values of its neighbors based on its observations and information provided from other nodes. Then, the node decides to forward or reject packets for its neighbors. Accordingly, the nodes have to take into account the future effects of their present actions. This implies that repeated games can be used to model the interaction among nodes in reputation-based multihop networks. 

General reputation-based networks, e.g., SORI~\cite{He2004SORI} and Catch~\cite{Mahajan2005Sustaining}, work based on the assumption that every node has to maintain the reputation of its neighbors, i.e., fraction of their forwarded packets. However, when collisions happen, a node can be treated as defecting, and thus its reputation will be degraded though it is cooperative. To deal with the collisions, the authors in~\cite{Milan2006Achieving} modeled the interaction between two nodes as a Prisoner's Dilemma with noise~\cite{Wu1995How} and used the generous TFT strategy to enforce nodes into the cooperation. Additionally, to detect and sustain the cooperation, the reputation-based mechanism is proposed. In the mechanism, each node evaluates its neighbors' reputation based on their packet dropping probabilities. Under appropriate conditions, it is proven that a subgame perfect equilibrium (SPE) is a mutual cooperation of the game. 

In~\cite{Milan2006Achieving}, to achieve a full cooperation between users, a perfect estimation of reputation is needed. Therefore, in~\cite{Jaramillo2007DARWIN}, the authors proposed a reputation strategy, namely distributed and adaptive reputation mechanism (DARWIN) that can achieve a full cooperation without the perfect estimation. With this strategy, each node estimates reputations of its neighbors, and then shares this information with others. Based on the reputation information received, each node evaluates whether its neighbor is cooperative or noncooperative. Then, by using the modification of the TFT strategy, called Contrite TFT~\cite{Wu1995How}, it is proven that DARWIN can also achieve a subgame perfect equilibrium.

Extended from~\cite{Milan2006Achieving}, Ji~\textit{et al.}~\cite{Ji2010ABelief} examined a belief-based packet forwarding approach which was developed in~\cite{Bhaskar2002Belief} to cope with not only noise, but also imperfect observation in reputation-based networks. In this approach, a node needs to maintain a belief probability distribution function with other nodes to estimate their actions. Then, based on this function, the node makes a decision to forward or reject a packet. After each game stage, the nodes will update their belief functions by using Bayes' rule~\cite{Osborne1994Acourse}. Then, their strategies are adjusted for the next game stage according to their beliefs. By using the proposed approach, it is proven that under some appropriate conditions, the proposed strategy is a sequential equilibrium. A sequential equilibrium~\cite{Osborne1994Acourse} is a well-defined counterpart of a subgame perfect equilibrium for repeated games with imperfect monitoring. The sequential equilibrium ensures that there is no incentive for deviating users. The simulations show that the proposed strategy achieves not only the sequential equilibrium, but also near-optimal performance. 

Similar to~\cite{Ji2010ABelief}, in~\cite{Wenjing2011Cooperation}, the authors also studied belief-based approaches to obtain sequential equilibria. However, in~\cite{Wenjing2011Cooperation}, the authors proposed a more efficient method to circumvent the complexity of updating beliefs as in~\cite{Ji2010ABelief}. The idea is based on a state machine. In particular, each node has two states (corresponding to two actions), called cooperate (C) or noncooperate (N). Based on signals observed from the environment, each node will make its decision to stay at the current state or switch to another state. The simulation results indicate that the performance achieved from the proposed strategy outperforms other noncooperative strategies. 

In~\cite{Kamhoua2010Belief}, the authors proposed a solution that even does not need to compute belief functions for the nodes as in~\cite{Ji2010ABelief} and~\cite{Wenjing2011Cooperation}. The solution is based on a belief-free equilibrium~\cite{Mailath2006Repeated, Yamamoto2009Alimit} that does not need to determine other nodes' private history, and hence the computation of the optimal strategy is not necessary. Consequently, the strategies considered in this paper do not depend on a node's own action and the equilibrium construction is not as complex as in~\cite{Wenjing2011Cooperation}. The authors considered a repeated game with random states. The game is modeled as a stochastic game instead of a simple repeated game. Since in each game stage, the node may or may not have packets to send. Thus, stages of the game can be random and different over time. Additionally, the authors used a monitoring technique (similar to a watchdog mechanism~\cite{Marti2000Mitigating}) for the nodes to monitor the behavior of their neighbors. The signals received from observing actions from the opponents are used to adjust the packet forwarding probability. If the nodes are observed to be cooperative, the packet forwarding probability will be one. Otherwise, the probability will be reduced. By using this strategy, it is proven that the nodes can achieve the belief-free equilibrium solution.

The authors in~\cite{Pandana2008Cooperation} also proposed a new method to overcome the limitations in~\cite{Ji2010ABelief} by using repeated games with communication and private monitoring (RGC\&PM)~\cite{Porath1996Communication}. In RGC\&PM, the nodes are assumed to be able to communicate with other nodes to exchange information about the behaviors that they can observe. The information can be used to identify the deviating node. Then, the distributed learning repeated game with communication framework was proposed to find and maintain cooperation among the nodes. The authors also showed that with the proposed repeated game framework, any cooperation equilibrium that is more efficient than the Nash equilibrium can be achieved by using some punishment strategies. Then, the learning algorithms are introduced to help the nodes achieve the efficient cooperation equilibria. The simulations demonstrate that the proposed framework can achieve 70\% to 98\% better performance compared to that of the centralized optimal solution. 

While in~\cite{Ji2010ABelief}, the authors examined repeated games with imperfect observation, in~\cite{Srivastava2006Equilibria}, the authors investigated repeated games with imperfect monitoring. There is a slight difference between them. In imperfect observation games, the players are not sure about the actions of others in each game stage. As a result, they need to maintain a belief function to evaluate opponents' actions. By contrast, in imperfect monitoring games, the players cannot monitor explicitly other nodes' actions, and thus they will use a random public signal to infer their actions. The notation ``public signal'' was introduced in~\cite{Abreu1990Toward}, and it has to satisfy some certain properties. By using the public signal it is proven in~\cite{Abreu1990Toward} that there exists a perfect public equilibrium for the repeated game with imperfect monitoring. In the context of wireless ad hoc networks, a public strategy of a node is a perfect public equilibrium if after every game stage the public strategy forms the Nash equilibrium from that stage onward. The authors then applied the grim trigger strategy for the nodes to punish deviators and enforce them to cooperate. 

In the above reviewed papers, when the node does not forward packet for others, it will be treated as a selfish node and will be punished. However, in many practical cases, nodes do not forward packets because they are unable to do so. For example, the nodes occasionally cannot forward packets due to energy depletion. Thus, if these nodes are treated as selfish nodes and are excluded from the cooperation, the performance of the network will be degraded severely. Therefore, the authors in~\cite{Sun2011APower} focused on designing a power control mechanism for the nodes with uncertainty. There is some subtle difference in defining actions and the payoffs of the nodes compared with other papers. Specifically, in each round, each node chooses not only the transmit power level, but also the forwarding probability. The node then broadcasts this information to its neighbors. The main goal of the node is to maximize its own throughput, while minimizing the energy consumption. Therefore, the payoff function is defined by the ratio between the successful self-transmission rate and the total power used (for both self-transmission and packet forwarding). In the game, each time slot is divided into two phases. In the first phase, the nodes need to make a decision on the amount of energy used to send packets (including its own packets). The node will forward packets for others if it believes that the energy consumption for forwarding is lower than the rewards to be received in the future. The beliefs of the nodes then are updated based on the Bayes' rule. In the second phase, the nodes have to determine the probability to forward packets for others in the next time slot. Similar to the first phase, the beliefs are used to infer the forwarding probability. The proposed strategy is shown to be the Pareto-dominate equilibrium numerically. However, theoretical proof is still open.


\subsubsection{Game-based Schemes}

In reputation-based approach, intermediate nodes decide to forward packets based on the reputation. Nevertheless, propagation of nodes' reputation increases an overhead in the network. Furthermore, the reputation can be lost, forgotten or manipulated and thus game-based methods have been introduced. In game-based methods, the node decides whether to forward packet according to its defined strategy, thereby avoiding the problem as in reputation-based approach. 

In~\cite{Srinivasan2003Cooperation}, repeated games are used to model the interaction of nodes. The nodes as the players need to choose the action to forward or reject when they receive a packet from other nodes. If the node accepts the packet, it has to spend a certain amount of energy to forward the packet. Thus, the node has to balance between the number of its own packets and the number of packets from other nodes to be forwarded. In the repeated game, the authors used a punishment strategy, namely, {\em generous} tit-for-tat (TFT) to enforce the node to cooperate with the aim to find a Pareto optimal point. To make the decision to accept or reject, the node has to monitor all packets that it sends and that the other nodes send to it. Based on this information, the node computes the acceptance probability and makes the decision to accept/reject incoming packets. 

A similar model and approach are also considered in~\cite{Urpi2003Modelling}. However, in this paper, instead of adjusting the acceptance probability, the action of the node is to determine the number of packets that it will send and the number of packets it will receive and forward. The TFT strategy is used. It is shown that the condition for the node to forward packets is that the amount of traffic forwarded by others is at least equal to that the node forwards. However, this can lead to unfairness because the deviating node will be punished severely forever.

To overcome the unfairness issue in~\cite{Srinivasan2003Cooperation}, the authors in~\cite{Bandyopadhyay2005AGame} proposed a new punishment scheme which can be seen as an improvement of generous TFT strategy used in~\cite{Srinivasan2003Cooperation}. At the beginning of the game, the nodes choose the strategy ``cooperate'' by forwarding packets that they receive. If the node defects from the cooperation, all the other nodes will continue monitoring and cooperating with this node for the next $p-1$ stages. If the deviating node continues noncooperating after $p$ stages, all nodes will punish the deviating node for $q$ stages by rejecting the packets from this node. During the punishment, if the deviating node regrets and wants to re-cooperate, it can help other nodes forward packets without sending its own packets for $r$ stages. After that, the deviating node can return to the cooperation. Otherwise, this node will be punished forever. By using this strategy, the nodes can avoid the unfair punishment as in~\cite{Srinivasan2003Cooperation} and the network performance can be improved significantly. 

Similar to~\cite{Bandyopadhyay2005AGame}, the authors in~\cite{Guan2010AModified} also proposed a fair solution, called a {\em restorative trigger strategy}. In this strategy, the nodes take a cooperative strategy by forwarding packets in the first two stages. In each game stage, if the cooperation is maintained, the node updates its estimation of the ``selfishness''. If the ``selfishness'' is lower than a predefined threshold, the node still cooperates in the next stage. Otherwise, it will switch to a defection state and report the selfish node to others. The selfish node will be removed from the cooperation and if the selfish node wants to re-cooperate, it has to forward packets for other nodes until the cooperation expectation of the selfish node is greater than a certain threshold. The similar approach of using a punishment strategy and learning algorithm is considered in~\cite{Han2005ASelf}. The action is the packet forwarding probability which is optimized to achieve better network performance.

In the same context, the authors of~\cite{Mohamed2012Cooperation} developed a new strategy to achieve better results than that of~\cite{Han2005ASelf}. This strategy is based on the idea of {\em the weakest link (TWL)} borrowed from a famous TV game show that encourages players to cooperate. The nodes form a route from a source to a destination and the nodes in this route are considered as the candidates of a chain in the TWL game. Each node has the payoff defined based on its cooperation level, i.e., its packet forwarding probability. To enforce the nodes to cooperate, an infinite repeated self-learning game model is developed. In each game stage, each node observes its payoff and the cooperation level of others. Accordingly, the node adjusts its forwarding probability. If the node defects from the cooperation, it will be excluded from the cooperation for a certain number of stages. If the node wants to return to the cooperation, it has to raise its forwarding probability and notify to the others. The simulation results show that the proposed solution of~\cite{Mohamed2012Cooperation} outperforms solutions of~\cite{Han2005ASelf}.

In the aforementioned strategies, there exist an infinite number of Nash equilibria and in general not all of them are efficient. Therefore, in~\cite{Wei2006OnOptimal}, the authors evaluated a few criteria with the aim to remove the Nash equilibria that are less robust, less rational, or less likely strategies. These criteria are subgame perfection, Pareto optimality, social welfare maximization, proportional fairness, and absolute fairness. Furthermore, in~\cite{Wei2006OnOptimal}, the authors considered the case where the nodes can report information dishonestly to gain their own benefits. The authors demonstrated that in the considered packet forwarding game, there is no incentive for the nodes to report honestly their private information. Therefore, the authors concluded that when cheating is possible, the node will not forward more packets than its opponent does for it. To address this problem, a {\em Cartel maintenance profit-sharing (CAMP)} strategy was proposed in~\cite{Zhu2008AGame}. With CAMP, the nodes start the game by cooperation, i.e., reporting the true forwarding cost. Then, in each round, the nodes check their payoffs. If their payoffs are lower than the expectation thresholds, they switch to play noncooperative game for a certain number of periods before returning to the cooperation state. 

In the above reviewed papers, the nodes can implement their strategies under the assumption that they are able to monitor perfectly the actions of all the other nodes in the network. However, this assumption is vulnerable due to the hardware limitation or noise/interference especially in a wireless environment. To address the imperfect monitoring problem, the authors in~\cite{Ng2010Game} proposed a repeated game model with communication. The key idea is based on the {\em Aoyagi's game}~\cite{Aoyagi2002Collusion}, which is a repeated game with private signals communicated among players. After receiving private signals at the end of each game stage, the nodes decide to cooperate or defect. However, selfish nodes can report fake information to gain their own profits. Therefore, the TFT strategy is applied to enforce the nodes into cooperation by revealing true information. The equilibrium solution obtained by the proposed strategy is a perfect public equilibrium. 

In previous work, the strategies applied in repeated games are uni-strategy, i.e., all nodes take the same strategy. However, in some cases, nodes may not follow and they can use strategies that they prefer. For example, one node can use the TFT strategy while others apply the grim strategy. Alternatively, the node can take the TFT strategy at the beginning of the game, but it switches to the grim strategy latter. The divergent strategies form a mixed-strategy repeated game that is very complex to analyze. Consequently, in~\cite{Yan2008Cooperative}, the authors used simulations to study the cooperation behaviors among nodes which are similar to an evolutionary game~\cite{Axelrod1984Theevolution}. Five strategies are studied, called, always cooperate, always defect, random, TFT, and gradual strategy. The core idea of the gradual strategy is based on the adaptiveness strategy~\cite{Beaufils1997Ourmeeting}. In particular, a node will start by cooperating and remain using this strategy as long as its opponents cooperate. Then, if this node detects someone deviating, it will defect for one stage and cooperate for two stages. After the $N$-$th$ defection, the node will defect for $N$ consecutive stages and cooperate for two stages. Many results are highlighted by comparing the performance of different strategies. For example, in the environment without noise, the gradual strategy achieves the best performance. However, with noise, the gradual strategy performs merely better than others when the noise ratio is high, e.g., 30\%. 

Similar to~\cite{Yan2008Cooperative}, an evolutionary game theory (EGT) was used in~\cite{Kamhoua2010Mitigating} to enforce selfish users into the cooperation. Some strategies studied are TFT, grim, Pavlov~\cite{Nowak1993Astrategy}, and pPavlov~\cite{Boerlijst1997Thelogic}. Through using the EGT framework, the authors showed that by using Pavlop or pPavlop strategies, no selfish nodes can gain benefits by playing a noncooperative strategy. The advantage of the Pavlop strategy is that it can be implemented in a distributed fashion, as it requires only local information. 

\subsubsection{Methods to Deal with Malicious Users} 

In the previous subsections, we review the methods to deal with selfish users and enforce them into the cooperation. In this section, we present the applications of repeated games to tackle with malicious users. While selfish users act only to gain their benefits noncooperatively, the malicious users aim to harm and degrade intentionally network performance. There are some papers from George Theodorakopoulos~\textit{et al.} studying the problem of malicious behaviors in wireless ad hoc networks. The first paper~\cite{Theodorakopoulos2006Enhancing} considers one malicious user in the network. Then the case with more than one malicious user is considered in~\cite{Theodorakopoulos2007Malicious}. Next, in~\cite{Theodorakopoulos2008Game}, a mechanism to detect malicious users is presented. 

In particular, the authors assumed that the users who want to disrupt the network are ``Bad'' users and the rest are the ``Good'' users. Bad users aim to degrade network performance, while Good users aim to maximize their benefits in a long-term basis. The actions for each node are to forward (cooperation) or reject (defection) packets. Good users can choose to forward or reject packets for their neighbors as long as their long-term payoffs are maximized. Moreover, the authors assumed that Bad users are intelligent. Specifically, instead of always choosing to reject the packets, Bad users can choose to forward packets randomly to deceive other Good users. Thus, it is difficult for the Good users to detect and punish the Bad users. The general scenario considered in these papers is as follows. Good users want to cooperate with other Good users, but not with Bad users. Thus, Good users try to find Bad users as soon as possible thereby reducing harmful interactions with Bad users. However, Good users do not know who Bad users are. The Good users can detect Bad users only if the game is played repeatedly. 

In~\cite{Theodorakopoulos2006Enhancing}, the authors considered only one malicious (Bad) user in the network. This Bad user is unknown by Good users in advance, but they can gradually detect the Bad behavior through repeated interactions. To detect the Bad user, the Good users will use a randomized policy, i.e., Good users will cooperate with probability $p$ independently at each round. The probability is set to maximize Good users' payoffs. With the randomized policy, it is proven that the Bad user will be detected by Good users after some periods of the interaction. The authors finally showed that the proposed strategy can achieve Nash equilibria and form the cooperation among Good users.

In~\cite{Theodorakopoulos2007Malicious}, the authors then extended for the case with multiple malicious users in the network. Instead of using the randomized policy as in~\cite{Theodorakopoulos2006Enhancing}, the authors applied a fictitious play model~\cite{Brown1951Iterative} and proposed an equilibrium algorithm for Good users. In the fictitious play, each Good user assumes that its neighbors' actions are chosen independently and identically distributed following the Bernoulli probability distribution. Hence, at each game stage, the Good users will choose the action that yields the highest payoff for them given the estimations of their neighbors' strategies. The main conclusion is that if the ratio of cost per benefit for the Good users is high, the achievable payoff of Bad users will be low. Finally, in~\cite{Theodorakopoulos2008Game} the authors proposed the mechanism to detect malicious users for the game model proposed in~\cite{Theodorakopoulos2007Malicious}. In this mechanism, a Good user will construct a star topology of connections with its neighbors where the Good user is a central. The central Good node is able to monitor all actions of its neighbors as well as obtained payoffs. From this information, the central Good user can identify the Bad user. 

In addition to the methods proposed by Theodorakopoulos, Tootaghaj~\textit{et al.}~\cite{Tootaghaj2011Game} also introduced a security mechanism that detects not only selfish, but also malicious users. In this mechanism, a node can generate their private/public keys and share them over the network by an independent public key infrastructure. After receiving packets, the node will send a confirmation message to its upstream hop. Based on the confirmation messages, the nodes can identify selfish or malicious nodes and perform punishment accordingly.

\subsection{Cooperative Transmission} 

Relay networks based on cooperative transmission typically work in two phases as illustrated in Fig.~\ref{fig:chap5_CompareNetworks2}(b). In the first phase, a source node broadcasts a message to the destination and relay nodes. In the second phase, the relay nodes help the source node transmit data to the destination. The destination node combines all signals received from the source and relay nodes to extract information. By using relay nodes, the quality of the signals received at the destination can be significantly improved. However, the relaying process consumes additional energy of the relay nodes, which may discourage relay nodes from the cooperation. In the following, we review the repeated game models developed for cooperative transmission in relay networks. In general, there are two methods used for cooperative transmission, namely, amplify-and-forward (AF) and decode-and-forward (DF)~\cite{Levin2012Amplify}. For the AF transmission, the relay nodes amplify signals received from the source node before transmitting to the destination. For the DF transmission, relay nodes decode the source information before forwarding to the destination. 

\subsubsection{Amplify-and-Forward Relaying Transmission} 

\begin{figure}[htb]
\centering
\includegraphics[scale=0.65]{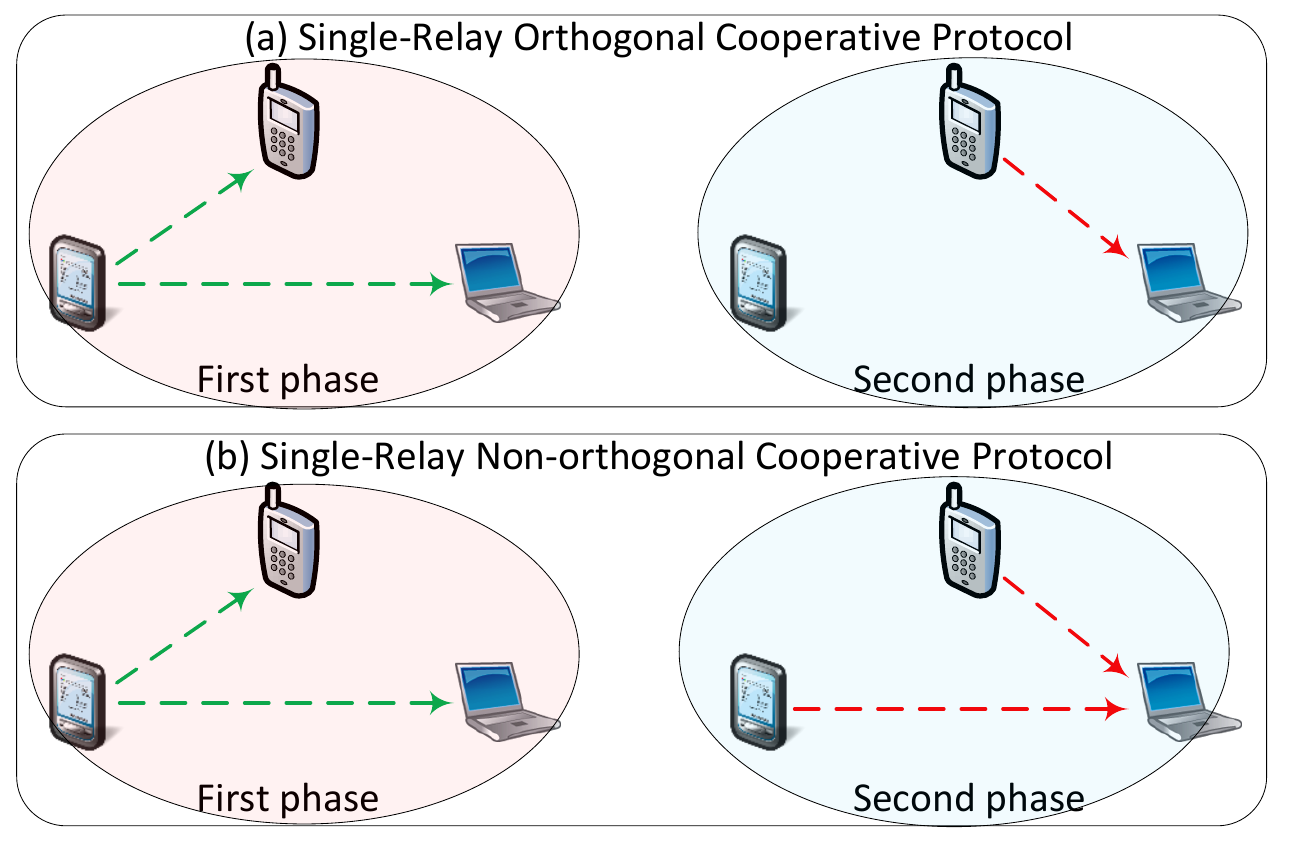}
\caption{The difference between (a) single-relay orthogonal cooperation protocol and (b) single-relay non-orthogonal cooperative protocol.}
\label{fig:chap5_Richard2007}
\end{figure}

In~\cite{Yang2007Energy}, Yang~\textit{et al.} considered a relay network with two source nodes along with two corresponding destination nodes. In this network, each source can use another source node as a relay node to transmit data to its destination through the half-duplex orthogonal cooperative AF protocol as shown in Fig.~\ref{fig:chap5_Richard2007}(a). 

The authors studied two cases for this model, namely, without and with fading channels. In the case of non-fading channels, similar to packet forwarding games, the authors modeled the interaction between two source nodes as a repeated game and showed that there exists a cooperative Nash equilibrium. In this game, the players are two source nodes. Their actions are to relay or not. Their payoff is the amount of energy saved. With fading channels, the repeated game becomes more complicated to find the cooperation points for source nodes. Specifically, at each game stage, the channel status is different, and thus the amount of energy saved also differs. Consequently, the payoff matrix is varied over game stages. To enforce the source nodes to cooperate, the authors proposed a conditional trigger strategy. In this strategy, the decision of each source node depends on the instantaneous value of the current payoff matrix and the statistic of its future payoffs that can be interpreted as the value of relay energy for other source nodes. The proposed conditional trigger strategy is proven to be a cooperative Nash equilibrium. Through simulations, it is shown that the proposed strategies achieve high energy efficiency, and it is close to that of the centralized optimization solution. A similar model is also presented in~\cite{Mohammed2014Cooperation}. However, in~\cite{Mohammed2014Cooperation}, the authors used the Q-learning algorithm to help nodes achieve the expected values. 

The authors in~\cite{Brown2011AGame} proposed an extension of~\cite{Yang2007Energy} that resolved two limitations in~\cite{Yang2007Energy}. Firstly, the authors introduced the solution that does not need to rely on centralized controlled energy allocations as presented in~\cite{Yang2007Energy, Mohammed2014Cooperation}. Secondly, the authors extended the model with more than two source nodes. Specifically, the authors used a bargaining solution to find efficient energy allocation strategies that can be performed locally for source nodes. Then the repeated game framework is used to determine sufficient conditions for the mutual cooperation between the source nodes. In the repeated game, a trigger strategy is introduced to encourage nodes to cooperate. The node has to choose action ``relay'' if the bargaining solution specifies that the required direct transmission energy for that node is greater than zero. Otherwise, this node will be punished forever. Under the proposed strategy and satisfied necessary conditions, it is proven that the mutual cooperation is possible in both fading and non-fading channel cases. Furthermore, to extend the model with more than two players, the ``stable roommates'' algorithm~\cite{Gale1962College} is applied to form partnership among players. Numerical results show that the proposed strategies can dramatically improve the energy efficiency for the relay network. 

In~\cite{Yang2007Energy} and~\cite{Brown2011AGame}, only one relay node is considered. However, multiple relay nodes are more commonly in practice that is impossible to use the half-duplex AF protocol. To address this issue, the authors in~\cite{Li2010RelayIET} used a non-orthogonal protocol which allows multiple relay nodes to transmit packets to the destination over the same frequency band at the same time as illustrated in Fig.~\ref{fig:chap5_Richard2007}(b). In the proposed system model, the relay network is modeled as an exchange market game in which the nodes trade their transmission power to gain data rate. The cooperation cycle formation algorithm is proposed to help nodes achieve the {\em strict core} of the game. The concept of ``strict core'' was introduced in~\cite{Shapley1974On, Roth1977Weak} with the aim to achieve the Pareto optimal, individual rational and strategy-proof solution~\cite{Jinpeng1994Strategy}. Additionally, to punish selfish users and revive the cooperation, a dynamic punishment strategy is presented which is proven that the full cooperation is a Pareto-dominant equilibrium.

\subsubsection{Decode-and-Forward Relaying Transmission} 

While the AF cooperative transmission is a better choice for uncoded systems, the DF relaying protocol is especially appropriate for the systems using powerful capacity-approaching codes. In DF relay networks, when a relay node receives a request from a source node, it can choose to cooperate by relaying the decoded (and then re-encoded) data, or dismiss the request. Similar to AF relaying systems, repeated games are also used to model and encourage the cooperation. In~\cite{Chen2008AGame}, the authors studied DF relaying networks under fading and non-fading channel scenarios. In the case of non-fading channel, the authors proved that there exists a mutually cooperative Nash equilibrium between two nodes when the game lasts for multiple periods. In the case of fading channels, the authors used a Markov chain to model the status of the channel with two states, namely, BAD or GOOD. Moreover, it is assumed that the players abide by the following strategy. If the player does not cooperate in the current period, it will be punished in the next $K-1$ periods. Since the payoff is a function of the uplink signal to noise ratio, the authors examined two types of function, i.e., convex and concave payoff functions. For convex payoff functions, it is possible to find the cooperative Nash equilibrium for the repeated game. However, for concave payoff functions, a mutually cooperative Nash equilibrium may not exist under the bad channel conditions. Then, two specific concave payoff functions are examined, namely, Shannon capacity and transmission success rate. The authors demonstrated that with these functions, cooperation can be achieved by setting an appropriate weight for the future payoff.

\subsection{Resource Sharing in Peer-to-Peer Networks} 

The goal of P2P networks is to share resources among participants without or with minimal support from a centralized server. In~\cite{Peng2008Enhancing}, the authors considered the interaction between two participants (i.e., players). Similar to relay networks, in P2P networks, a player uses a service provided by other players. However, providing the service also incurs a certain cost. Using a one-shot game to model this situation, it is shown that all players will not cooperate. However, with repeated games, the cooperation can be sustainable. In~\cite{Peng2008Enhancing}, the authors highlighted that noise or link failure can cause misbehaviors for players, and thus they proposed an improvement, called, proportion increment TFT strategy. With this improvement, after a player is treated as a deviator because of noise, the cooperation probability of peers will be updated based on a cooperation ratio function and the probability will converge to the cooperation point between two peers if both peers continue cooperating. The results show the efficiency of the proposed strategy compared with the classical TFT strategy, and it can even reduce the adverse effect from the malicious behavior in P2P networks.

In~\cite{Peng2008Enhancing}, the authors assumed that the requests from participants are identical, i.e., the same quality-of-service and the same cost and profit. Differently, the authors in~\cite{Lin2009Cheat} studied a resource sharing problem in a wireless live streaming social network for users with heterogeneous types. Each user has a type of $\{laptop,PDA,cellphone\}$ and a buffer to store content. Video streaming from a service provider is divided into chunks and the users with different types have different costs of sharing and different gains of the received chunks. In each round, the users report their buffer information of each other, and then they send their chunk requests. After receiving the request, the users will decide how many chunks that they will transfer. The payoff is determined by the gain minus the cost of transferring. The users do not know, but they have beliefs about the opponent's types. By using a Bayesian repeated game, the authors showed that there exist an infinite number of Bayesian-Nash equilibria (BNE). However, not all the BNE are efficient and thus the authors proposed different solutions including Pareto optimality, bargaining solution, and cheat-proof cooperation strategies with the aim to encourage users to cooperate for better performance. The authors finally concluded that to maximize the payoffs and to circumvent cheating behaviors, the players should always agree to send chunks as shown in time-restricted bargaining strategy quota. 

In~\cite{Lin2010Cooperation}, the authors extended the two-player game model proposed in~\cite{Lin2009Cheat} for multiple players. When a user requests chunks from the other users at different time slots to maximize its payoff, the requests for chunks may not be simultaneously received. Therefore, a repeated game is inapplicable and the solutions used in the two-player game cannot be used directly. The authors proposed a multiuser cheat-proof cooperation strategy, and in~\cite{Lin2009Incentive}, it is proven that this strategy is a subgame perfect and Pareto-optimal Nash equilibrium. The request-answer and chunk-request algorithms were proposed. They will reward more for the users who share more video chunks. This will encourage users to cooperate. The results demonstrate that by using the proposed strategy, the users have incentive to cooperate and they can achieve cheat-free and attack-resistance for the networks. 

Similar to~\cite{Lin2009Cheat} and \cite{Lin2010Cooperation}, the buffer cheating attacks in P2P wireless live video streaming systems were considered in~\cite{Chen2012Analysis}. Instead of using bargaining cheat-proof cooperation strategy, the authors in~\cite{Chen2012Analysis} considered proportional fairness optimality criteria and proposed a cheat-proof strategy to deter buffer cheating attackers. While the cheat-proof strategy in~\cite{Lin2009Cheat} is based on hiding private information and thus users have to bargain the amount of chunks that they exchange, the cheat-proof strategy in~\cite{Chen2012Analysis} is based on a fairness solution. In particular, the user gains nothing if it does not share any chunk. With the proposed strategy, the authors concluded that the user will not cheat if its opponent does not refuse to cooperate in the previous round or if there is no useful chunk in the opponent's buffer.

While in the above papers, repeated games are used to model the interaction between users in P2P networks, in~\cite{Wu2012AGame}, the authors used a repeated game to study the relation between a content server and users. The content server needs to assign a reward to users, while the users have to decide the transmission rate to forward packets received from the content server. It is assumed that the content provider and users follow a Stackelberg game~\cite{Osborne2004AnIntroduction} in which the content server (i.e., a leader) assigns the reward first, and then the users (i.e., followers) choose the transmission rate based on the reward. The authors showed that for the one-shot Stackelberg game, the Stackelberg equilibrium can be achieved through using the backward induction technique~\cite{Osborne2004AnIntroduction}. The authors then extended the static Stackelberg game model to the repeated Stackeberg game model. It is found that the user may have motivations to deviate from the Stackelberg equilibrium. Specifically, the users may deny to forward packets for the content server if it does not offer higher rewards than that of the Stackeberg equilibrium. With the repeated game, if the potential benefit is large or if the users care much about the future reward, they will have higher incentive to urge the content server to offer higher rewards than that of the Stackelberg equilibrium. Moreover, a cheating prevention mechanism~\cite{Ng2008Game} is used to deal with the cheating problem from users.

\subsection{Miscellaneous Issues} 

Besides the use of repeated games for the aforementioned problems in different types of wireless ad hoc networks, there are also many other applications of repeated games that will be discussed in this section.

\subsubsection{Multiple Accesses} 

In~\cite{Tembine2007Multiple}, the authors studied a multiple access game in ad-hoc networks. The authors proposed the modified multiple access game (MMAG) by adding a regret cost when no user transmits. The MMAG possesses two pure Nash equilibria that are also Pareto optimal. If the MMAG is played repeatedly, there exists a subgame perfect Nash equilibrium. Additionally, to improve the network performance, the authors studied the relation between the regret cost and the transmission probability. The results demonstrate that based on this cost, we can adjust other parameters to achieve efficient equilibria.

A similar scenario is also studied in~\cite{Chen2007Selfishness}. However, to refine Nash equilibria to an efficient equilibrium for better fairness, social welfare maximization, and Pareto optimality, the authors in~\cite{Chen2007Selfishness} proposed a searching algorithm. The TFT strategy and generous TFT strategy are also proposed to detect selfish nodes and attract them to the cooperation. The authors then extended the game model to a multihop network scenario. Then, it is shown that although the globally optimal solution may not be achieved, we still can obtain the Pareto optimal equilibrium through the proposed game model. 

Qu~\textit{et al.}~\cite{Qu2012Efficient} examined the multiple access control scheme in wireless sensor networks using a TDMA protocol. A repeated game is applied to determine the transmission strategy. The players are the sensors and they send information to the datacenter. Specifically, each sensor will report its energy level and queue state to the datacenter. Based on the received information, the datacenter selects only one sensor which has the highest measured state to transmit all packets in the data queue to the datacenter in the rest of the time slot. The payoff of the sensor is the number of packets successfully transmitted. However, the sensors may report fake information to obtain additional benefits. The authors designed a mechanism to detect selfish sensors and used a punishment strategy to enforce selfish users to cooperate. If the defecting sensor is detected, all sensors will play a noncooperative game by reporting their highest state to the datacenter for the next $L$ steps before they re-cooperate. Nevertheless, the priority of different sensors should be taken into account in the mechanism.

\subsubsection{Clustering} 
In wireless ad hoc networks, due to a large number of wireless nodes, one of the efficient ways to reduce energy consumption of the nodes is using a clustering technique. In~\cite{Sun2011ARepeated}, the authors considered a clustering problem in mobile ad hoc networks. The authors first proposed an algorithm to cluster the nodes into groups. Then, the algorithm finds a cluster head for each group. The cluster head is able to communicate with all the nodes in the cluster. However, since the cluster head will consume more energy than others, the nodes may not be willing to send true information to avoid becoming a cluster head. To resolve this issue, the authors used a repeated game to model the interaction among nodes. In the game, the nodes are the players, and they can choose one of two actions, i.e., ``honestly'' or ``dishonestly'' reporting information. The payoff of the node is the gain that is received from successfully transmitting packets. To detect selfish nodes and enforce the nodes to cooperate, the authors proposed a limited punishment mechanism. It is shown that the nodes have no incentive to deviate from the cooperation, and thus they have to report information honestly. 

\subsubsection{Security} 
In the context of wireless sensor network, the authors in~\cite{Agah2007Preventing} used a repeated game to prevent denial-of-service (DoS) attacks from malicious sensor nodes. The repeated game is to model the interaction between the intrusion detection (ID) node and a set of sensor nodes. The ID node acts as a base station with the objective to monitor behaviors of sensors and punish them when they defect through using reputation as presented in~\cite{Michiardi2002Core}. Therefore, the ID node will have two actions, namely, punish or not punish. Differently, the sensor will have two actions, i.e., cooperate or defect. When the sensor cooperates by forwarding packets, it will not be punished, and thus it will gain rewards. However, if the sensor defects and it is detected by the ID node, it will be punished forever and will be out of the cooperation. Consequently, this sensor will gain nothing after leaving the cooperation, while other sensors still in the cooperation gain rewards. The advantage of this technique is that it handles not only selfish players, but also malicious users who are aiming to damage others.

Similar to~\cite{Agah2007Preventing}, the authors in~\cite{Estiri2010AGame} also considered the interaction between the ID node and a set of sensors. However, while in~\cite{Agah2007Preventing} after a selfish/malicious node is detected, it will be punished forever, in~\cite{Estiri2010AGame}, the node will be punished for an appropriate number of periods. Then, this node can re-participate in the cooperation again if it accepts to forward packets from the other nodes. This scheme will help selfish nodes to have an opportunity to re-cooperate in order to improve the network performance. 

\begin{figure*}[htb]
\centering
\includegraphics[scale=0.65]{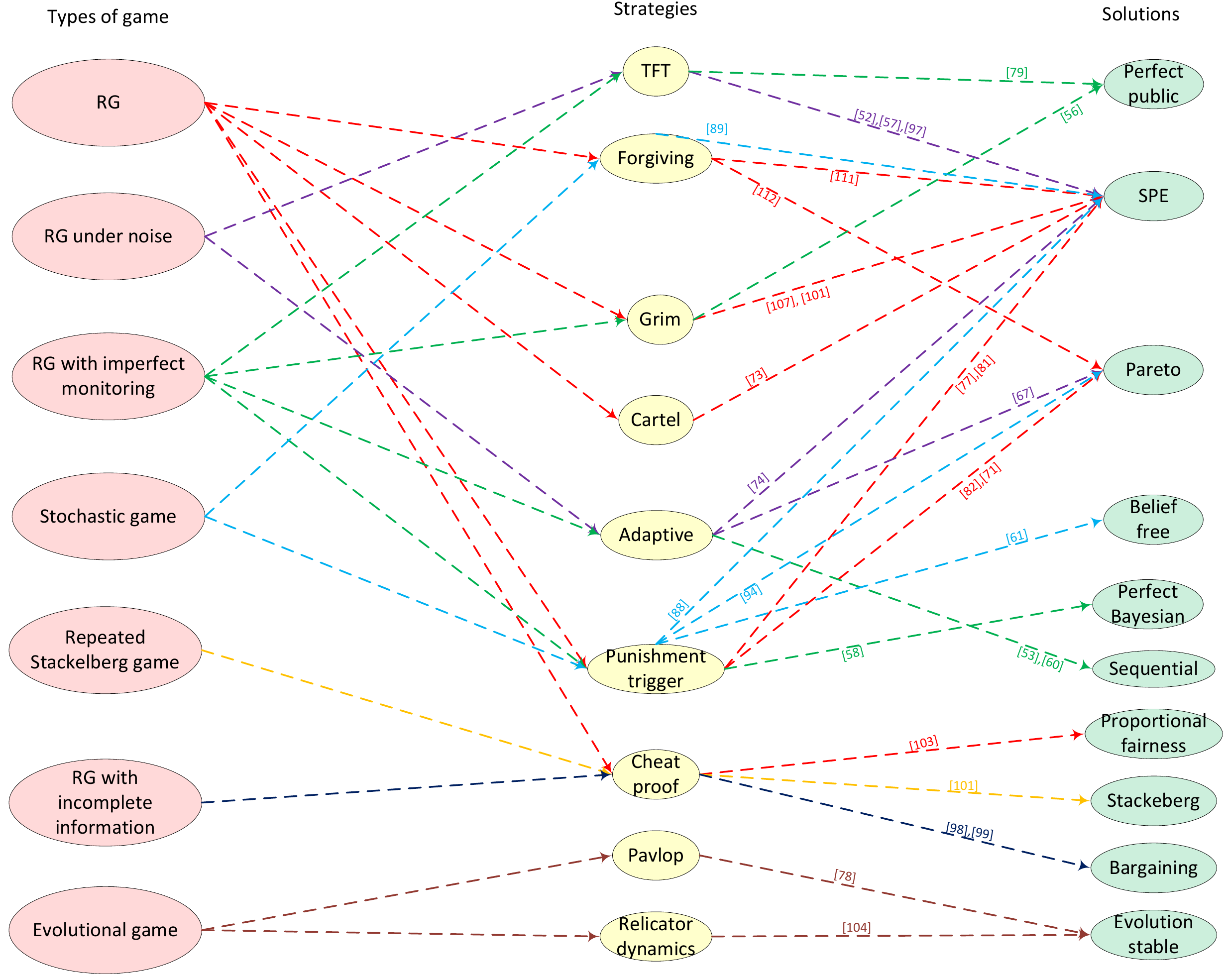}
\caption{Relation among types of repeated games, strategies, and solutions (RG = repeated game, SPE = subgame perfect equilibrium, TFT = Tit-for-Tat) in wireless ad hoc networks.}
\label{fig:chap5_WAHN_sum2}
\end{figure*}

\subsubsection{Multi-network Cooperation} 

The authors in~\cite{Quer2013Inter} considered the interaction between two wireless ad hoc networks through using a repeated game. In this game, the players are two wireless operators. Each operator has a set of wireless nodes, and if two operators cooperate by sharing nodes for relaying traffic, they can help to increase QoS performance for the entire network. In the cooperation, the operator determines how many nodes and which nodes should be shared. The payoff is defined as a function of its cost (e.g., bandwidth usage). In the repeated game, with a trigger strategy (that is similar to the grim strategy) and some appropriate conditions of the discount factor, there exists the subgame perfect equilibrium. Furthermore, to find the optimal cooperation action profile for both networks, the Nash bargaining solution~\cite{Owen2001Game} is applied. The results show that the proposed solution can achieve the performance close to that of a fully cooperative approach. The system model can be extended by considering different types of networks, e.g., a relay cellular network and multihop WiFi network.

\textbf{Summary:} From Table~\ref{table_WAHNs_sum}, we observe that there are many repeated game models developed for wireless ad hoc networks. Majority of them are for solving the packet forwarding problem. However, in wireless ad hoc networks, there are also many important problems, e.g., quality-of-service, multicasting, security, as presented in~\cite{Rubinstein2006Asurvey} and \cite{Sesay2004Asurvey} which are not well examined using repeated games. Moreover, from Fig.~\ref{fig:chap5_WAHN_sum2}, we observe that compared with the repeated game models for cellular and WLANs, those for wireless ad hoc networks use more diverse variations of repeated games and solution concepts.

\begin{table*}
\caption{Applications of repeated games in wireless ad-hoc networks} 
\label{table_WAHNs_sum}

\begin{centering}
\begin{tabular}{|>{\centering\arraybackslash}m{1.1cm}|>{\centering\arraybackslash}m{0.9cm}|>{\centering\arraybackslash}m{2.2cm}|>{\centering\arraybackslash}m{3.5cm}|>{\centering\arraybackslash}m{2.8cm}|>{\centering\arraybackslash}m{2.3cm}|>{\centering\arraybackslash}m{1.7cm}|}
\hline 
\textbf{Problem} & \textbf{Article} & \textbf{Players} & \textbf{Actions} & \textbf{Payoff} & \textbf{Strategy} & \textbf{Solution}\tabularnewline
\hline 
\hline 
\parbox[t]{2mm}{\multirow{30}{*}{\rotatebox[origin=c]{90}{Packet forwarding problem in multihop networks}}} 
& \cite{Milan2006Achieving} & wireless nodes & dropping probability & benefit minus cost & Generous TFT & SPE		\tabularnewline 	\cline{2-7} 
& \cite{Jaramillo2007DARWIN} & wireless nodes & cooperate or defect & benefit minus cost & Contrite TFT & SPE		\tabularnewline 	\cline{2-7} 
& \cite{Ji2010ABelief}~\cite{Wenjing2011Cooperation} & wireless nodes & forward or dropping & benefit minus cost & adaptive & sequential 		\tabularnewline 	\cline{2-7} 
& \cite{Kamhoua2010Belief} & wireless nodes & cooperate or defect & benefit minus cost & belief-free equilibrium & belief-free		\tabularnewline 	\cline{2-7} 
& \cite{Pandana2008Cooperation} & wireless nodes & packet forwarding probabilities & data transmission efficiency & punishment trigger & perfect Bayesian 
\tabularnewline 	\cline{2-7} 
& \cite{Srivastava2006Equilibria} & wireless nodes & proportion of packets forwarded & benefit minus cost & Grim & perfect public		\tabularnewline 	\cline{2-7} 
& \cite{Sun2011APower} & wireless nodes & transmission power level and forwarding probability & the ratio of achieve throughput and power consumption & adaptive & Pareto optimal		\tabularnewline 	\cline{2-7} 
& \cite{Srinivasan2003Cooperation} & wireless nodes & forward or reject & benefit minus cost & Generous TFT & Pareto optimal		\tabularnewline 	\cline{2-7} 
& \cite{Urpi2003Modelling} & wireless nodes & the number of packets sent and forwarded & benefit minus cost & TFT & SPE 	\tabularnewline 	\cline{2-7} 
& \cite{Bandyopadhyay2005AGame} & wireless nodes & forward or reject & benefit minus cost & improvement of Generous TFT & SPE 	\tabularnewline 	\cline{2-7} 
& \cite{Guan2010AModified} & wireless nodes & forward or drop & benefit minus cost & restorative trigger & SPE 	\tabularnewline 	\cline{2-7} 
& \cite{Han2005ASelf} & wireless nodes & forward probability & benefit minus cost & punishment trigger & Pareto optimal 	\tabularnewline 	\cline{2-7} 
& \cite{Mohamed2012Cooperation} & wireless nodes & forward probability & cooperation level & The Weakest Link scheme & SPE 	\tabularnewline 	\cline{2-7} 
& \cite{Zhu2008AGame} & wireless nodes & cooperate or defect & benefit minus cost & Cartel maintenance profit sharing & SPE 	\tabularnewline 	\cline{2-7} 
& \cite{Ng2010Game} & wireless nodes & report information & benefit minus cost & T-segmented TFT & perfect public 	\tabularnewline 	\cline{2-7} 
& \cite{Yan2008Cooperative} & wireless nodes & forward or drop & benefit minus cost & adaptive  & SPE 	\tabularnewline 	\cline{2-7} 
& \cite{Kamhoua2010Mitigating} & wireless nodes & cooperate or defect & benefit minus cost & Pavlop & Evolutionary Stable 	\tabularnewline 	\cline{2-7} 
\hline 
\parbox[t]{2mm}{\multirow{5}{*}{\rotatebox[origin=c]{90}{Energy}}} 
\parbox[t]{2mm}{\multirow{5}{*}{\rotatebox[origin=c]{90}{efficiency}}}
& \cite{Yang2007Energy} & source nodes & relay or not relay & energy conservation & conditional trigger   & SPE	\tabularnewline \cline{2-7} 
& \cite{Mohammed2014Cooperation} & source nodes & relay or not relay & energy conservation & punishment trigger  & Pareto optimal	\tabularnewline \cline{2-7}
& \cite{Brown2011AGame} & source nodes & relay or not relay & energy conservation & punishment trigger  & Pareto optimal	\tabularnewline \cline{2-7} 
& \cite{Li2010RelayIET} & source nodes & cooperate or defect & achievable throughput & punishment trigger  & Pareto optimal	\tabularnewline \cline{2-7} 
& \cite{Chen2008AGame} & source nodes & cooperate or defect & a function of SNR & forgiving & SPE \tabularnewline \cline{2-7} 
\hline 
\parbox[t]{2mm}{\multirow{10}{*}{\rotatebox[origin=c]{90}{Media streaming}}} 
& \cite{Peng2008Enhancing} & wireless peers & accept or reject & benefit minus cost &  proportion increment TFT & SPE		\tabularnewline 	\cline{2-7} 
& \cite{Lin2009Cheat} & wireless peers & report buffer information and the number shared chunks & gain minus cost &  cheat-proof & time-restricted bargaining	\tabularnewline 	\cline{2-7}
& \cite{Lin2010Cooperation} & wireless peers & report buffer information and the number shared chunks & gain minus cost &  cheat-proof  & time-sensitive bargaining 	\tabularnewline 	\cline{2-7}
& \cite{Chen2012Analysis} & wireless peers & relay or not relay & gain minus cost &  cheat-proof & proportional fairness optimality	\tabularnewline 	\cline{2-7}
& \cite{Wu2012AGame} & content server and peers & the content server (CS): rewards to nodes, nodes: transmission rates & the CS: minimizes the total cost, peers: maximizes their rewards &   cheat-proof &  Stackelberg equilibrium 	\tabularnewline 	\cline{2-7}
\hline 
\multirow{2}*{\parbox{\linewidth}{\centering Multiple Access}}
& \cite{Tembine2007Multiple} & wireless nodes & transmit or stay quiet & reward minus cost & replicator dynamics and imitate better dynamics  & Evolutionary stable	\tabularnewline 	\cline{2-7} 
& \cite{Chen2007Selfishness} & wireless nodes & transmit or stay quiet & reward minus cost & TFT & Pareto optimal		\tabularnewline 	\cline{2-7} 
& \cite{Qu2012Efficient} & wireless sensor nodes & report energy level and number of packets &  the amount of data successfully transmitted & forgiving & Pareto optimal		\tabularnewline 	\cline{2-7} 
\hline 
\multirow{0}*{\parbox{\linewidth}{\centering Clustering}}
& \cite{Sun2011ARepeated} & wireless nodes & report honestly or dishonestly & gain minus cost & Generous TFT & SPE		\tabularnewline 	\cline{2-7} 
\hline 
\multirow{0}*{\parbox{\linewidth}{\centering Security}}
& \cite{Agah2007Preventing} & intrusion detection (ID) and sensors & ID: miss or catch, sensors: regular or malicious & gain minus cost & Grim & SPE		\tabularnewline 	\cline{2-7} 
& \cite{Estiri2010AGame} & intrusion detection (ID) and sensors & ID: miss or catch, sensors: regular or malicious & gain minus cost & forgiving & 	SPE	\tabularnewline 	\cline{2-7} 
\hline 
\multirow{0}*{\parbox{\linewidth}{\centering Operators Coop}}
& \cite{Quer2013Inter} & wireless ad-hoc operators & cooperate or defect & a function of cost metric & Grim & SPE	\tabularnewline 	\cline{2-7} 
\hline 
\end{tabular}
\par\end{centering}

\end{table*}

\section{Applications of Repeated Games in Cognitive Radio Networks}
\label{sec:CRNs}
This section discusses the applications of repeated games in cognitive radio networks (CRNs). CRNs are intelligent communication networks that have been designed to improve the spectrum utilization and transmission efficiency. CRNs allow unlicensed users, called secondary users, to access opportunistically available spectrum allocated to licensed users, called primary users~\cite{Hossain2009Dynamic}. In CRNs, before accessing a licensed channel, unlicensed users need to check the channel state and then decide whether to access the channel or not. Therefore, the applications of repeated games in CRNs mainly focus on solving two major problems, namely, {\em spectrum sensing} and {\em spectrum usage}. Additionally, there are some applications in pricing competition problem, called {\em spectrum trading}, that is also reviewed in this section. 

\subsection{Spectrum Sensing} 

Spectrum sensing is a task of secondary users (SUs) to detect the presence of primary users (PUs) through many sensing techniques such as energy detection, matched filter detection, and cyclostationary feature detection as presented in~\cite{Yucek2009Asurvey}. The effectiveness of spectrum sensing is limited by geographic separation or channel fading, and thus SUs need to cooperate to overcome this problem. However, SUs may not be interested in cooperating since they have to exchange sensing results that consumes a certain amount of resources. Consequently, a repeated game is a useful tool to motivate the SUs to cooperate with the aim to improve the quality of spectrum sensing thereby enhancing the network performance. 

In~\cite{Song2009Achieving}, the authors proposed using repeated games to model the interaction among secondary users (SUs) when they want to cooperate to sense a common spectrum allocated by a primary user. In this game, the SUs have two actions, namely, cooperate or not cooperate. If the SUs are cooperative, they will share their spectrum sensing results with others. However, this consumes energy due to broadcasting information. By contrast, if they are not cooperative, they will gain and lose nothing. When the profit obtained from sharing sensing results (i.e., $b$) is lower than the cost for broadcasting information (i.e., $c$), the SUs will never be cooperative. Otherwise, for $b>c$, there is an incentive for the SUs to cooperate if the interaction among SUs is repeated for sufficiently many periods. In the repeated game, the authors investigated two strategies, namely, grim and carrot-and-stick, for two scenarios, i.e., with and without transmission loss. With the transmission loss, the SUs may not receive information from others. Consequently, the SUs will not cooperate and their performance will be detrimental. It is shown that as the packet loss probability increases, the ratio $b/c$ must be increased to guarantee the cooperation among SUs for both the strategies. Furthermore, the authors concluded that the value of $b$ from the carrot-and-stick strategy must be twice larger than that of the grim strategy to make SUs cooperate. 

In~\cite{Kondareddy2011Enforcing}, Kondareddy~\textit{et al.} extended the cooperative sensing model to packet relay in CRNs. The authors formulated the cross-layer game which is a combination of a cooperative sensing game and a packet forwarding game. In this cross-layer game, the SUs choose actions from a set of four independent actions \{Share (S), Do not Share (DS), Forward (F), and Do not Forward (DF)\}. Therefore, there are four possible actions for each player to choose from, i.e., S-F, S-DF, DS-F, DS-DF. The payoff is the profit. It is shown that for the one-shot cross-layer game, the Nash equilibrium is the mutual defection, i.e., the players will choose actions (DS-DF) and this leads to the poor network performance. Nevertheless, if the cross-layer game is repeated, by using the TFT strategy and under some conditions for the discount factor, the authors proved that the subgame perfect Nash equilibrium is the mutual cooperation strategy for the SUs. The most noticeable achievement of the repeated cross-layer game is that the players can achieve cooperation in both spectrum sensing and packet forwarding by using punishments for packet forwarding action only. Moreover, the authors considered imperfect observations that may happen by mistake or due to noise. The game is modeled as the Prisoner's Dilemma with noise~\cite{Wu1995How}. By using the generous TFT strategy~\cite{Wu1995How}, the authors showed that the mutual cooperation strategy for the SUs is the subgame perfect Nash equilibrium. 

In~\cite{Song2009Achieving} and~\cite{Kondareddy2011Enforcing}, the authors presented the solutions to deal with selfish SUs. However, there can be malicious SUs sharing wrong information. For example, when the primary channel is idle, the malicious SU can report a busy channel, and hence no other SUs can use this channel. In~\cite{Zhendong2013Anovel}, the repeated game model was developed for this problem. The game is presented in Table~\ref{table_sec3_Song}. $U_h$ and $U_d$ are the profits when the SUs choose $co$ and $no$ (i.e., cooperate and not cooperate) actions, respectively. $C_s$ is the cost when SUs participate in spectrum sensing and $C_r$ is the cost for overhearing the sensing information from other SUs. From this payoff matrix, a malicious SU will gain a negative profit if it is noncooperative. Consequently, the malicious SU has no incentive to defect. A distributed trusted model is used to identify the malicious SU through the evaluation of SUs' reputation. Based on the action history of SUs, the SU can update the global reputation for its all neighbors. If the global reputation of any SU reaches a predefined limit, this SU will be identified to be a malicious and will be punished forever. 

\begin{table}[h!]
\caption{Payoff Matrix} 
\label{table_sec3_Song}
\begin{tabular}{l|l|c|c|c|c|}
\cline{3-4}
\multicolumn{2}{c|}{} & cooperate (co)  & not cooperate (no) \\		\cline{2-4}
\multirow{2}{*}
& cooperate (co) & $U_h$-$C_r$-$C_s$, $U_h$-$C_r$-$C_s$ & -$C_s$-$C_r$, $U_d$-$C_r$  \\	\cline{2-4}
& not cooperate (no) & $U_d$-$C_r$, -$C_s$-$C_r$ & -$C_r$,-$C_r$ \\	\cline{2-4}
\end{tabular}

\end{table}

\subsection{Spectrum Usage Management} 

\begin{figure}[htb]
\centering
\includegraphics[scale=0.5]{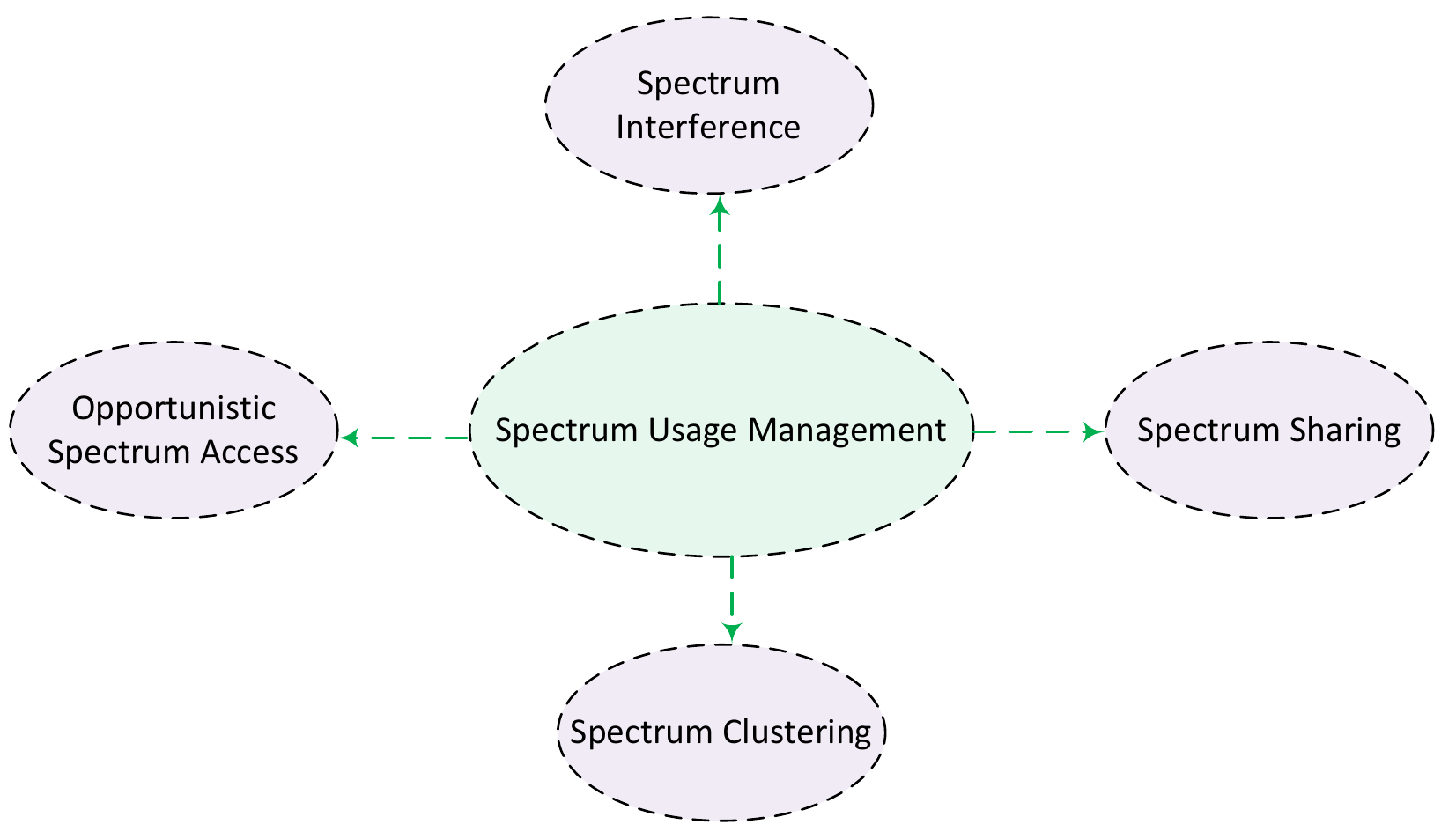}
\caption{Spectrum usage issues.}
\label{fig:sec4_CRN_spectrum_usage_management}
\end{figure}

After spectrum sensing, spectrum usage or spectrum access will be performed. In this section, we review the applications of repeated games to help secondary users (SUs) to access available channels. We discuss four major problems which are illustrated in Fig.~\ref{fig:sec4_CRN_spectrum_usage_management}.

\subsubsection{Opportunistic Spectrum Access} 

Unlike multiple access control in cellular and WLANs presented in Section~\ref{subsection:MAC}, in CRNs when an SU accesses a licensed channel, it takes not only the existence of other SUs, but also the presence of PUs into account. Thus, the spectrum access in CRNs is more dynamic and more complex than that of cellular and WLANs networks. One of the popular solutions is using a Markov model for analysis and optimization. Then a repeated game is used as a tool to encourage the cooperation for SUs. 

In~\cite{Wang2007Self}, the authors developed the distributed dynamic spectrum access scheme for SUs and modeled non-cooperation relation among SUs as an infinitely repeated game. The players are the SUs, the actions are to access the idle channel or not, and the payoff is the average throughput. Since the SUs can access only an idle channel, the authors used a primary-prioritized Markov dynamic spectrum access~\cite{Wang2007Primary} to analyze the impact of PUs to SUs and derive optimal SU's channel access probabilities. Then, by using the Folk theorem, it is shown that with the discount factor close enough to one, the game can achieve an efficient Nash equilibrium. To do so, the authors introduced the distributed self-learning algorithm for the SUs to obtain the optimal access probabilities. In the learning algorithm, the SUs start with low access channel probabilities. In each iteration, the SUs increase the access channel probability and observe the payoff. If the payoff is improved, then the SU will continue increasing the channel access probability. Otherwise, the SU stops. However, the convergence of the learning algorithm is open for further study.

In~\cite{Wang2007Self}, a Markov model was developed to optimize channel access probabilities for SUs. However, as the number of SUs is large, the computation complexity increases exponentially which causes difficulties in finding optimal policies for the SUs. Therefore, the authors in~\cite{Li2010Dynamic} proposed a reduced-complexity suboptimal Markov model. The method is to separate users' behaviors, thereby significantly reducing the state space of the Markov model. To motivate the SUs to cooperate, a repeated game is used. Additionally, the authors used the learning algorithm based on the policy gradient method to update the channel access probability for the SUs and proposed the solutions to detect and enforce deviating users into the cooperation which were not addressed in~\cite{Wang2007Self}. 

While~\cite{Wang2007Self} and~\cite{Li2010Dynamic} used learning algorithms to control channel access probabilities, in~\cite{Yan2011Game}, the authors examined the CSMA/CA protocol to resolve collisions when SUs access channels. In the CSMA/CA protocol, when the player transmits data and collision occurs, it selects a random number $w$ and retransmits after $w$ time slots. Selfish users can take advantage by reducing the parameter $w$ thereby increasing the channel access probability. Thus, the authors used the detection method based on the evaluation of average throughput. In particular, within a monitoring duration, if a player $i$ detects that player $j$ has the average throughput higher than $i$'s throughput, the player $i$ will notify all other players in the network. Then, the honest players will punish the selfish player $j$ by jamming the $j$'s transmission in the next certain number of game stages. The authors then established the conditions for the monitoring duration and punishment period such that the punishment will be effective to deter the deviation.

\subsubsection{Interference Control} 

In underlay spectrum access, i.e., SUs access a common channel simultaneously, mutual interference occurs. To control the impacts of interference, the transmission power levels at the transmitters have to be carefully optimized. There are two types of networks studied in the literature, namely, unlicensed and licensed band networks. In the unlicensed networks, SUs needs to control the interference only with other SUs. Conversely, in licensed networks, SUs have to avoid the interference with PUs as well.  

\paragraph{Unlicensed band networks}

In~\cite{Etkin2005Spectrum}, Etkin~\textit{et al.} studied the spectrum sharing problem among SUs for interference-constrained networks. The authors considered the scenario with $M$ systems (SUs) coexisting in the same area. Each system includes a pair of a transmitter and a receiver. The systems share the same unlicensed channel, and this is modeled as a Gaussian interference channel. If the systems are not cooperative, they will transmit data with high power levels that leads to severe interference and reducing significantly network throughput. Therefore, a repeated game is used to encourage cooperation. In this game, the players are the SUs, the actions are transmission rates and the payoff is the sum rates of SUs. The TFT strategy is used to punish selfish users. Based on the Folk theorem, it is proven that the proposed strategy achieves the subgame perfect Nash equilibrium when the discount factor is set sufficiently close to one. 

To deal with liars who may gain benefit by communicating fake information (e.g., fake channel measurement),  Etkin~\textit{et al.} extended their work in~\cite{Etkin2005Spectrum} by introducing a truth telling method~\cite{Etkin2007Spectrum}. This method is based on the protocol with detection mechanisms using test messages exchanged among SUs together with energy detection. Thus, the repeated game is modified by adding an initial stage to exchange and verify the channel measurement in all stages of the repeated game. If there is any deviation detected, a punishment is triggered in which all the transmitters spread their power over the total bandwidth in the rest of the game. The proposed method not only detects deviations, but also liars who send fake information. However, this method requires extensive information exchange, which could be a shortcoming for energy-limited networks.

In~\cite{Etkin2005Spectrum} and~\cite{Etkin2007Spectrum}, after a deviation is detected, all users will play a noncooperative game forever. This strategy can be inefficient. To overcome this drawback, the authors in~\cite{Wu2009Repeated} adopted different strategy, namely, publish-and-forgive. In this strategy, after detecting the deviation, the noncooperative game will be played for the next $T-1$ periods, and then the players will cooperate again. Besides the cooperation criterion which maximizes the total throughput (MTT) for the network as presented in~\cite{Etkin2005Spectrum} and \cite{Etkin2007Spectrum}, the authors in \cite{Wu2009Repeated} considered different cooperation criterion, i.e., approximated proportional fairness criterion (APF) to help the players with poor channel conditions have an equal opportunity to access a channel. The method to detect liars in \cite{Wu2009Repeated} is proven to be less complex and easier to implement that that in~\cite{Etkin2007Spectrum}. 

To reduce interference and improve the network performance for SUs, the authors in~\cite{Bennis2009Ahierarchical} considered using multi-carrier transmission and studied the interactions among SUs communicating over the same frequency band composing of multiple carriers. In the multi-carrier system, the SUs have to choose the transmission power levels on each carrier to maximize their own payoff. Again, a repeated game is used to encourage the SUs to cooperate. In this game, the SUs cooperate by using Pareto transmission strategies. If any SU deviates from the cooperation, then the other SUs will choose the noncooperative Nash equilibrium strategy in the remaining of the game. 

In all above work, to control interference, SUs have to adjust their transmission power levels. In~\cite{Xiao2012Repeatedgames}, the authors used an intervention scheme~\cite{Park2012thetheory} for repeated games with the aim to enlarge the limit set of equilibrium payoffs and loosen the conditions for the discount factor that needs to be close to one in general repeated games. In the intervention scheme, an intervention device is introduced to observe and intervene interactions among SUs. Specifically, the intervention device monitors the actions of SUs and then makes decisions (e.g., punishment) to deter deviation and enforce cooperation. With the intervention device, in the repeated game, the players include SUs and the intervention device. The actions are transmission power levels and the payoff is throughput. The protocol is proposed to maximize a joint objective function (e.g., total throughput of the network). This protocol is shown to achieve the subgame perfect equilibrium of the repeated game. The author proposed a trigger policy that makes players cooperate. If any user deviates from the cooperation, it will be punished for certain number of periods. Simulation results show good performance of using intervention in terms of sum payoff and max-min fairness. Additionally, with the support of the intervention, the set of equilibrium payoffs are extended and the discount factor condition can be reduced by around 50\%.

\paragraph{Licensed band networks}

\begin{figure}[htb]
\centering
\includegraphics[scale=0.47]{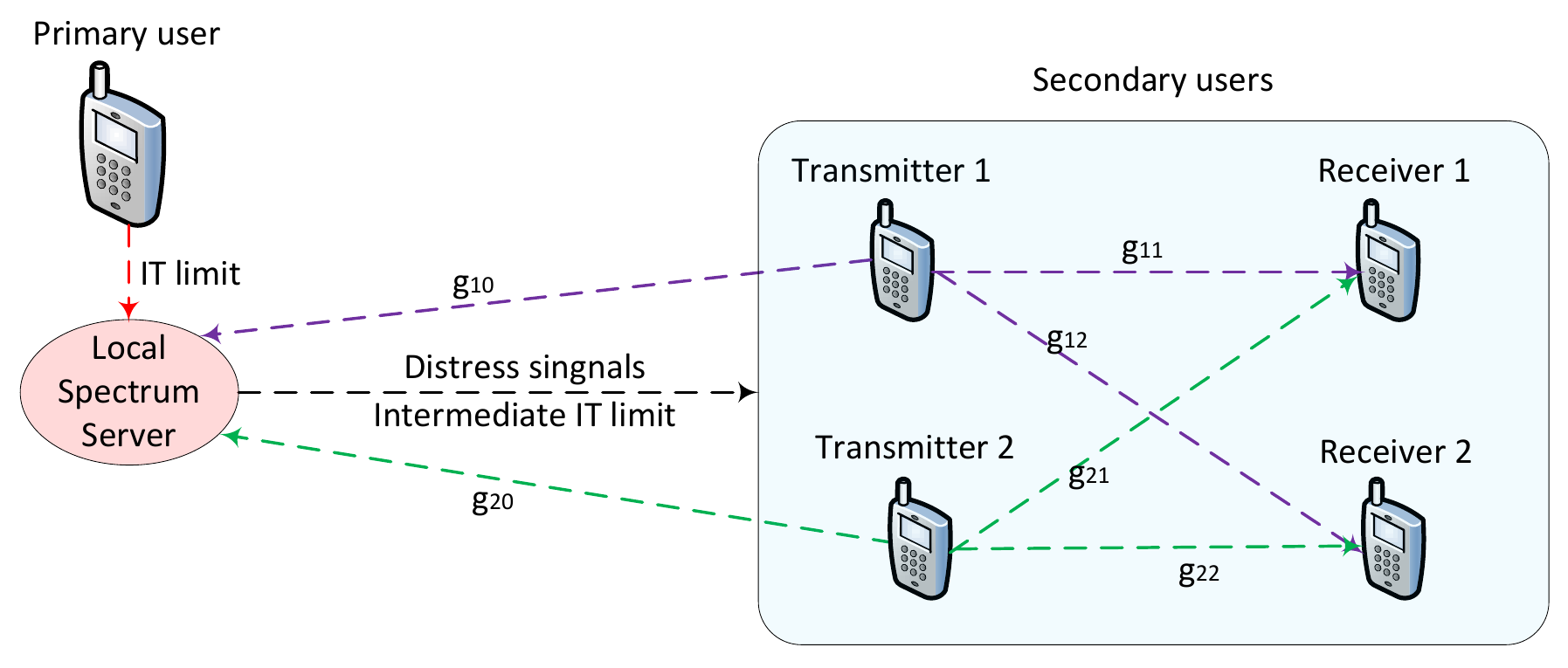}
\caption{An example for using local spectrum server. $g_{ij}$ is the channel gain between entities $i$ and $j$, where $0$ is the local spectrum server.}
\label{fig:chap4_Xiao2012Dynamic}
\end{figure}

To reduce the interference with PUs on a licensed band, the authors in~\cite{Xiao2012Dynamic} adopted a local spectrum server (LSS) as presented in Fig.~\ref{fig:chap4_Xiao2012Dynamic}. The LSS is to control the interference between SUs and PU. The PU specifies an acceptable interference temperature to the LSS. The LSS also monitors the activities of SUs in which the monitoring is imperfect. Thus, a repeated game with imperfect monitoring~\cite{Fudenberg1994thefolk} is used. After the SUs transmit data, the LSS measures and compares the interference with a predefined threshold. If the interference is higher than the threshold, the LSS will send a warning message to the SUs. Based on the received information from the LSS, the SUs will adjust power levels accordingly. The authors then introduced a deviation-proof policy implemented in a distributed fashion for the SUs. In the policy, the SUs compute indexes that measure ``urgency'' for their data transmission. The SU with the highest index is chosen to transmit data in a time slot. It is proven that with the proposed policy, the SUs will achieve an optimal operating point that will give them no incentive to deviate and the proposed policy converges to a perfect public equilibrium.

\subsubsection{Spectrum Sharing} 

Spectrum sharing is another method used to prevent collision and interference. The idea is to divide the spectrum into non-overlap sub-channels and then allocate them to users. However, different from bandwidth allocation in cellular and WLANs networks which have a central node (e.g., an access point or base station) to control the allocation, in CRNs, the SUs have to unify or self-define the amount of used spectrum without any support from a central node. Thus, SUs may be selfish and they may excessively use spectrum, thereby significantly degrading the network performance. In this section we review the applications of repeated games for spectrum sharing. 

In~\cite{Berlemann2005Strategies}, the authors considered the spectrum sharing problem between two WLAN networks when they coexist in the same area and share a common unlicensed spectrum to serve their users. The WLANs need to compete with each other to use the spectrum. A repeated game is used to encourage WLANs as the players to cooperate. The actions of WLANs are choosing the amount of spectrum. The action is defection if it maximizes the payoff. Alternatively, the action is cooperation if the MAC parameters of the WLAN are less aggressive and take opponents' actions into account. The payoff is QoS measure of the players in terms of achievable throughput, period length, and delay. The authors considered static strategies, i.e., always cooperate, always defect, random policy (50\% cooperate and 50\% defect), and dynamic strategies, i.e., grim and TFT. Numerical results show that different strategies are appropriate for different circumstances depending on the QoS requirements. For example, when the QoS requirements of player 1 is (throughput=0.1, period length=0.05, and delay=0.02) and for player 2 is (throughput=0.4, period length=0.031, and delay=0.02), the authors showed that the best strategy for player 1 is always cooperate. However, the best strategy for player 2 is TFT. 

In~\cite{Jia2013Asymmetric}, the authors studied a cognitive radio network with two SUs sharing a common unlicensed channel. The SUs can cooperate by using half of bandwidth through the frequency division multiplexing (FDM) technique or not cooperate by spreading their transmit power, called STP, over the entire bandwidth. The authors proposed a scheme to detect the selfishness of a rival player. The authors then designed a guard interval for the SUs, if a player starts by cooperating and after the guard interval, the rival player still does not cooperate, the cooperative player will choose to perform STP (i.e., not cooperate). Through the analysis and simulation, it is shown that the proposed strategy can achieve mutual cooperation and the equilibrium is stable. 

In previous papers, the spectrum usage schemes are random, i.e., the demand parameters are selected randomly at the beginning of the game~\cite{Berlemann2005Strategies}, or they work under some predefined assumption, for example, SUs choose FDM or STP~\cite{Jia2013Asymmetric}. In~\cite{Liu2010Deviation}, the authors proposed an alternative method for the use of bandwidth through exchanging information among SUs. In particular, before making decisions to access the channels, an access point reports its traffic demands to all other access points in the same network. Based on the information received from the access points and its demand, the access point will make channel usage decision. If the interaction among access points takes place one time, the access points will play a noncooperative game by claiming the highest traffic demand. However, when the interaction lasts for multiple periods, a repeated game can be used to model and enforce the access points to cooperate by informing the true information. In~\cite{Liu2010Deviation}, it is shown that by using trigger punishment strategies~\cite{Osborne2004AnIntroduction}, there exists a subgame perfect equilibrium for the proposed repeated game. The length of the punishment phase is determined in a similar way to that in~\cite{Wu2009Repeated} and with the punishment phase, the selfish access point will have no incentive to deviate. 

Extended from~\cite{Liu2010Deviation}, in~\cite{Liu2010CheatProof}, the authors used two specific cheat-proof mechanisms along with a method to identify the information from access points which was not provided in~\cite{Liu2010Deviation}. The first mechanism is based on the ``transfer'' in Bayesian theory~\cite{Fudenberg1991GameThery}. With this mechanism, the access points are asked to pay a tax when they use high traffic demands and the access points with low bandwidth demand will be compensated from the tax. The tax will be increased according to the declaration of traffic demand, and thus the access points have to balance between traffic demand and monetary budget. Then, the payoff of the access points is defined such that it can gain the highest payoff if the access points report true information. In the second mechanism, the authors studied a statistical cheat-proof strategy for the repeated game. Here, the access point is considered as a deviator if it overuses the allocated spectrum and after being detected, the selfish access point will not be allocated bandwidth until its average usage is under a safety range.

In~\cite{Xie2012Unlicensed}, Xie~\textit{et al.} studied the spectrum sharing problem between satellites and terrestrial cognitive radio networks (CRNs). The authors considered a scenario where there is a satellite in an area with two other CRNs (e.g., two base stations), and they compete for a common unlicensed spectrum band allocated by primary users. The satellite can interfere with both CRNs; however, there is no interference between two CRNs. Since the movement of a satellite is often very fast, the satellite is in the interference area for very short time period. Due to the short interaction, the repeated game considered in this paper is a finitely repeated game. The players of the game are satellite (BS1), and two CRNs (BS2, BS3). The authors assumed that the size of the band is totally $2W$, and thus the players can choose to occupy one band (i.e., $W$) or all two bands (i.e., $2W$). The payoff function is the achievable rate. The players know the end of the game, and thus there is a Nash equilibrium, which is the noncooperative strategy where the players always spread their power over all 2W bands. Additionally, the authors addressed the issue of partially blind observations. Specifically, BS2 and BS3 cannot observe the actions of each other. Moreover, BS2 and BS3 may not know the existence of the satellite and cannot observe its actions. To resolve partially blind observations, the authors proposed using refreshing TFT (R-TFT). The R-TFT is an extension of the TFT strategy where the cooperation is refreshed quickly once one of the players cooperates. When the observations of the players are affected by noise, the TFT strategy becomes an ineffective strategy. Therefore, the authors investigated two schemes, namely, refreshing contrite TFT (RC-TFT) and refreshing generous TFT (RG-TFT), which are based on the contrite TFT and generous TFT strategy in~\cite{Wu1995How}. Here, the refreshing process is applied when a satellite appears in the area. After an efficient equilibrium is obtained between the satellite and the BSs, the G-TFT and C-TFT are applied to circumvent an adverse impact from noise. 

\subsubsection{Cooperative Clustering} 

Recently, there have been some research work considering clustering problems in CRNs. In clustered CRNs, some SUs are grouped together to access a channel cooperatively. This can avoid an over-accessing problem when the number of SUs is large. 

Clustering is an effective energy saving solution for SUs in CRNs. Traditionally, SUs must transmit data directly to distant base stations, causing contention and consuming excessive energy. With clustering, some SUs can use short range communication to transfer data, thereby reducing congestion and energy consumption. However, when clusters are formed, there must be an effective way to motivate SUs to cooperate by transmitting data for each other. In~\cite{Zhang2012Acontext}, the authors developed a repeated packet forwarding game to model an interaction among SUs in the same cluster. Each SU has two actions, i.e., cooperate by forwarding packet for others or not. The payoff is the profits minus the cost for forwarding packets. It is shown that the grim strategy can lead to the Pareto-optimal Nash equilibrium where the players want to cooperate. 

\begin{figure}[htb]
\centering
\includegraphics[scale=0.5]{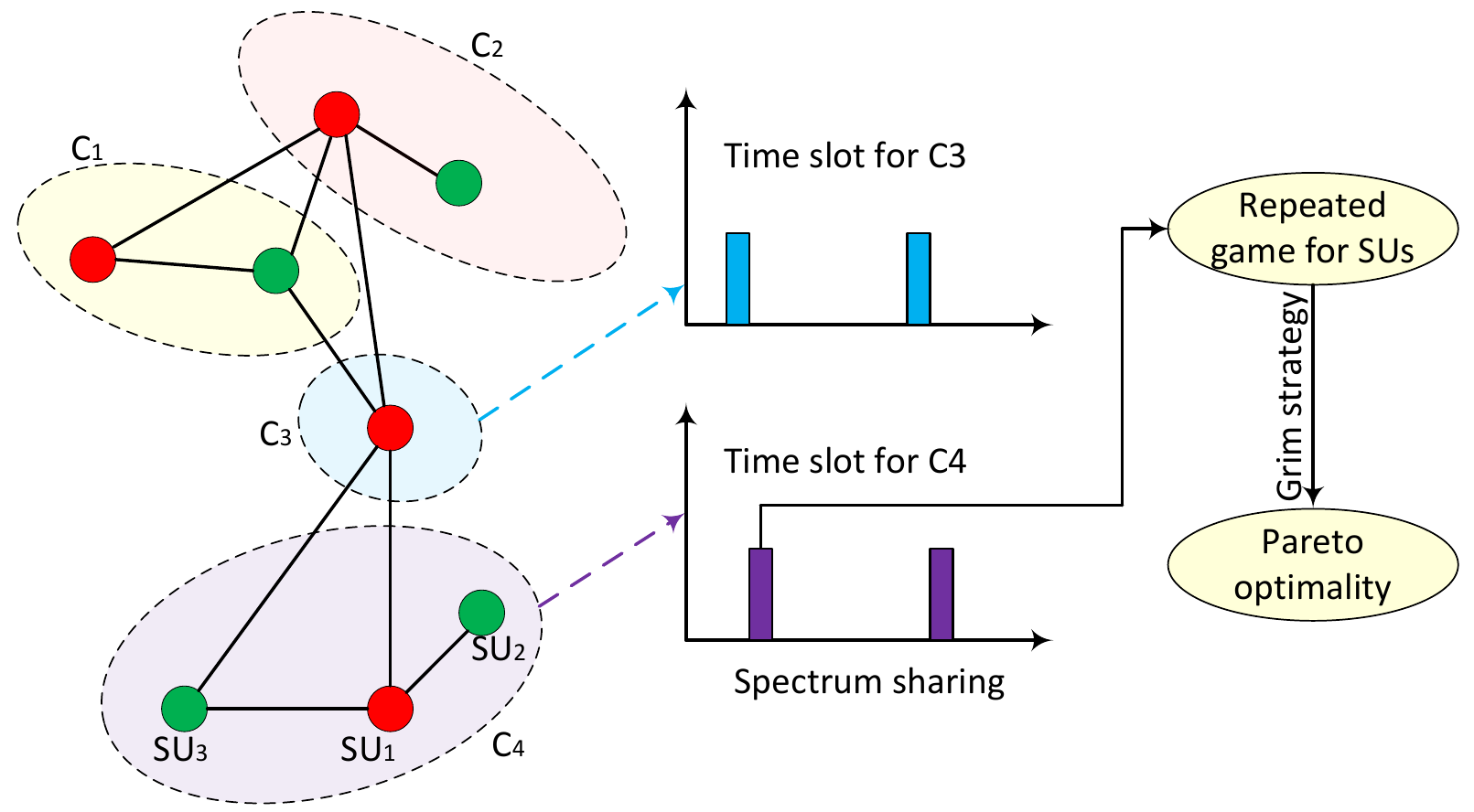}
\caption{A repeated game for clustered cognitive radio networks.}
\label{fig:chap4_Li2013Repeated}
\end{figure}

Li~\textit{et al.}~\cite{Li2013Repeated} extended the clustering model in CRNs proposed in~\cite{Zhang2012Acontext} by using cluster-head SU. In particular, each cluster will have a cluster-head SU which is responsible to receive packets from other SUs in the same cluster and transmit them over primary channels as shown in Fig.~\ref{fig:chap4_Li2013Repeated}. With the centralized model through the cluster-head SU, the SUs can avoid frequent collision and reduce delay of packet transmission. The idea of clustering and nomination cluster-head SU is based on the use of weight metrics, including the ideal degree, transmission power, and battery power as introduced in~\cite{Chatterjee2002WCA}. A repeated game is used to model the interaction between SUs in the same cluster, and through using the grim strategy, the authors proved that the game can achieve Pareto optimality. 

\begin{table*}
\caption{Applications of repeated games in cognitive radio networks} 
\label{table_CRNs_sum}

\begin{centering}
\begin{tabular}{|>{\centering\arraybackslash}m{1cm}|>{\centering\arraybackslash}m{1.2cm}|>{\centering\arraybackslash}m{2.6cm}|>{\centering\arraybackslash}m{3cm}|>{\centering\arraybackslash}m{2.8cm}|>{\centering\arraybackslash}m{2.5cm}|>{\centering\arraybackslash}m{1.8cm}|}
\hline 
\textbf{Problem} & \textbf{Article} & \textbf{Players} & \textbf{Actions} & \textbf{Payoff} & \textbf{Strategy} & \textbf{Solution}\tabularnewline
\hline 
\hline 
\parbox[t]{2mm}{\multirow{5}{*}{\rotatebox[origin=c]{90}{Spectrum}}} 
\parbox[t]{2mm}{\multirow{5}{*}{\rotatebox[origin=c]{90}{Sensing}}} 
& \cite{Song2009Achieving} & Secondary users & sharing spectrum results & profit minus cost & Carrot-and-Stick & SPE 	\tabularnewline 	\cline{2-7} 
& \cite{Kondareddy2011Enforcing} & Secondary users & sharing spectrum results and forwarding packets & profit minus cost & TFT and Generous-TFT & SPE	\tabularnewline	\cline{2-7} 
& \cite{Zhendong2013Anovel} & Secondary users and malicious uses & sharing spectrum results & profit minus cost & cheat-proof & SPE		\tabularnewline 	\cline{2-7} 
\hline 
\parbox[t]{2mm}{\multirow{30}{*}{\rotatebox[origin=c]{90}{Spectrum Usage Management}}} 
& \cite{Wang2007Self},~\cite{Li2010Dynamic} & Secondary users & channel access probability & achievable throughput & punishment trigger & SPE		\tabularnewline \cline{2-7} 
& \cite{Yan2011Game} & Secondary users & backoff counter  & achievable throughput & cheat-proof & Pareto optimal	\tabularnewline \cline{2-7} 
& \cite{Etkin2005Spectrum},~\cite{Etkin2007Spectrum} & pairs of transmitter-receiver & power levels & transmission rates & TFT & SPE 	\tabularnewline \cline{2-7} 
&\cite{Wu2009Repeated} & pairs of transmitter-receiver & power levels & achievable throughput & cheat-proof & SPE 	\tabularnewline \cline{2-7} 
& \cite{Bennis2009Ahierarchical} & pairs of transmitter-receiver & power levels on multi-carrier & achievable rates & Cartel maintain & Pareto optimal 	\tabularnewline \cline{2-7} 
& \cite{Xiao2012Dynamic} & pairs of transmitter-receiver & power levels & achievable throughput & cheat-proof &  perfect public	\tabularnewline \cline{2-7} 
& \cite{Xiao2012Repeatedgames} & pairs of transmitter-receiver and an intervention  & power levels & achievable throughput & forgiving & SPE 	\tabularnewline \cline{2-7} 
& \cite{Berlemann2005Strategies} & WLANs & MAC parameters & a function of QoS & Grim, TFT, and static strategies & SPE 	\tabularnewline \cline{2-7} 
& \cite{Jia2013Asymmetric} & secondary users & FDM or STP & achievable transmission rate & adaptive & SPE 	\tabularnewline \cline{2-7} 
& \cite{Liu2010Deviation} & access points & exchange information  & spectrum usage & punishment trigger & SPE 	\tabularnewline \cline{2-7} 
& \cite{Liu2010CheatProof} & access points & exchange information  & spectrum usage & cheat-proof & SPE	\tabularnewline \cline{2-7} 
& \cite{Xie2012Unlicensed} & a satellite and two base stations & spectrum usage & achievable rate & variations of TFT & SPE 	\tabularnewline \cline{2-7} 
& \cite{Zhang2012Acontext},~\cite{Li2013Repeated} & SUs in the same cluster & forward or reject & profits minus cost & Grim & Pareto optimal 	\tabularnewline \cline{2-7} 
\hline 
\multirow{0}*{\parbox{\linewidth}{\centering Spectrum Trading}}
& \cite{Niyato2008Competitive} & primary users & prices & profits & Grim & SPE 	\tabularnewline \cline{2-7} 
& \cite{Kasbekar2010Spectrum} & primary users & prices & profits & Grim &  SPE	\tabularnewline \cline{2-7} 
\hline
\end{tabular}
\par\end{centering}

\end{table*}

\subsection{Spectrum Trading} 

Repeated games are also used for spectrum trading between SUs and primary users (PUs). In CRNs, PUs can sell their spectrum to SUs to gain revenue and improve spectrum utilization. However, when there are multiple PUs selling spectrum, they have to choose appropriate offer prices to attract SUs and compete with each other. A repeated game is used to help PUs in pricing. 

In~\cite{Niyato2008Competitive}, the authors considered an interaction among primary services when they provide and sell opportunistic spectrum accesses for SUs. The players are primary services and the strategy is to offer the price per unit of spectrum. The payoff is the profit from selling bandwidth to SUs minus the cost for spectrum sharing, e.g., due to QoS degradation of PUs. The authors first formulated the competitive pricing problem among PUs as the Bertrand game~\cite{Bertrand1883Bookreview}. If this game is played only one time, then the solution of the game is the noncooperative and inefficient Nash equilibrium. However, if the game is played repeatedly, there is a motivation for the primary services to cooperate by offering collusive prices that maximize their profits. A trigger strategy is used to force players to cooperate. 

In~\cite{Niyato2008Competitive}, the authors assumed that PUs always have available spectrum to sell to SUs. However, in reality, such licensed spectrum may not be available when PUs are using it. The authors in~\cite{Kasbekar2010Spectrum} considered this problem by assuming the probability to have a unit of available bandwidth in each time slot. The authors showed that there is no pure Nash equilibrium for the one-shot game, and thus the authors considered a special class of a Nash equilibrium, called a symmetric Nash equilibrium~\cite{Cheng2004Notes}. Then, the one-shot game is proven to possess the unique symmetric Nash equilibrium. For the repeated game, the subgame perfect Nash equilibrium (SPNE) is adopted. However, the efficiency of the SPNE is shown to be low. Thus, the authors proposed a Nash reversion strategy~\cite{Colell1995Microeconomic} to make players cooperate and improve the efficiency of the SPNE. The Nash reversion is stated as follows. At the beginning of the game, the players select a maximum price and this selection will be remained as long as no player deviates. If there is at least one player deviating, all players will play the symmetric Nash equilibrium strategy as in the one-shot game. Finally, the efficiency of the above SPNE is proven to be maximal if the discount factor satisfies a certain condition.

\begin{figure*}[htb]
\centering
\includegraphics[scale=0.65]{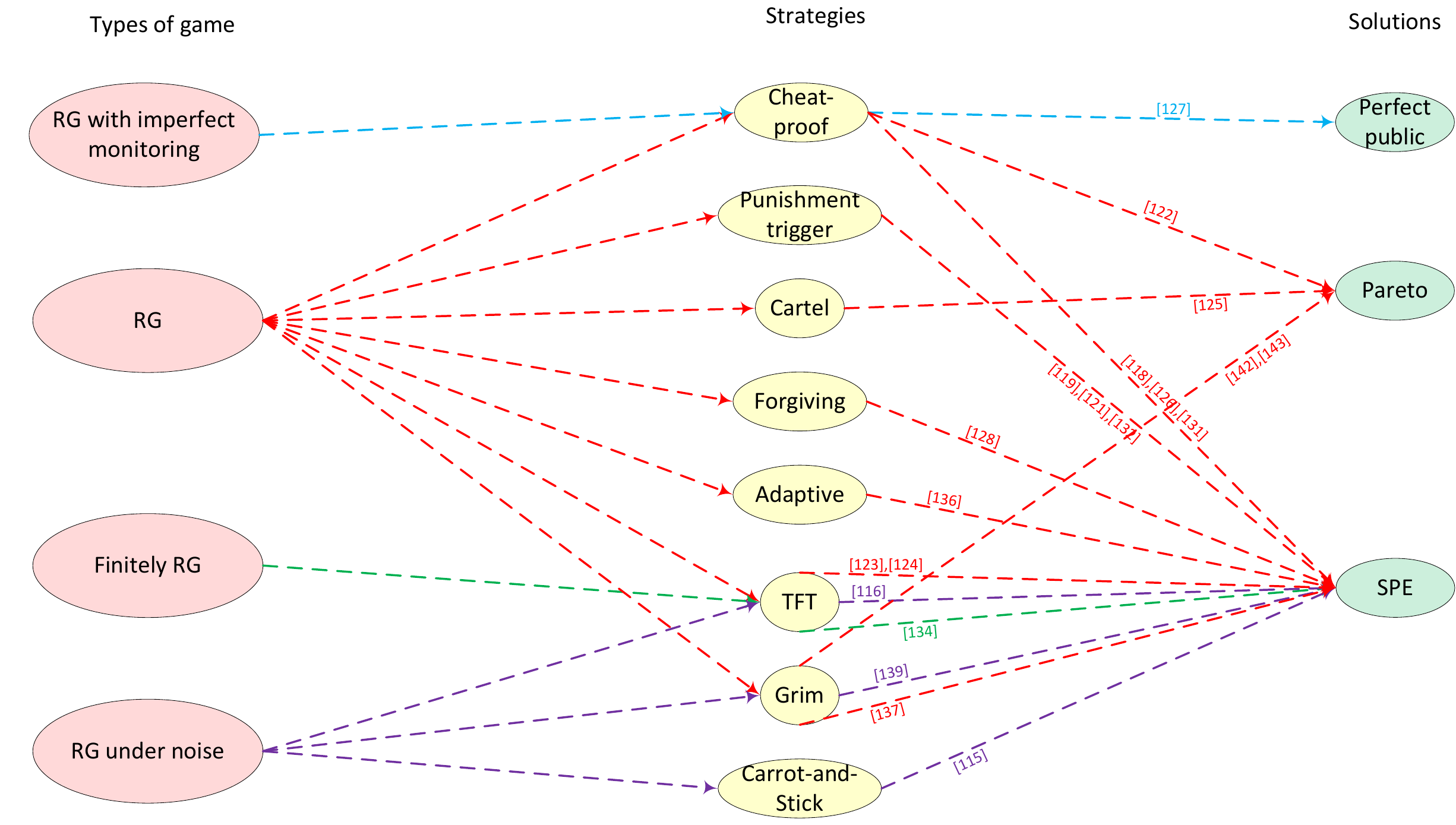}
\caption{Relation among types of repeated games, strategies, and solutions (RG = repeated game, SPE = subgame perfect equilibrium, TFT = Tit-for-Tat) in cognitive radio networks.}
\label{fig:chap5_CRNs_sum2}
\end{figure*}

\textbf{Summary:} From Table~\ref{table_CRNs_sum} and Fig.~\ref{fig:chap5_CRNs_sum2}, we observe that most of repeated game models developed for CRNs are conventional repeated games. Their main aim is to solve spectrum management problems. Nevertheless, there are few repeated game models for the cooperative spectrum sensing and trading. Moreover, there is only one finitely repeated game model~\cite{Xie2012Unlicensed}. 

\section{Applications of Repeated Games in Other Networks}
\label{sec:ONs}

Apart from the traditional wireless networks, in this section, we review some important applications of repeated games in special types of networks including network coding, fiber wireless access, and multicast networks.


\subsection{Wireless Network Coding} 
\begin{figure}[htb]
\centering
\includegraphics[scale=0.5]{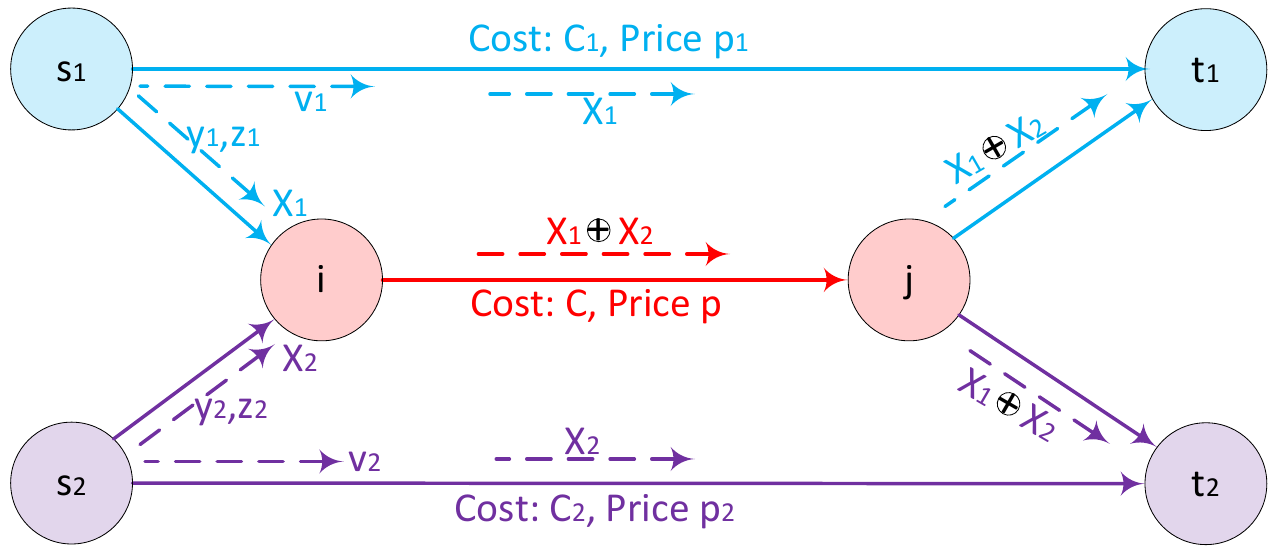}
\caption{Wireless network coding.}
\label{fig:chap6_Rad2010Bargaining}
\end{figure}
The core idea of network coding is to combine received packets and transmitting them together in the same flow instead of simply forwarding each packet separately~\cite{Ahlswede2000Network}. In~\cite{Rad2010Bargaining} and~\cite{Rad2014Repeated}, the authors introduced using network coding in a butterfly network including two source nodes $s_1$ and $s_2$ and two destination nodes $t_1$ and $t_2$ as shown in~Fig.~\ref{fig:chap6_Rad2010Bargaining}. Two types of packets from $s_1$ and $s_2$ are ``routing'' or ``network coding''. A routing packet $y_n$ will be forwarded immediately after it is sent to node $i$. A network coding packet $z_n$ will be encoded together with others using XOR encoding~\cite{Ho2006On}. When source nodes $s_1$ and $s_2$ send network coding packets to node $i$, they also have to send ``remedy data'' $v_n$ to their destinations $t_1$ and $t_2$, through their side links $(s_1,t_2)$ and $(s_2,t_1)$, respectively. The transmission on each link incurs a certain cost. Thus, the source nodes need to make a decision to choose the transmission rates, i.e., $y_n$, $z_n$, and $v_n$. For a noncooperative one-shot game, it is proven that the nodes will choose the noncooperative strategy ($z_n$=$v_n$=0). In particular, the nodes will not use the network coding to avoid payments over the links ($s_1$,$t_2$) and ($s_2$,$t_1$),  and this strategy is the unique Nash equilibrium of the game. However, if the game is played repeatedly and by using the grim trigger strategy, the nodes can achieve the subgame perfect equilibrium in which the nodes have an incentive to cooperate. Additionally, to find the efficient common network coding rate for the nodes, the bargaining game model is introduced. The results show that by using the proposed solution, the worst-case efficiency compared with the optimal network performance can be upper-bounded by 48\% as shown in~\cite{Chen2010INPAC} and \cite{Chen2014AnEnforceable}. In~\cite{Chachulski2007Trading}, the repeated game for network coding considers not only a forwarding action, but also the number of transmissions, the set of upstream nodes, and the set of downstream nodes. The authors showed that there exists a strategy profile that is a subgame perfect equilibrium.

\begin{figure*}[htb]
\begin{center}
$\begin{array}{ccc} 
\epsfxsize=2.3 in \epsffile{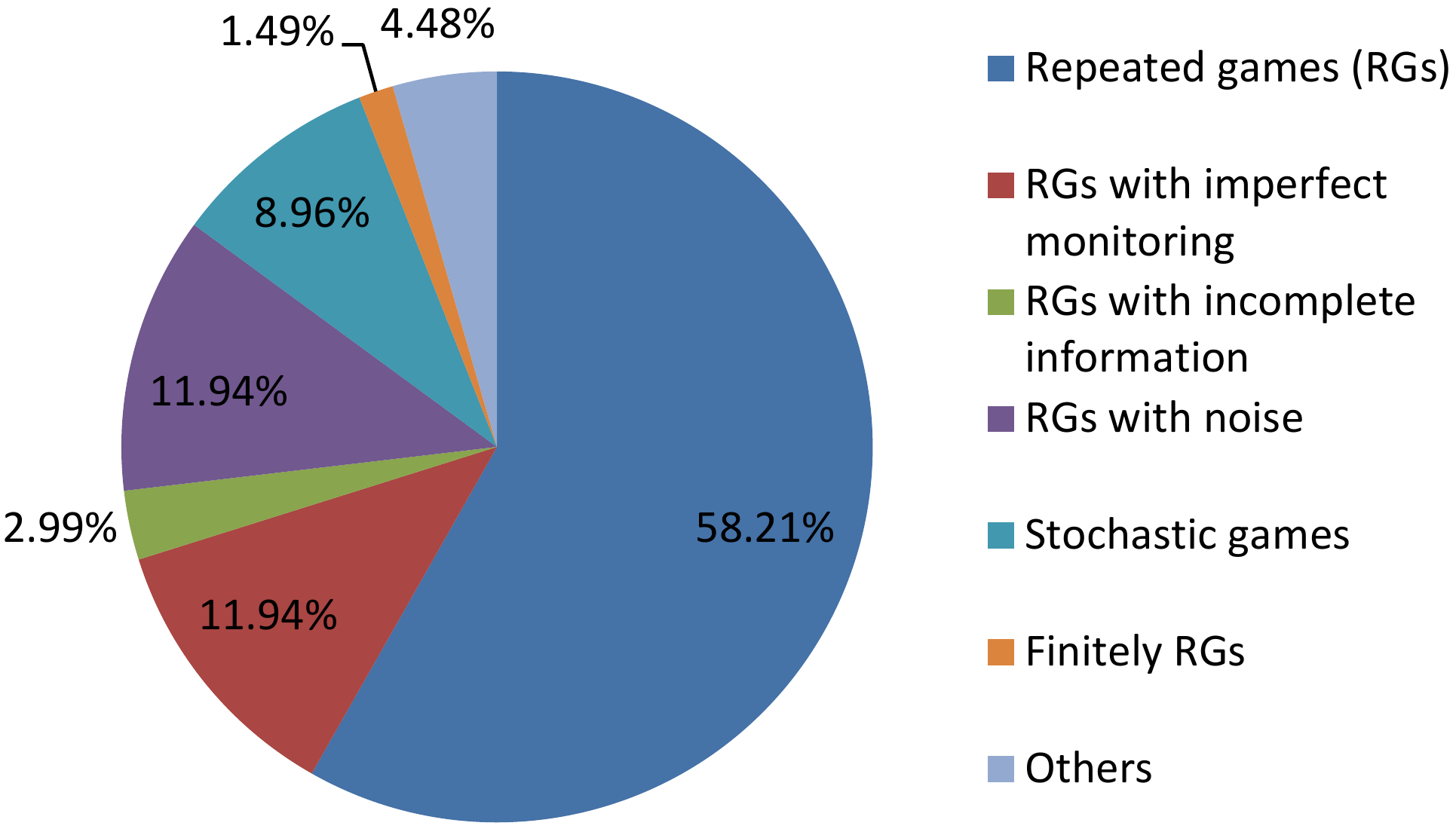}	&
\epsfxsize=2.3 in \epsffile{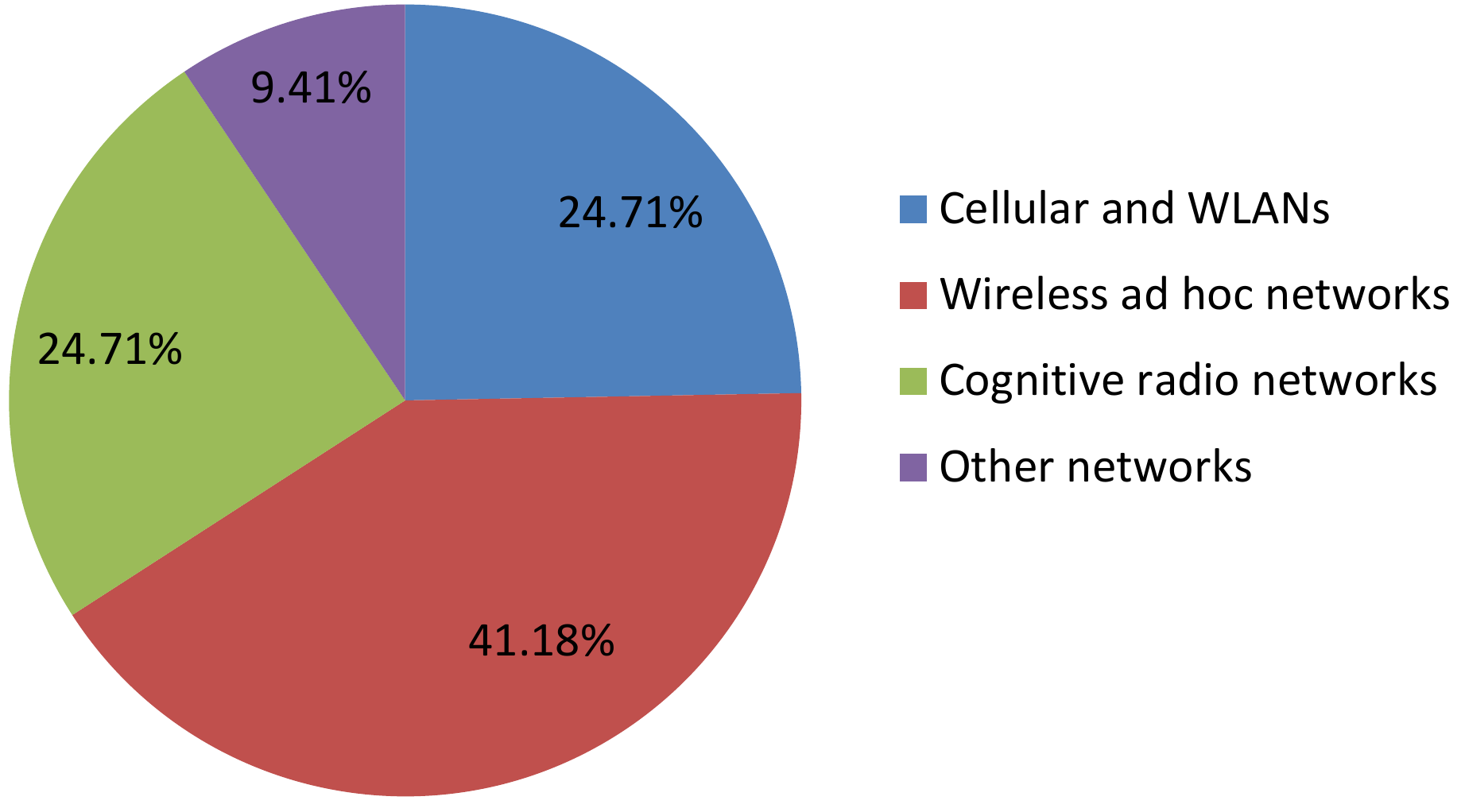}	&
\epsfxsize=2.3 in \epsffile{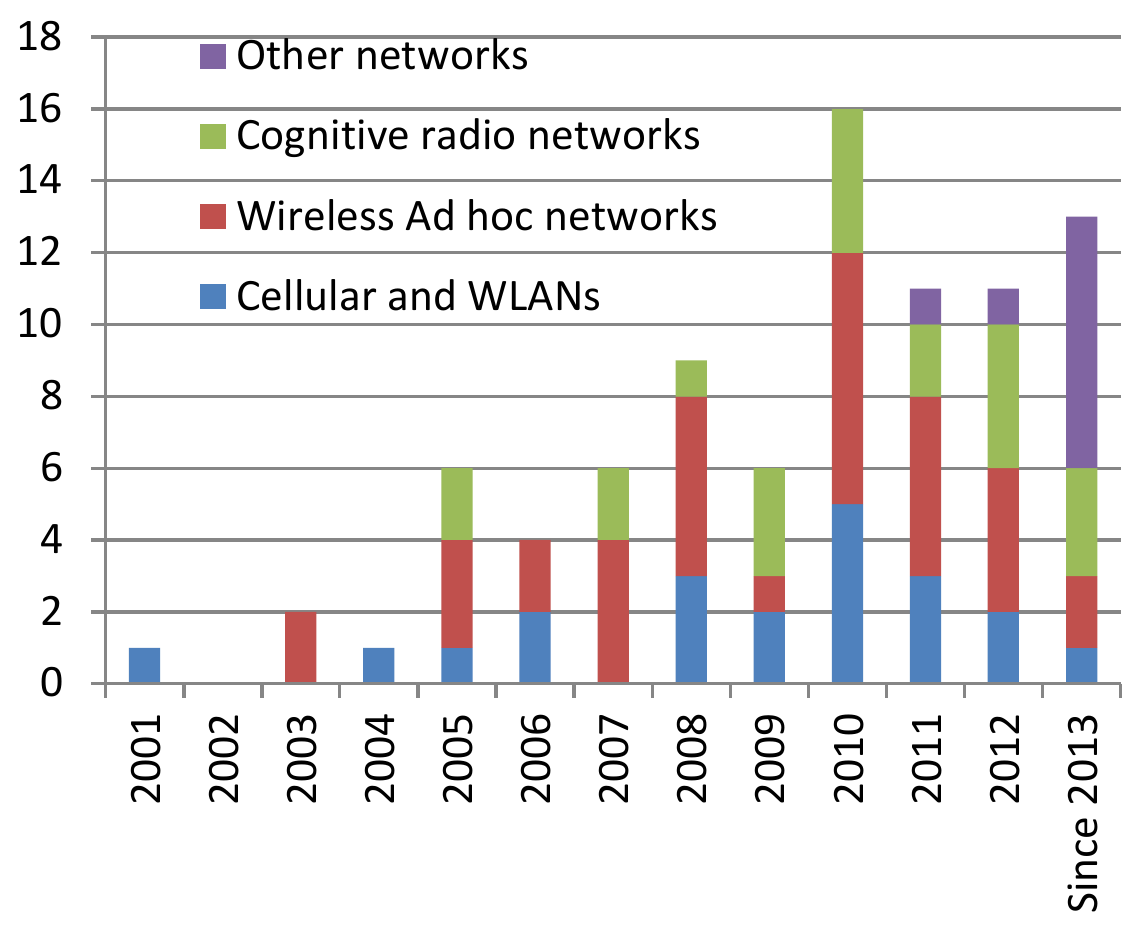}		\\ [-0.2cm]
(a)	& (b) & (c) 
\end{array}$
\caption{Summary information (a) the percentage of repeated game models, (b) the percentage of wireless network models, and (c) research trends.}
\label{fig:chap8_Summaries}
\end{center}
\end{figure*}

\subsection{Fiber Wireless Access Networks} 
In~\cite{Coimbra2010Forwarding} and~\cite{Coimbra2013Agame}, the authors applied a repeated game to Fiber-wireless (FiWi) access mesh networks (AMNs). There is a difference between traditional wireless mesh networks (TWMNs) and FiWi access mesh networks (AMNs). In TWMNs, packets are sent to any node in the network. By contrast, in FiWi AMNs, packets are always transmitted through gateways. Therefore, FiWi AMNs concern about the interaction between forwarding and gateway nodes. In the repeated game, the players are the forwarding and gateway nodes. The actions are to set the amount of foreign and local traffics. The payoff of the forwarding nodes is a gain function minus a cost function. The payoff for the gateway nodes is only a gain function. The forwarding node can defect by forwarding small amount of foreign traffic and high amount of local traffic. However, the gateway can punish the defecting node and thus curtail its payoff in the future. Through the repeated game model, there is an incentive for the nodes to cooperate resulting in the better network performance.

\subsection{Wireless Multicast Networks} 

The authors in~\cite{Niu2011ACooperation} examined the application of repeated games in wireless multicast networks. The authors considered a single hop multicast network with a base station and a group of nodes receiving packets from the base station with two phases. In the first phase, the base station transmits multicast packets to the group of nodes and selects a relay node that successfully receives the packets from the base station. In the second phase, the relay node will rebroadcast the packets to other nodes. The relay node can gain some benefit from the other nodes, but it incurs a certain cost for the relay node. Therefore, selfish nodes may not be interested in forwarding packets. The repeated game model is developed to encourage the node to cooperate. In this game, the nodes are the players and if the node is selected as a relay node, it has to choose transmission power levels to rebroadcast packets. The payoff is the reward from the number of successfully forwarded packets minus the cost. The worst behavior TFT incentive strategy is proposed to enforce the nodes into the cooperation. By using the punishment strategy, the authors proved that the nodes can achieve the subgame perfect equilibrium in both the cases of perfect monitoring and imperfect monitoring. 

\subsection{Other Networks}

Besides the aforementioned networks, there are some applications of repeated games in miscellaneous wireless networks such as vehicular ad hoc networks (VANET)~\cite{Wahab2013ADempster}, small cell networks~\cite{Mhiri2013Onthe}, wireless network virtualization~\cite{Fu2013Stochastic}, and wireless mesh networks~\cite{Vuuren2014Optimal}. In~\cite{Wahab2013ADempster}, a repeated game is used to model packet forwarding strategies of vehicular nodes based on QoS optimized link state routing (OLSR) protocol. In~\cite{Mhiri2013Onthe}, the power control problem of small cell networks is modeled as a repeated game. The setting is similar to other typical repeated game-based power control schemes except that there is co-channel interference among densely deployed small cells. 

\section{Summaries, Open Issues and Future Research Directions} 
\label{sec:FutureResearch}

\subsection{Summary} 

In Fig.~\ref{fig:chap8_Summaries}, we summarize key information from this survey. From Fig.~\ref{fig:chap8_Summaries}~(a), we observe that almost all papers focus on using conventional repeated games (approximately 60\%), while other extensions have not been much used. This is a potential research opportunity. Furthermore, we summarize the percentage of repeated game models for different types of wireless networks in Fig.~\ref{fig:chap8_Summaries}~(b). We also show the research trend for each network in Fig.~\ref{fig:chap8_Summaries}~(c). From Fig.~\ref{fig:chap8_Summaries} (b), we observe that majority of applications are for wireless ad hoc networks, followed by cognitive radio networks, and cellular and WLANs, subsequently. However, from Fig.~\ref{fig:chap8_Summaries} (c), while the applications of repeated games in traditional networks subside, those of emerging networks have increasingly grown. 

\subsection{Challenges and Open Issues of Repeated Games in Wireless Networks}

\subsubsection{Applications with finitely repeated games} 

Through this survey, most repeated game models are for infinite time horizon and there is only one application of finitely repeated game~\cite{Xie2012Unlicensed}. However, there are many issues in wireless networks that require the models developed for a short-term context such as~\cite{Shengbo2011Finite, Hong2014Opportunistic, Khalid2010Finite}. Therefore, the study of effective solutions to the finitely repeated games is essential. 

\subsubsection{Defining payoff functions} 

Similar to other game models, the most challenging part of developing repeated game models for wireless networks is defining a payoff. The payoff function must capture and balance network performance and individual player's incentive. More importantly, the physical meaning of the function must be well justified. Nevertheless, some important properties of repeated games (e.g., existence of an equilibrium) depend on the payoff function. Therefore, it is important to determine a formal approach to define the payoff function of the players in the game.

\subsubsection{Sequential decisions} 

In repeated games, the actions of players are assumed to be taken simultaneously. However, in some wireless networks, the players may not take action precisely at the same time. One of the solutions to such a situation, specifically sequential decision making, is a Stakelberg game. The combination of repeated game and Stakelberg game allows us to solve the non-simultaneous decision making problems. In~\cite{Wu2012AGame}, the authors considered this game setting.

\subsubsection{The mixed-strategy} 

Repeated games assume that all players follow the same strategy. For example, if one player uses the TFT strategy, all other players also have to play the TFT strategy. This assumption may not be true in some cases in which different players can adopt different strategies. This leads to the difficulty in finding an equilibrium of the game. In~\cite{Yan2008Cooperative}, the authors studied this problem through using simulations and stated that evolutionary games can be one of the potential solutions.

\subsection{Potential Research Directions of Repeated Games in Next Generation Wireless Communication Systems} 

\subsubsection{Mobile cloud computing} 

Mobile cloud computing (MCC) is a convergence model of mobile devices, service providers, wireless networks and cloud computing to bring benefits to mobile users, mobile applications developers, and cloud providers~\cite{Hoang2013ASurvey}. In MCC, different entities can cooperate to achieve better performance and profit. For example, cloud providers can cooperate with mobile operators. To attract them to cooperate, a repeated game can be applied.

\subsubsection{Wireless powered communication networks} 

The remarkable advancement of wireless energy harvesting and transfer technologies has created many new research directions recently~\cite{Priya2008Energy, Xiao2014Wireless}. The introduction of wireless energy harvesting techniques has solved inherent problems of wireless networks, i.e., energy-constraint problem of wireless nodes, thereby bringing many new applications such as wireless body area networks~\cite{Movassaghi2014Wireless} or wireless charging~\cite{Kurs2007Wireless}. However, as energy becomes the resource that can be shared efficiently, incentive mechanisms for wireless energy harvesting and transfer nodes have to be developed. In this case, repeated games can be adopted for such mechanisms. 

\subsubsection{Machine-to-machine communication in 5G networks} 

With the emerging of Internet-of-Things (IoT) and 5G networks, the number of wireless devices is expected to dramatically increase~\cite{CVNI2014}. Through the concept of machine-to-machine (M2M) communication, the wireless devices has to contend for limited radio resource, and cooperation is construed as desirable behavior to achieve efficient network operation. However, due to large number of devices, enforcing the cooperation becomes a complex task, and a conventional repeated game cannot be simply applied. An extension of the game to support a large number of players is worthwhile for studying.

\section{Conclusion}
\label{sec:Conclusion}

A repeated game is a powerful tool to model and resolve an interaction and conflict in wireless networks. In this paper, we have presented a comprehensive survey of the repeated game models developed for a variety of wireless networks. We have provided basic and fundamental of the game as well as demonstrated its advantages. Basically, the applications of repeated games can be divided based on the types of networks, i.e., cellular and WLANs, wireless ad-hoc networks, and cognitive radio networks. In addition to detailed reviews of the related work, we have also provided analysis, comparisons and summaries of the literature. Furthermore, some open issues and future research directions of repeated games in wireless networks have been highlighted. In conclusion, this paper will be a keystone for understanding repeated games in wireless communication systems.



\begin{thebibliography}{100}
\bibliographystyle{IEEEtranS}

\bibitem{Han2012BookGame}
Z.~Han, D.~Niyato, W.~Saad, T.~Basar, and A.~Hjorungnes, \emph{Game Theory in Wireless and Communication Networks: Theory, Models, and Applications}, Cambridge University Press, Cambridge, 2012. 

\bibitem{Charilas2010Asurvey}
D.~E.~Charilas, and A.~D.~Panagopoulos, ``A survey on game theory applications in wireless networks,'' \emph{The International Journal of Computer and Telecommunications Networking Computer Networks}, vol. 54, issue 18, pp. 3421-3430, Dec. 2010. 

\bibitem{Hoang2013ASurvey}
D.~T.~Hoang, C.~Lee, D.~Niyato, and P.~Wang, ``A survey of mobile cloud computing: architecture, applications, and approaches,'' \emph{Journal of Wireless Communications and Mobile Computing}, vol. 13, issue 18, pp. 1587-1611, Dec. 2013.



\bibitem{Fudenberg1994thefolk}
D.~Fudenberg, D.~K.~Levine, and E.~Maskin, ``The Folk theorem with imperfect public information,'' \emph{Econometrica}, vol. 62, no. 5, pp. 997-1039, Sep. 1994.

\bibitem{M.2002Kandori}
M. Kandori, ``Introduction to repeated games with private monitoring," \emph{Journal of Economic Theory}, vol. 102, no. 1, pp. 1-15, Jan. 2002.

\bibitem{R.1986Radner}
R. Radner, ``Repeated partnership games with imperfect monitoring and no discounting," \emph{Rev. Econ. Stud.} vol. 53, pp. 43-58, Jan. 1986.

\bibitem{O.1998Compte}
O. Compte, ``Communication in repeated games with imperfect private monitoring," {\em Econometria}, vol. 66, pp. 597-626, May. 1998.

\bibitem{puterman94}
M.~L.~Puterman, {\em Markov Decision Processes: Discrete Stochastic Dynamic Programming}, John Wiley \& Son, New York, April. 1994.

\bibitem{J.1993White}
D. J. White, ``A survey of applications of Markov decision processes," \emph{Journal of the Operational Research Society}, vol. 44, no. 11, pp. 1073-1096, Nov. 1993.

\bibitem{A.2003Neyman}
A. Neyman, and S. Sorin, editors. {\em Stochastic Games and Applications}, vol. 570 of NATO Science Series C. Springer-Verlag, 2003.

\bibitem{W.2009Saad}
W. Saad, Z. Han, M. Debbah, A. Hj\o rungnes, and T. Basar, ``Coalitional Game Theory for Communication Networks: A Tutorial," \emph{IEEE Signal Processing Magazine, Special Issue on Game Theory}, vol. 26, no. 5, pp. 77-97, Sep. 2009.

\bibitem{basar1999}
T. Basar and G. J. Olsder, {\em Dynamic Noncooperative Game Theory}, Society for Industrial and Applied Mathematics, Jan. 1999.

\bibitem{Abreu1990Toward}
D.~Abreu, D.~Pearce, and E.~Stacchetti, ``Toward a theory of discounted repeated games with imperfect monitoring,'' \emph{Econometrica}, vol. 58, pp. 1041-1063, Sep. 1990.



\bibitem{MacKenzie2004Gametheory}
Allen B.~MacKenzie, and Stephen B.~Wicker, ``Game theory in communications: motivation, explanation, and application to power control,'' in \emph{IEEE Global Telecommunications Conference}, pp. 821-825, San Antonio, TX, USA, Nov. 2001. 

\bibitem{Han2004Dynamicdistributed}
Z.~Han, Z.~Ji, and K.~J.~Ray Liu, ``Dynamic distributed rate control for wireless networks by optimal cartel maintenance strategy,'' in \emph{IEEE Global Telecommunications Conference}, pp. 3454-3458, Dallas, TX, USA, Dec. 2004. 

\bibitem{Zhong2003Sprite}
S.~Zhong, J.~Chen, and Y.~R.~Yang, ``Sprite: a simple, cheat-proof, credit-based system for mobile ad-hoc networks,'' \emph{IEEE INFOCOM}, vol. 3, pp. 1987-1997, San Francisco, Calif, USA, April. 2003. 

\bibitem{Han2008Acartel}
Z.~Han, Z.~Ji, and K.~J.~Ray Liu, ``A cartel maintenance framework to enforce cooperation in wireless networks with selfish users,'' \emph{IEEE Transactions on Wireless Communications}, vol. 7, no. 5, pp. 1889-1899, May. 2008. 

\bibitem{Fudenberg1991GameThery}
D.~Fundeburg, and J.~Tirole, {\em Game Theory}, MIT Press, Cambridge, MA 1991. 

\bibitem{Auletta2008Interferencegames}
V.~Auletta, L.~Moscardelli, P.~Penna, and G.~Persiano, ``Interference games in wireless networks,'' in \emph{Proceedings of the 4th International Workshop on Internet and Network Economics}, pp. 278-285, Shanghai, China, Dec. 2008. 

\bibitem{Treust2010Implicitcooperation}
M.~L.~Treust, S.~Lasaulce, and M.~Debbah, ``Implicit cooperation in distributed energy-efficient networks,'' in \emph{IEEE 4th International Symposium on Communications, Control and Signal Processing}, pp. 1-6, Cyprus, Mar. 2010. 

\bibitem{Treust2010Arepeated}
M.~L.~Treust, and S.~Lasaulce, ``A repeated game formulation of energy-efficient decentralized power control,'' \emph{IEEE Transactions on Wireless Communications}, vol. 9, no. 9, pp. 2860-2869, Sep. 2010. 

\bibitem{Horner2010Recursive}
J.~Horner, T.~Sugaya, S.~Takahashi, and N.~Vieille, ``Recursive methods in discounted stochastic games: An algorithm for $\delta \rightarrow 1$ and a Folk theorem,'' Econometrica, vol. 79, issue 4, pp. 1277-1318, Jul. 2011.

\bibitem{Cagalj2005Onselfish}
M.~Cagalj, S.~Ganeriwal, I.~Aad, and J-P.~Hubaux, ``On selfish behavior in CSMA/CA networks,'' in \emph{Proceedings IEEE INFOCOM}, pp. 2513-2524, Miami, FL, Mar. 2005. 

\bibitem{LANMANstandards1999}
LAN/MAN Standards Committee, \emph{ANSI/IEEE Std 802.11: Wireless LAN Medium Access Control (MAC) and Physical Layer (PHY) Specifications,} IEEE Computer Society, 1999. 

\bibitem{Konorski2006Agame}
Jerzy Konorski, ``A game-theoretic study of CSMA/CA under a backoff attack,'' \emph{IEEE/ACM Transactions on Networking}, vol. 14, no. 6, pp. 1167- 1178, Sep. 2009. 

\bibitem{Knoblauch1994Computable}
V.~Knoblauch, ``Computable strategies for repeated Prisoner's Dilemma,'' \emph{Games and Economic Behavior}, vol. 7, issue 3, pp. 381-389, Nov. 1994. 

\bibitem{Riedel2006Achieving}
A.~Riedel, and T.~Fischer, ``Achieving Pareto-efficient bandwidth allocations using a non-monetary mechanism,'' in \emph{First International Conference on Innovative Computing, Information and Control}, pp. 405-409, Beijing, China, Sep. 2006. 

\bibitem{Liao2003Wireless}
R.~Liao, R.~Wouhaybi, and A.~Campbell, ``Wireless incentive engineering,'' \emph{IEEE Journal on Selected Areas in Communications}, vol. 21, issue 10, pp.  1764-1779, Dec. 2003.

\bibitem{Tran2010Onselfish}
D.~T.~Tran, Z.~Chen, and A.~Farago, ``On selfish behavior in TDMA-based bandwidth sharing protocols in wireless networks,'' \emph{Journal of Telecommunications}, vol. 4, issue 1, pp. 1-9, Aug. 2010. 

\bibitem{Hajj2011SIRA}
Ahmad M.~El-Hajj, M.~Awad, and Z.~Dawy, ``SIRA: A socially inspired game theoretic uplink/downlink resource aware allocation in OFDMA systems,'' in \emph{IEEE International Conference on Systems, Man, and Cybernetics}, pp. 884-890, Alaska, USA, Oct. 2011. 

\bibitem{Kong2010Efficient}
Z.~Kong, and Y-K.~Kwok, ``Efficient wireless packet scheduling in a noncooperative environment,'' \emph{Journal of Parallel And Distributed Computing}, vol.~70, issue 8, pp. 790-799, Aug. 2010.

\bibitem{A.1995Nowak}
M. A. Nowak, K. Sigmund and E. El-Sedy, ``Automata, repeated games and noise." \emph{Journal of Mathematical Biology} vol. 33, no. 7, pp. 703-722, Aug.1995. 

\bibitem{Shen2014Universalnon}
F.~Shen, and E.~Jprswoecl, ``Universal non-linear cheat-proof pricing framework for wireless multiple access channels,'' \emph{IEEE Transactions on Wireless Communications}, vol. 13, no. 3, pp. 1436-1448 , Mar. 2014.

\bibitem{Lai2008Thewater}
L.~Lai and H.~E.~Gamal, ``The water-filling game in fading multiple-access channels,'' \emph{IEEE Transactions on Information Theory}, vol. 54, no. 5, pp. 2110-2122, May. 2008.

\bibitem{Tse1998Multiaccess}
D.~N.~C.~Tse, and S.~V.~Hanly, ``Multiaccess fading channels. I. Polymatroid structure, optimal resource allocation and throughput capacities,'' \emph{IEEE Transactions on Information Theory}, vol. 44, issue 7, pp. 2796-2815, Nov. 1998.

\bibitem{Fudenberg1986Thefolk}
D.~Fudenberg and E.~Maskin,``The Folk theorem in repeated games with discounting or with incomplete information,'' \emph{Econometrica}, vol. 54, no. 3, pp. 533-554, May. 1986. 

\bibitem{Hu2009Atransaction}
R.~Hu, D.~Yang, and R.~Qi, ``A transaction management mechanism for building trust in mobile commerce,'' in \emph{IEEE International Conference on Management of e-Commerce and e-Government}, pp. 408-411, Nanchang, China, Sep. 2009. 

\bibitem{Antoniou2011Networkselection}
J.~Antoniou, V.~Papadopoulou, V.~Vassiliou, and A.~Pitsillides, \emph{Network selection and handoff in wireless networks: a gmae theoretic approach}, Book Chapter, Game Theory for Wireless Communications and Networking, CRC Press, Taylor and Francis Group 2011.

\bibitem{Cho2011Agame}
J.~P.~Cho, Y.-W.~P.~Hong, and C.-C. Jay Kuo, ``A game theoretic approach to eavesdropper cooperation in MISO wireless networks,'' in \emph{IEEE International Conference on Acoustics, Speech and Signal Processing}, pp. 3428-3431, Prague, Czech Republic. May. 2011. 

\bibitem{Felegyhazi2006Wireless}
M.~Felegyhazi, and J-P.~Hubaux, ``Wireless Operators in a shared spectrum,'' in \emph{Proceedings IEEE INFOCOM}, pp. 1-11, Barcelona, Spain. Apr. 2006.

\bibitem{Treust2010Converage}
M.~L.~Treust, H.~Tembine, and M.~Debbah, ``Coverage games in small cells networks,'' in \emph{Proceedings of Future Network and Mobile Summit}, pp. 1-8, Florence, Italy, Jun. 2010. 

\bibitem{Kalai1975Other}
E.~Kalai and M.~Smorodinsky, ``Other solutions to Nash's bargaining problem,'' \emph{Econometrica}, vol. 43, no. 3, pp. 513-518, May. 1975.

\bibitem{Kalai1985Solutions}
E.~Kalai, ``Solutions to the bargaining problem,'' in \emph{Social Goals and Social Organization}, L.~Hurwicz , D.~Schmeidler and H.~Sonnenschein (eds.), Cambridge University Press, Cambridge, UK 1985.

\bibitem{Kamhoua2012Gametheoretic}
C.~A.~Kamhoua, N.~Pissinou, K.~Kwiat, and S.~Iyengar, ``Game theoretic analysis of users and providers behavior in network under scare resources,'' in \emph{International Conference on Computing, Networking and Communications}, pp. 1149-1155, Maui, Hawaii, Feb. 2012. 

\bibitem{Chaabane2012Anew}
I.~B.~Chaabane, S.~Hamouda, S.~Tabbane, and J.~L.~Vicario, ``A new PRB sharing scheme in dual-hop LTE-advanced system using game theory,'' in \emph{IEEE 23rd International Symposium on Personal, Indoor and Mobile Radio Communications}, pp. 375-379, Sydney, Australia, Sep. 2012. 



\bibitem{Rubinstein2006Asurvey}
M.~G.~Rubinstein, I.~M.~Moraes, M.~E.~M.~Campista, L.~H.~M.~K.~Kosta, and O.~C.~M.~B.~Duarte, ``A Survey on Wireless Ad Hoc Networks,'' in \emph{Mobile and Wireless Communication Networks}, vol. 211, pp. 1-33, Aug. 2006. 

\bibitem{Sesay2004Asurvey}
S.~Sesay, Z.~Yang, and J.~He, ``A survey on mobile Ad hoc wireless network,'' \emph{Information Technology Journal}, vol. 3, pp. 168-175, 2004.

\bibitem{Butt2003Stimulating}
L. Buttyan and J.-P.~Hubaux, ``Stimulating cooperation in self-organizing mobile ad hoc networks,'' \emph{ACM/Kluwer Mobile Networks and Applications}, vol. 8, no. 5, pp. 579-592, Oct. 2003.

\bibitem{Bereby-Meyer2006}
Y. Bereby-Meyer, and A. E. Roth, ``The speed of learning in noisy games: partial reinforcement and the sustainability of cooperation," \emph{The American Economic Review}, vol. 96, no. 4, pp. 1029-1042, Sep. 2006.

\bibitem{He2004SORI}
Q.~He, D.~Wu and P.~Khosla, ``SORI: A secure and objective reputation-based incentive scheme for ad hoc networks,'' in \emph{IEEE Wireless Communications and Networking Conference}, pp. 825-830, Atlanta, USA, Mar. 2004.

\bibitem{Mahajan2005Sustaining}
R.~Mahajan, M.~Rodrig, D.~Wetherall, and J.~Zahorjan, ``Sustaining cooperation in multihop wireless networks,'' in \emph{Proceedings of second USENIX Symposium on Networked System Design and Implementation}, pp. 231-244, Boston, USA, May. 2005.

\bibitem{Milan2006Achieving}
F.~Milan, J.~J.~Jaramillo, and R.~Srikant, ``Achieving cooperation in multihop wireless networks of selfish nodes,'' in \emph{GameNets '06 Proceeding from the 2006 Workshop on Game Theory for Communications and Networks}, Article 3, Pisa, Italy, Oct. 2006.

\bibitem{Ji2010ABelief}
Z.~Ji, W.~Yu, and K.~J.~Ray Liu, ``A belief evaluation framework in autonomous MANETs under noisy and imperfect observation: vulnerability analysis and cooperation enforcement,'' \emph{IEEE Transactions on Mobile Computing}, vol.~9, no.~9, pp. 1242-1254, Sep. 2010.

\bibitem{Bhaskar2002Belief}
V.~Bhaskar and I.~Obara, ``Belief-based equilibria in the repeated prisoners' dilemma with private monitoring,'' \emph{Journal of Economic
Theory}, vol. 102, issue 1, pp. 40-69, Jan. 2002.

\bibitem{Osborne1994Acourse}
M.~J.~Osborne and A.~Rubinstein, ``A Course in Game Theory,'' the MIT Press, Cambridge, MA, 1994.

\bibitem{Srivastava2006Equilibria}
V.~Srivastava, and L.~A.~DaSilva, ``Equilibria for node participation in ad hoc networks - an imperfect monitoring approach,'' in \emph{IEEE International Conference on Communications}, pp. 3850-3855, Istanbul, Turkey, Jun. 2006.

\bibitem{Jaramillo2007DARWIN}
J.~J.~Jaramillo, and R.~Srikant, ``DARWIN: distributed and adaptive reputation mechanism for wireless ad-hoc networks,'' in \emph{Proceedings of the 13th annual ACM International Conference on Mobile Computing and Networking}, pp. 87-98, Montreal, Canada, Sep. 2007.

\bibitem{Pandana2008Cooperation}
V.~Pandana, Z.~Han, and K.~J.~Ray Liu, ``Cooperation enforcement and learning for optimizing packet forwarding in autonomous wireless networks,'' \emph{IEEE Transactions on Communications}, vol.~7, no.~8, pp. 3150-3163, Aug. 2008.

\bibitem{Porath1996Communication}
E.~Ben-Porath and M.~Kahneman, ``Communication in repeated games with private monitoring,'' \emph{Journal of Economic Theory}, vol. 70, issue 2, pp. 281-297, Aug. 1996.

\bibitem{Wenjing2011Cooperation}
W.~Wang, M.~Chatterjee, and K.~Kwiat, ``Cooperation in wireless networks with unreliable channels,'' \emph{IEEE Transactions on Communications}, vol.~19, no.~10, pp. 2808-2817, Oct. 2011.

\bibitem{Kamhoua2010Belief}
C.~A.~Kamhoua, N.~Pissinou, A.~Busovaca, and K.~Makki, ``Belief-free equilibrium of packet forwarding game in ad hoc networks under imperfect monitoring,'' in \emph{IEEE 29th International Performance Computing and Communications Conference}, pp. 315-324, Albuquerque, USA, Dec. 2010.

\bibitem{Nowak1993Astrategy}
M.~Nowak and K.~Sigmund, ``A strategy of win-stay, lose-shift that outperforms tit-for-tat in the Prisoner's Dilemma game,'' \emph{Nature}, vol. 364, pp. 56-58, Jul. 1993.

\bibitem{Boerlijst1997Thelogic}
M.~Boerlijst, M.~Nowak, and K.~Sigmund, ``The Logic of Contrition,'' \emph{Journal of Theoretical Biology}, vol. 185, no. 3, pp. 281-293, Apr. 1997.

\bibitem{Mailath2006Repeated}
G.~Mailath, L.~Samuelson, \emph{Repeated Games and Reputations, Long-run relationships}, Oxford University Press, Oxford, UK 2006.

\bibitem{Yamamoto2009Alimit}
Y.~Yamamoto, ``A limit characterization of belief-free equilibrium payoffs in repeated games,'' \emph{Journal of Economic Theory}, vol. 144, issue 2, pp. 802-824, Mar. 2009.

\bibitem{Marti2000Mitigating}
S.~Marti, T.~J.~Giuli, K.~Lai, M.~Baker, ``Mitigating routing misbehavior in mobile ad hoc networks,'' in \emph{Proceedings of the Sixth Annual International Conference in Mobile Computing and Networking}, pp. 255-265, Boston, USA, Aug. 2000.

\bibitem{Sun2011APower}
Y.~Sun, S.~Lu, Y.~Ge, Z.~Li, and E.~Dutkiewicz, ``A power control mechanism for non-cooperative packet forwarding in ad hoc networks,'' in \emph{36th Annual IEEE Conference on Local Computer Networks}, pp. 123-126, Bonn, Germany, Oct. 2011.



\bibitem{Srinivasan2003Cooperation}
V.~Srinivasan, P.~Nuggehalli, C.~F.~Chiasserini, and R.~R.~Rao, ``Cooperation in wireless ad hoc networks,'' in \emph{IEEE INFOCOM}, pp.  808-817, San Franciso, USA. Apr. 2003. 

\bibitem{Urpi2003Modelling}
A.~Urpi, M.~Bonuccelli, and S.~Giordano, ``Modelling cooperation in mobile ad hoc networks: a formal description of selfishness,'' in \emph{Proceedings of Modelling and Optimization in Mobile, Ad Hoc and Wireless Networks}, pp. 1-10, Sophia-Antipolis, France, Mar. 2003.

\bibitem{Bandyopadhyay2005AGame}
Seema Bandyopadhyay, and Subhajyoti Bandyopadhyay, ``A game-theoretic analysis on the conditions of cooperation in a wireless ad hoc network,'' in \emph{Third International Symposium on Modeling and Optimization in Mobile, Ad Hoc, and Wireless Networks}, pp. 54-58, Trentino, Italy, Apr. 2005.

\bibitem{Han2005ASelf}
Z.~Han, C.~Pandana, and K.~J.~Ray Liu, ``A self-learning repeated game framework for optimizing packet forwarding networks,'' in \emph{IEEE Wireless Communications and Networking Conference}, pp. 2131-2136, New Orleans, USA, Mar. 2005.

\bibitem{Wei2006OnOptimal}
W.~Yu, and K.~J.~Ray Liu, ``On optimal and cheat-proof packets forwarding strategies in autonomous ad hoc networks,'' in \emph{40th Annual Conference on Information Sciences and Systems}, pp. 1455-1460, Princeton, USA, Mar. 2006.

\bibitem{Zhu2008AGame}
Z.~Ji, W.~Yu, and K.~J.~Ray Liu, ``A game theoretical framework for dynamic pricing-based routing in self-organized MANETs,'' \emph{Journal on Selected Areas in Communications}, vol.~26, no.~7, pp. 1204-1217, Sep. 2008.

\bibitem{Yan2008Cooperative}
L.~Yan, and S.~Hailes, ``Cooperative packet relaying model for wireless ad hoc networks,'' in \emph{Proceedings of the 1st ACM International Workshop on Foundations of Wireless Ad Hoc and Sensor Networking and Computing}, pp. 93-100, Hong Kong, China, May. 2008.

\bibitem{Axelrod1984Theevolution}
R.~Axelrod, ``The evolution of cooperation,'' \emph{Journal of Economic Behavior \& Organization}, vol. 5, no. 3-4, pp. 406-409. 1984.

\bibitem{Beaufils1997Ourmeeting}
B.~Beaufils, J.-P.~Delahaye and P.~Mathieu, ``Our meeting with gradual: a good strategy for the iterated prisoner's dilemma,'' in \emph{Proceedings of the Fifth International Workshop on the Synthesis and Simulation of Living Systems}, pp. 202-209, Nara, Japan, May. 1996.

\bibitem{Guan2010AModified}
G.~Wang, Y.~Sun, and J.~Liu, ``A modified cooperation stimulation mechanism based on game theory in ad hoc networks,'' in \emph{International Conference on Information Theory and Information Security}, pp. 379-383, Beijing, China, Dec. 2010.

\bibitem{Kamhoua2010Mitigating}
C.~A.~Kamhoua, N.~Pissinou, J.~Miller, and S.~K.~Makki, ``Mitigating routing misbehavior in multi-hop networks using evolution game theory,'' in \emph{IEEE Globecom 2010 Workshop on Advances in Communications and Networks}, pp. 1957-1962, Miami, USA, Dec. 2010.

\bibitem{Ng2010Game}
S.-K.~Ng and W.~K.~G.~Seah, ``Game-theoretic approach for improving cooperation in wireless multihop networks,'' \emph{IEEE Transactions on Systems, Man, and Cybernetics}, vol.~40, no.~3, pp. 559-574, Jun. 2010.

\bibitem{Aoyagi2002Collusion}
M.~Aoyagi, ``Collusion in dynamic bertrand oligopoly with correlated private signals and communication,'' \emph{Journal of Economic Theory}, vol. 102, no. 1, pp. 229-248, Jan. 2002.

\bibitem{Mohamed2012Cooperation}
M-H.~Zayani, and D.~Zeghlache, ``Cooperation enforcement for packet forwarding optimization in multi-hop ad-hoc networks,'' in \emph{IEEE Wireless Communications and Networking Conference: Mobile and Wireless Networks}, pp. 1915-1920, Paris, France, Apr. 2012.


\bibitem{Theodorakopoulos2006Enhancing}
G.~Theodorakopoulos, and J.~S.~Baras, ``Enhancing benign user cooperation in the presence of malicious adversaries in ad hoc networks,'' in \emph{Securecomm and Workshops}, pp. 1-6, Baltimore, USA, Sep. 2006.

\bibitem{Theodorakopoulos2007Malicious}
G.~Theodorakopoulos, and J.~S.~Baras, ``Malicious users in unstructured networks,'' in \emph{IEEE INFOCOM}, pp. 884-891, Anchorage, USA, May. 2007.

\bibitem{Theodorakopoulos2008Game}
G.~Theodorakopoulos, and J.~S.~Baras, ``Game theoretic modeling of malicious users in collaborative networks,'' \emph{IEEE Journal on Selected Areas in Communications}, vol.~26, issue 7, pp. 1317-1327, Sep. 2008.

\bibitem{Brown1951Iterative}
G.~W.~Brown, ``Iterative solution of games by fictitious play,'' in \emph{Activity Analysis of Production and Allocation}, T. Koopmans, Ed. New York: John Wiley and Sons, pp. 374-376, 1951.

\bibitem{Tootaghaj2011Game}
D.~Z.~Tootaghaj, F.~Farhat, M-R.~Pakravan, and M-R.~Aref, ``Game-theoretic approach to mitigate packet dropping in wireless ad-hoc networks,'' in \emph{IEEE Consumer Communications and Networking Conference}, pp. 163-165, Las Vegas, USA, Jan. 2011.


\bibitem{Levin2012Amplify}
G.~Levin and S.~Loyka, ``Amplify-and-forward versus decode-and-forward relaying: which is better?,'' in \emph{International Zurich Seminar on Communications}, pp. 123-126, Zurich, Germany, Mar. 2012.

\bibitem{Yang2007Energy}
Y.~Yang, and D.~R.~Brown III, ``Energy efficient relaying games in cooperative wireless transmission systems,'' in \emph{Conference Record of the Forty-First Asilomar Conference on Signals, Systems and Computers}, pp. 835-839, Pacific Grove, USA, Nov. 2007.

\bibitem{Chen2008AGame}
Y.~Chen, and S.~Kishore, ``A game-theoretic analysis of decode-and-forward user cooperation,'' \emph{IEEE Transactions on Wireless Communications}, vol.~7, no.~5, pp. 1941-1951, May. 2008.

\bibitem{Li2010RelayIET}
D.~Li, Y.~Xu, J.~Liu, and J.~Zhang, ``Relay assignment and cooperation maintenance in wireless networks: a game theoretical approach,'' \emph{IET Communications}, vol. 4, issue 17, pp. 2133-2144, Nov. 2010. 

\bibitem{Shapley1974On}
L.~Shapley and H.~Scarf, ``On cores and indivisibility,'' \emph{Journal of Mathematical Economics} vol.1, no. 1, pp. 23-28, 1974.

\bibitem{Roth1977Weak}
A.~E.~Roth and A.~Postlewaite, ``Weak versus strong domination in a market with indivisible goods,'' \emph{Journal of Mathematical
Economics}, vol.4, issue 2, pp. 131-137, Aug. 1977.

\bibitem{Jinpeng1994Strategy}
J.~Ma, ``Strategy-proofness and the strict core in a market with indivisibilities,'' \emph{International Journal of Game Theory}, vol.23, issue 1, pp. 75-83, Mar. 1994.

\bibitem{Brown2011AGame}
D.~R.~Brown III, and F.~Fazel, ``A game theoretic study of energy efficient cooperative wireless networks,'' \emph{Journal of Communications and Networks}, vol.~13, issue 3, pp. 266-276, Jun. 2011.

\bibitem{Gale1962College}
D.~Gale and L.~Shapley, ``College admissions and the stability of marriage,'' \emph{The American Mathematical Monthly}, vol. 69, no. 1, pp. 9-15, Jan. 1962.

\bibitem{Mohammed2014Cooperation}
M.~W.~Baidas, ``Cooperation in wireless networks: a game-theoretic framework with reinforcement learning,'' in \emph{IET Communications}, vol. 8, issue 5, pp. 740-753, Mar. 2014.


\bibitem{Peng2008Enhancing}
D.~Peng, W.~Liu, C.~Lin, Z.~Chen, and X.~Peng, ``Enhancing tit-for-tat strategy to cope with free-riding in unreliable P2P networks,'' in \emph{The Third International Conference on Internet and Web Applications and Services}, pp. 336-341, Athens, Greek, Jun. 2008.

\bibitem{Lin2009Cheat}
W.~S.~Lin, H.~V.~Zhao, and K.~J.~Ray Liu, ``Cheat-proof cooperation strategies for wireless live streaming social networks,'' in \emph{IEEE International Conference on Acoustics, Speech and Signal Processing}, pp. 3469-3472, Taipei, Taiwan, Apr. 2009.

\bibitem{Lin2010Cooperation}
W.~S.~Lin, H.~V.~Zhao, and K.~J.~Ray Liu, ``Cooperation stimulation strategies for peer-to-peer wireless live video-streaming social networks,'' \emph{IEEE Transactions on Image Processing}, vol. 19, no. 7, pp. 1768-1784, Jul. 2010.

\bibitem{Lin2009Incentive}
W.~S.~Lin, H.~V.~Zhao, and K.~J.~R.~Liu, ``Incentive cooperation strategies for peer-to-peer live streaming social networks,'' \emph{IEEE Transactions on Multimedia}, vol. 11, no. 3, pp. 396-412, Apr. 2009.

\bibitem{Wu2012AGame}
W.~Wu, John C.~S.~Lui, and Richard T.~B.~Ma, ``A game theoretic analysis on incentive mechanisms for wireless ad hoc VoD systems,'' in \emph{Proceedings of 10th International Symposium on Modeling and Optimization in Mobile, Ad Hoc and Wireless Networks}, pp. 177-184, Paderborn, Germany, May. 2012.

\bibitem{Ng2008Game}
S.~Ng and W.~Seah. ``Game-theoretic model for collaborative protocols in selfish, tariff-free, multihop wireless networks,'' in \emph{IEEE INFOCOM}, pp. 762-770, Phoenix, USA, Apr. 2008.

\bibitem{Chen2012Analysis}
J.~Chen, L.~Li, Z.~Zhang, and X.~Dong, ``Game theory analysis for message dropping attacks prevention strategy in mobile P2P live streaming system,'' in \emph{IEEE 3rd International Conference on Software Engineering and Service Science}, pp. 359-363, Beijing, China, Jun. 2012.


\bibitem{Tembine2007Multiple}
H.~Tembine, E.~Altman, Rachid El-Azouzi, and Y.~BarHayel, ``Multiple access game in ad-hoc network,'' in \emph{Proceedings of the 2nd International Conference on Performance Evaluation Methodologies and Tools}, pp. 1-7, Nantes, France, Oct. 2007.

\bibitem{Chen2007Selfishness}
L.~Chen, and J.~Leneutre, ``Selfishness, not always a nightmare: modeling selfish MAC behaviors in wireless mobile ad hoc networks,'' in \emph{Proceedings  of 27th International Conference on Distributed Computing Systems}, pp. 1809-1818, Toronto, Canada, Jun. 2007.

\bibitem{Sun2011ARepeated}
L-H.~Sun, H.~Sun, B-Q.~Yang, and G-J.~Xu, ``A repeated game theoretical approach for clustering in mobile ad hoc networks,'' in \emph{IEEE International Conference on Signal Processing, Communications and Computing}, pp. 1-6, Xian, China, Sep. 2011.

\bibitem{Quer2013Inter}
G.~Quer, F.~Librino, L.~Canzian, L.~Badia, and M.~Zorzi, ``Inter-network cooperation exploiting game theory and Bayesian networks,'' \emph{IEEE Transactions on Communications}, vol.~61, issue 10, pp. 4310-4321, Oct. 2013.

\bibitem{Owen2001Game}
G.~Owen, \emph{Game Theory}, 3rd ed. New York: Academic, 2001.

\bibitem{Agah2007Preventing}
A.~Agah, and S.~K.~Das, ``Preventing DoS attacks in wireless sensor networks: a repeated game theory approach,'' \emph{International Journal of Network Security}, vol. 5, no. 2, pp. 145-153, Sep. 2007.

\bibitem{Michiardi2002Core}
P.~Michiardi, R.~Molva, ``Core: A Collaborative Reputation mechanism to enforce node cooperation in mobile ad hoc networks,'' in \emph{Communications and Multimedia Security Conference}, pp. 107-121, Portoroz, Slovenia, Sep. 2002.

\bibitem{Estiri2010AGame}
M.~Estiri, and A.~Khademzadeh, ``A game-theoretical model for intrusion detection in wireless sensor networks,'' in \emph{IEEE International Symposium on Parallel and Distributed Processing}, pp. 1-5, Calgary, Canada, May. 2010.

\bibitem{Qu2012Efficient}
Z.~Qu, D.~Chen, G.~Sun, X.~Wang, X.~Tian, and J.~Liu, ``Efficient wireless sensor networks scheduling scheme: game theoretic analysis and algorithm,'' in \emph{IEEE International Conference on Communications}, pp. 356-360, Ottawa, Canada, Jun. 2012.




\bibitem{Hossain2009Dynamic}
E.~Hossain, D.~Niyato, and Z.~Han, \emph{Dynamic Spectrum Access and Management in Cognitive Radio Networks}, Cambridge University Press, Cambridge, UK, 2009. 

\bibitem{Yucek2009Asurvey}
T.~Yucek and H.~Arslan, ``A survey of spectrum sensing algorithms for cognitive radio applications,'' \emph{IEEE Communications Surveys \& Tutorials}, vol. 11, issue 1, pp. 116-130, Mar. 2009.

\bibitem{Song2009Achieving}
C.~Song and Q.~Zhang, ``Achieving cooperative spectrum sensing in wireless cognitive radio networks,'' \emph{ACM SIGMOBILE Mobile Computing and Communications Review}, vol. 13, issue 2, pp. 14-25 , Apr. 2009.

\bibitem{Kondareddy2011Enforcing}
Y.~Kondareddy and P.~Agrawal, ``Enforcing cooperative spectrum sensing in cognitive radio networks,'' in \emph{IEEE Global Telecommunications Conference}, pp. 1-6, Houston, USA, Dec. 2011.

\bibitem{Wu1995How}
J.~Wu and R.~Axelrod, ``How to cope with noise in the iterated prisoner's dilemma,'' \emph{The Journal of Conflict Resolution}, vol. 39, no. 1, pp. 183-189, Mar. 1995.

\bibitem{Zhendong2013Anovel}
W.~Zhendong, W.~Huiqiang, F.~Guangsheng, Lv Hongwu, and Z.~Qiang, ``A novel cooperative spectrum sensing approach against malicious users in cognitive radio networks,'' in \emph{IEEE International Conference on Computer Sciences and Applications}, pp. 163-166, Wuhan, China, Dec. 2013.

\bibitem{Wang2007Self}
B.~Wang, Z.~Ji, and K.~J.~Ray Liu, ``Self-learning repeated game framework for distributed primary-prioritized dynamic spectrum access,'' in \emph{4th Annual IEEE Communications Society Conference on Sensor, Mesh and Ad Hoc Communications and Networks}, pp. 631-638, San Diego, USA, Jun. 2007.

\bibitem{Wang2007Primary}
B.~Wang, Z.~Ji, and K.~J.~Ray Liu, ``Primary-prioritized Markov approach for dynamic spectrum access,'' in \emph{2nd IEEE International Symposium on New Frontiers in Dynamic Spectrum Access Networks}, pp. 507-515, Dublin, Ireland, Apr. 2007.

\bibitem{Li2010Dynamic}
H.~Li, Y.~Liu, and D.~Zhang, ``Dynamic spectrum access for cognitive radio systems with repeated games,'' in \emph{IEEE International Conference on Wireless Communications, Networking and Information Security}, pp. 59-62, Beijing, China, Jun. 2010. 

\bibitem{Yan2011Game}
M.~Yan, L.~Du, L.~Huang, L.~Xiao, and J.~Tang, ``Game-theoretic approach against selfish attacks in cognitive radio networks,'' in \emph{10th IEEE/ACIS International Conference on Computer and Information Science}, pp. 58-61, Sanya, China, May. 2011.

\bibitem{Etkin2005Spectrum}
R.~Etkin, A.~Parekh, and D.~Tse, ``Spectrum sharing for unlicensed bands,'' in \emph{First IEEE International Symposium on New Frontiers in Dynamic Spectrum Access Networks}, pp. 251-258, Baltimore, USA, Nov. 2005.

\bibitem{Etkin2007Spectrum}
R.~Etkin, A.~Parekh, and D.~Tse, ``Spectrum sharing for unlicensed bands,'' \emph{IEEE Journal on Selected Areas in Communications}, vol. 25, no. 3, pp. 517-528, Apr. 2007.

\bibitem{Bennis2009Ahierarchical}
M.~Bennis, M.~Debbah, S.~Lasaulce, and A.~Anpalagan, ``A hierarchical game approach to inter-operator spectrum sharing,'' in \emph{IEEE Global Telecommunications Conference}, pp. 1-6, Honolulu, USA, Dec. 2009. 

\bibitem{Wu2009Repeated}
Y.~Wu, B.~Wang, K.~J.~Ray Liu, and T.~C.~Clancy, ``Repeated open spectrum sharing game with cheat-proof strategies,'' \emph{IEEE Transactions on Wireless Communications}, vol. 8, no. 4, pp. 1922-1933, Apr. 2009.

\bibitem{Xiao2012Dynamic}
Y.~Xiao, and Mihaela van der Schaar, ``Dynamic spectrum sharing among repeatedly interacting selfish users with imperfect monitoring,'' \emph{IEEE Journal on Selected Areas in Communications}, vol. 30, no. 10, pp. 1890-1899, Nov. 2012.

\bibitem{Xiao2012Repeatedgames}
Y.~Xiao, J.~Park, and Mihaela van der Schaar, ``Repeated games with intervention: theory and applications in communications,'' \emph{IEEE Transactions on Communications}, vol. 60, no. 10, pp. 3123-3132, Oct. 2012. 

\bibitem{Park2012thetheory}
J.~Park and M.~van~der~Schaar, ``The theory of intervention games for resource sharing in wireless communications,'' \emph{IEEE Journal on Selected Areas in Communications}, vol. 30, no. 1, pp. 165-175, Jan. 2012.

\bibitem{Berlemann2005Strategies}
L.~Berlemann, G.~R.~Hiertz, B.~Walke, and S.~Mangold, ``Strategies for distributed QoS support in radio spectrum sharing,'' in \emph{IEEE International Conference on Communications}, pp. 3271-3277, Seoul, Korea, May. 2005. 

\bibitem{Liu2010CheatProof}
Y.~Liu, J.~Jing, and J.~Yang, ``Cheat-Proof Traffic-Aware Dynamic Spectrum Access,'' in \emph{6th International Conference on Wireless Communications Networking and Mobile Computing}, pp. 1-5, Chengdu, China, Sep. 2010.

\bibitem{Liu2010Deviation}
Y.~Liu, J.~Jing, and J.~Yang, ``Deviation-resistant traffic-aware dynamic spectrum access,'' in \emph{6th International Conference on Wireless Communications Networking and Mobile Computing}, pp. 1-4, Chengdu, China, Sept. 2010.

\bibitem{Osborne2004AnIntroduction}
M.~J.~Osborne, \emph{An Introduction to Game Theory}, Oxford University Press, Oxford, UK, 2004.

\bibitem{Xie2012Unlicensed}
Z.~Xie, M.~Ma, and X.~Liang, ``Unlicensed spectrum sharing game between LEO satellites and terrestrial cognitive radio networks ,'' \emph{Chinese Journal of Aeronautics}, vol. 25, issue 4, pp. 605-614, Aug. 2012. 

\bibitem{Francoise_Forges}
F.~Forges and A.~Salomon, ``Bayesian repeated games and reputation,'' \emph{CESifo Working Paper Series No. 4700}, Mar. 2013.

\bibitem{Jia2013Asymmetric}
Y.~Jia, Z.~Zhang, X.~Tan, and X.~Liu, ``Asymmetric active cooperation strategy in spectrum sharing game with imperfect information,'' \emph{International Journal of Communication Systems}, doi: 10.1002/dac.2667, article first published online Sep. 2013.

\bibitem{Niyato2008Competitive}
D.~Niyato, and E.~Hossain, ``Competitive pricing for spectrum sharing in cognitive radio networks: dynamic game, inefficient of Nash equilibrium, and collusion,'' \emph{IEEE Journal on Selected Areas in Communications}, vol. 26, no. 1, pp. 192-202, Jan. 2008.

\bibitem{Bertrand1883Bookreview}
L.~Bertrand, ``Book review of theorie mathematique de la richesse sociale and of recherches sur les principles mathematiques de la theorie des richesses,'' \emph{Journal de Savants}, 67: 499-508, 1883. 

\bibitem{Kasbekar2010Spectrum}
G.~S.~Kasbekar, and S.~Sarkar, ``Spectrum pricing games with bandwidth uncertainty and spatial reuse in cognitive radio networks,'' in \emph{Proceedings of the eleventh ACM international symposium on Mobile ad hoc networking and computing}, pp. 251-260, Chicago, USA, Sep. 2010.

\bibitem{Cheng2004Notes}
S.~F.~Cheng, D.~M.~Reeves, Y.~Vorobeychik, and M.~P.~Wellman, ``Notes on Equilibria in Symmetric Games,'' in \emph{Proceedings of the 6th International Workshop On Game Theoretic And Decision Theoretic Agents}, pp. 23-28, New York, USA, Aug. 2004.

\bibitem{Colell1995Microeconomic}
A.~Mas-Colell, M.~Whinston, and J.~Green, \emph{Microeconomic Theory,} Oxford University Press, Oxford, UK, 1995.

\bibitem{Zhang2012Acontext}
Q.~Zhang, and K.~Moessner, ``A context aware architecture for energy efficient cognitive radio,'' in \emph{Mobile Multimedia Communications}, pp. 452-462, Lisbon, Portugal, Sep. 2012.

\bibitem{Li2013Repeated}
C.~Li, and J.~Xie, ``Repeated game-inspired spectrum sharing for clustering cognitive Ad hoc networks,'' \emph{International Journal of Distributed Sensor Networks}, DOI: 10.1155/2013/514037, 2013.

\bibitem{Chatterjee2002WCA}
M.~Chatterjee, S.~K.~Das, and D.~Turgut, ``WCA: a weighted clustering algorithm for mobile Ad Hoc networks,'' \emph{Journal of Clustering Computing}, vol. 5, no. 2, pp. 193-204, Apr. 2002.



\bibitem{Ahlswede2000Network}
R.~Ahlswede, N.~Cai, S.~Li, and R.~Yeung, ``Network information flow,'' \emph{IEEE Transactions on Information Theory}, vol. 46, pp. 1204-1216, Apr. 2000.

\bibitem{Ho2006On}
T.~Ho, Y.~Chang, and K.~Han, ``On constructive network coding for multiple unicasts,'' in \emph{Proceedings of Annual Allerton Conference on Communications, Control, and Computing}, pp. 1-10, Champaign, USA. Sep. 2006.

\bibitem{Rad2010Bargaining}
A.-H.~Mohsenian-Rad, J.~Huang, Vincent W.~S.~Wong, and R.~Schober, ``Bargaining and price-of-anarchy in repeated inter-session network coding games,'' in \emph{IEEE INFOCOM}, pp. 1927-1935, San Diego, USA, Mar. 2010.

\bibitem{Rad2014Repeated}
A.-H.~Mohsenian-Rad, J.~Huang, Vincent W.~S.~Wong, and R.~Schober, ``Repeated intersession network coding games: efficiency and min-max bargaining solution,'' \emph{IEEE/ACM Transactions on Networking}, vol. 22, issue 4, pp. 1121-1135, Jul. 2013.

\bibitem{Coimbra2010Forwarding}
J.~Coimbra, G.~Schutz, and N.~Correia, ``Forwarding repeated game for end-to-end QoS support in fiber-wireless access networks,'' in \emph{IEEE Global Telecommunications Conference}, pp. 1-6, Miami, USA, Dec. 2010.

\bibitem{Coimbra2013Agame}
J.~Coimbra, G.~Schutz, and N.~Correia, ``A game-based algorithm for fair bandwidth allocation in Fibre-Wireless access networks,'' \emph{Journal of Optical Switching and Networking}, vol. 10, issue 2, pp. 149-162, Apr. 2013.

\bibitem{Chen2010INPAC}
T.~Chen, and S.~Zhong, ``INPAC: An enforceable incentive scheme for wireless networks using network coding,'' in \emph{IEEE INFOCOM}, pp. 1828-1836, San Diego, USA, Mar. 2010.

\bibitem{Chen2014AnEnforceable}
T.~Chen, and S.~Zhong, ``An enforceable scheme for packet forwarding cooperation in network-coding wireless networks with opportunistic routing,'' \emph{IEEE Transactions on Vehicular Technology}, vol. 63, issue 9, pp. 4476-4491, Mar. 2014.

\bibitem{Chachulski2007Trading}
S.~Chachulski, M.~Jennings, S.~Katti, and D.~Katabi, ``Trading structure for randomness in wireless opportunistic routing,'' in \emph{Proceedings of ACM SIGCOMM}, pp. 169-180, Kyoto, Japan, Aug. 2007.

\bibitem{Niu2011ACooperation}
B.~Niu, H.~Vicky~Zhao, and H.~Jiang, ``A cooperation stimulation strategy in wireless multicast networks,'' \emph{IEEE Transactions on Signal Processing}, vol.~59, issue 5, pp. 2355-2369, Apr. 2011.

\bibitem{Wahab2013ADempster}
O. A. Wahab, H.~Otrok, A.~Mourad, ``A dempster - Shafer based Tit-for-Tat strategy to regulate the cooperation in VANET using QoS-OLSR protocol,'' \emph{Journal of Wireless Personal Communication}, vol. 75, issue 3, pp. 1635-1667, Apr. 2014. 

\bibitem{Mhiri2013Onthe}
M.~Mhiri, V. S. Varma, M. L. Treust, S.~Lasaulce, and A.~Samet, ``On the benefits of repeated game models for green cross-layer power control in small cells,'' in \emph{First International Black Sea Conference on Communications and Networking}, pp. 137-141, Batumi, Adjara, Jul. 2013.

\bibitem{Fu2013Stochastic}
F.~Fu, and U. C. Kozat, ``Stochastic game for wireless network virtualization,'' \emph{IEEE/ACM Transactions on Networking}, vol. 21, issue 1, pp. 84-97, Apr. 2012. 

\bibitem{Vuuren2014Optimal}
P.~A.~Jansen van Vuuren, A. S. Alfa, and B. T. Maharaj, ``Optimal and fair rate adaptation in wireless mesh networks based on mathematical programming and game theory,'' in \emph{IEEE Vehicular Technology Conference}, pp. 1-6, Vancouver, Canada, Sep. 2014. 



\bibitem{Shengbo2011Finite}
C.~Shengbo, P.~Sinha, N.~B.~Shroff, and J.~Changhee, ``Finite-horizon energy allocation and routing scheme in rechargeable sensor networks,'' \emph{IEEE INFOCOM}, pp. 2273-2281, Shanghai, China, Apr. 2011.

\bibitem{Hong2014Opportunistic}
S.~G.~Hong, J.~Widmer, and B.~Rengarajan, ``Opportunistic beamforming for finite horizon multicast,'' \emph{IEEE International Symposium on A World of Wireless, Mobile and Multimedia Networks}, pp. 1-9, Sydney, Australia, Jun. 2014.

\bibitem{Khalid2010Finite}
M.~Khalid, Le.~X.~Hung, In-ho Ra, and R.~Sankar, ``Finite horizon scheduling in wireless ad hoc networks,'' \emph{IEEE GLOBECOM workshops}, pp. 876-881, Miami, USA, Dec. 2010.

\bibitem{Smith1988The}
J. M. Smith, ``The limitations of evolution theory,'' in \emph{Did Darwin Get It Right?}, chapter 20, Springer US, 1988. 

\bibitem{Priya2008Energy}
S.~Priya and D.~J.~Inman, \emph{Energy Harvesting Technologies}, Springer Publishing Company, Incorporated 2008.

\bibitem{Xiao2014Wireless}
L.~Xiao, W.~Ping, D.~Niyato, and D.~I.~Kim, ``Wireless networks with RF energy harvesting: a contemporary survey,'' \emph{IEEE Communications Surveys \& Tutorials}, Nov. 2014.

\bibitem{Movassaghi2014Wireless}
S.~Movassaghi, M.~Abolhasan, J.~Lipman, and D.~Smith, ``Wireless body area networks: a survey,'' \emph{IEEE Communications Surveys \& Tutorials}, vol. 16, issue 3, pp. 1658-1686, Jan. 2014.

\bibitem{Kurs2007Wireless}
A.~Kurs, A.~Karalis, R.~Moffatt, J.~D.~Joannopoulos, P.~Fisher, and M.~Soljacic, ``Wireless power transfer via strongly coupled magnetic resonances,'' \emph{Science}, vol. 317, no. 5834, pp. 83-86, Jun. 2007

\bibitem{CVNI2014}
``Global Mobile Data Traffic Forecast Update 2013-2018,'' Cisco Visual Networking Index, Feb. 2014. 




\bibitem{H.1997Nachbar}
J. H. Nachbar, ``Prediction, optimization, and learning in repeated games," \emph{Econometrica: Journal of the Econometric Society}, vol. 65, no. 2, pp. 275-309, 1997.



\end{thebibliography}
\end{document}